\documentclass{aa}  

\usepackage{graphicx}
\usepackage{txfonts}
\begin{document}

   \title{The $^{12}$CO/$^{13}$CO ratio in AGB stars of different chemical type}

   \subtitle{Connection to the $^{12}$C/$^{13}$C ratio and the evolution along the AGB}

   \author{S. Ramstedt
          \inst{1}
          \and
          H. Olofsson\inst{2}}

   \institute{Department of Physics and Astronomy, Uppsala University,
              Box 516, SE-751 20 Uppsala, Sweden\\
              \email{sofia.ramstedt@physics.uu.se}
         \and
             Onsala Space Observatory, Dept. of Earth and Space Sciences, Chalmers University of Technology, SE-43992 Onsala, Sweden \\
             }

   \date{Received ; accepted }

  \abstract
   {}
   {The aim of this paper is to investigate the evolution of the $^{12}$C/$^{13}$C ratio along the AGB through the circumstellar $^{12}$CO/$^{13}$CO abundance ratio. This is the first time a sample including a significant number of M- and S-type stars is analysed together with a carbon-star sample of equal size, making it possible to investigate trends among the different types and establish evolutionary effects.}
   {The circumstellar $^{12}$CO/$^{13}$CO abundance ratios are estimated through a detailed radiative transfer analysis of single-dish radio line emission observations. Several different transitions have been observed for each source to ensure that a large extent of the circumstellar envelope is probed and the radiative transfer model is well constrained. The radiative transfer model is based on the Monte Carlo method and has been benchmarked against a set of similar codes. It assumes that the radiation field is non-local and solves the statistical equilibrium equations in full non-LTE. The energy balance equation, determining the gas temperature distribution, is solved self-consistently, and the effects of thermal dust radiation (as estimated from the spectral energy distribution) are taken into account. First, the $^{12}$CO radiative transfer is solved, assuming an abundance (dependent on the chemical type of the star), to give the physical parameters of the gas, i.e. mass-loss rate, $\dot{M}$, gas expansion velocity, $\upsilon_{\rm{e}}$, and gas temperature distribution. Then, the $^{13}$CO radiative transfer is solved using the results of the $^{12}$CO model giving the $^{13}$CO abundance. Finally, the $^{12}$CO/$^{13}$CO abundance ratio is calculated.}
   {The circumstellar $^{12}$CO/$^{13}$CO abundance ratio differs between the three spectral types. This is consistent with what is expected from stellar evolutionary models assuming that the spectral types constitute an evolutionary sequence; however, this is the first time this has been shown observationally for a relatively large sample covering all three spectral types. The median value of the $^{13}$CO abundance in the inner circumstellar envelope is 1.6$\times$10$^{-5}$, 2.3$\times$10$^{-5}$, and 3.0$\times$10$^{-5}$ for the M-type, S-type, and carbon stars of the sample, respectively, corresponding to $^{12}$CO/$^{13}$CO abundance ratios of 13, 26, and 34, respectively. The spread in the $^{13}$CO abundance, quantified by the ratio between the 90th and 10th percentile, is 4, 3, and 15 for the M-type, S-type, and carbon stars, respectively. Interestingly, the abundance ratio spread of the carbon stars is much larger than for the M- and S-type stars, even when excluding J-type carbon stars, in line with what could be expected from evolution on the AGB. We find no correlation between the isotopologue ratio and the mass-loss rate, as would be expected if both increase as the star evolves.}
   {}


   \maketitle
%

\section{Introduction}
The evolution of low- and intermediate-mass stars on the Asymptotic Giant Branch (AGB) is characterized by an intense mass loss. The stellar wind creates a circumstellar envelope (CSE), that carries gas and dust from the star into the interstellar medium, hence these stars contribute to the chemical evolution of and the dust production in the universe \citep[see e.g.][]{matsetal09}. The molecular setup, and also the grain types, in the CSEs of AGB stars are to a large extent determined by the C/O ratio of the central star. Depending on the C/O ratio, the AGB stars are divided into three different spectral types: the M-type stars, with C/O$<$1, the S-type stars, with C/O$\approx$1, and the carbon stars, with C/O$>$1. 

Chemical evolution models, based on calculated stellar yields \citep[e.g.][]{kara10} and assumptions about initial stellar mass functions and star formation histories, predict the evolution of individual elements and isotopic ratios through the combined action of different stellar types and populations \citep[e.g.][]{kobaetal11}. It is for instance concluded that the $^{12}$C/$^{13}$C ratio evolves as a consequence of nucleosynthesis in AGB stars. There is also observational evidence that the ejecta from AGB stars dominate the evolution of the $^{12}$C/$^{13}$C ratio in the local interstellar medium (ISM) \citep{greaholl97}. 

Therefore, measurements of the $^{12}$C/$^{13}$C ratio can be used to trace the past star formation history and stellar mass function. However, to do so, the expected change needs to be constrained, and thus it is important to know the $^{12}$C/$^{13}$C ratio in the stellar sources and constrain the change as the stars ascend the AGB. Presently, there is good evidence that the evolution on the first red giant branch (RGB) will lower the $^{12}$C/$^{13}$C ratio considerably, in particular for the lower-mass stars \citep{tsui07}. In fact, the ratio is so low that non-standard processes must be invoked to explain the observational results, e.g. cool bottom processing \citep{bootsack99}. The subsequent evolution on the AGB will gradually increase the $^{12}$C/$^{13}$C ratio, at least for the carbon-rich AGB stars \citep{lambetal86}. 

Furthermore, the evolution beyond the AGB is not well understood. The $^{12}$C/$^{13}$C ratio derived for planetary nebulae (PNe) do not always agree with the ratios measured in AGB stars \citep{balsetal02,gustwahl06}. Measured $^{12}$C/$^{13}$C ratios of a large sample of AGB stars with different chemistries and mass-loss rates could bring us closer to answering the question of which stars progress to become PNe, and which do not.

Previous estimates of the $^{12}$C/$^{13}$C ratio in AGB stars have been made, mainly for carbon stars, both through measurements of photospheric molecular near-IR lines \citep{lambetal86,ohnatsuj96} and circumstellar CO radio lines \citep[e.g.][]{knapchan85,schoolof00,woodetal03}. From their detailed study of a carbon star sample using near-IR molecular lines, \citet{lambetal86} concluded that carbon stars have large $^{12}$C/$^{13}$C ratios, typically in the range $30-70$, with a few notable exceptions with ratios as low as $\approx$4 (classified as J-type stars). The ratios obtained by Lambert et al. were questioned by \citet{ohnatsuj96} who derived lower ratios for the same sample of stars, by about a factor of two. Based on optical data \citet{abiaiser97} estimated $^{12}$C/$^{13}$C ratios in the range $20-35$ for the majority of their 44 carbon stars. \citet{greaholl97} estimated $^{12}$C/$^{13}$C ratios in the range $12-57$ for 10 carbon stars stars using the circumstellar CO($J$\,=\,$2-1$) line. \citet{schoolof00} used several circumstellar CO radio lines and a radiative transfer model to estimate $^{12}$C/$^{13}$C ratios for a sample of carbon stars, the ratios fall in the range $20-90$ (the J-type stars excluded). They also compared their results to the photospheric ratios reported by \citet{lambetal86} for the same stars and found a good correlation, indicating that the circumstellar $^{12}$CO/$^{13}$CO ratio accurately measures the stellar $^{12}$C/$^{13}$C ratio. In addition, they concluded that a detailed non-LTE radiative transfer treatment is necessary to obtain reliable isotopologue ratios (for carbon stars) from circumstellar CO lines, mainly due to optical depth effects in the $^{12}$CO lines. 

Estimates of the $^{12}$C/$^{13}$C ratio for M-type AGB stars exist for only a limited number of stars. \citet{knapchan85} detected circumstellar $^{13}$CO in nine M-type AGB stars and they all appear to have $^{12}$C/$^{13}$C\,$<$\,20.  More recently, \citet{milaetal09} estimated $^{12}$C/$^{13}$C ratios in the range 10\,--\,35 for four M-type AGB stars using circumstellar $J$\,=\,1\,--\,0 and 2\,--\,1 data for $^{12}$CO and $^{13}$CO. Interestingly, \citet{delfetal97} estimated remarkably low$^{12}$C/$^{13}$C ratios, in the range 3\,--\,4, for four OH/IR stars using the same lines, suggesting that these are massive stars where the hot bottom burning process has converted $^{12}$C into $^{14}$N \citep{bootetal93}.

For S-type AGB stars the study by \citet{walletal11} was an important step forward. They obtained $^{12}$C/$^{13}$C ratios of $35\pm6$ for seven S-stars using circumstellar $J$\,=\,$1-0$ and $2-1$ data for $^{12}$CO and $^{13}$CO. Comparison with the results for the M-stars and carbon stars suggests that, in terms of $^{12}$C/$^{13}$C ratio, the S-stars are more similar to the carbon stars. 

In this paper we present a major study of comparable samples of M- and S-type AGB stars and carbon stars. Compared to previous publications performing similar analysis, this paper covers a larger sample including all chemical types. Several different rotational transitions are observed for each CO isotopologue resulting in a comprehensive dataset and a large coverage of the circumstellar envelopes, which are modelled using detailed radiative transfer. Circumstellar $^{12}$CO and $^{13}$CO lines up to $J$\,=\,$6-5$, combined with a radiative transfer analysis, are used to derive circumstellar $^{12}$CO/$^{13}$CO ratios. The sample and new observational data, as well as archival data, are presented in Sect.~\ref{s:obs}. The radiative transfer modelling and the results are presented in Sects~\ref{s:radtrans} and \ref{s:results}, respectively. The results, their uncertainties, and their implications for stellar and chemical evolution are discussed in Sect.~\ref{s:disc}. 


\section{Observations}
\label{s:obs}

\subsection{The sample and distances}

The sample of AGB stars is presented in Table~\ref{sample} with spectral and variable type (V), period ($P$), stellar luminosity ($L_{\star}$), parallax ($p$), and distance estimate ($D$), for each star. The stars are selected from the samples of \citet[][carbon stars]{schoolof01}, \citet[][M-type stars]{delgetal03}, and \citet[][S-type stars]{ramsetal06} and contains 19 M-, 17 S-, and 19 carbon stars on the AGB. The purpose is to have statistically relevant samples from the three different chemical types covering a range in wind properties, i.e. mass-loss rate and wind velocity, in order to avoid strong selection effects. Stars with CSEs that are known to exhibit strong asymmetries, and known detached shell sources are not included in the analysis. The binary fraction of the sample is not known. A handful of the stars have been suggested to have a binary companion, e.g.~TX~Cam \citep{castetal10}, RW~LMi \citep{monnetal00}, and AFGL~3068 \citep{maurhugg06}, but W~Aql is, to our knowledge, the only source in the sample where the presence of a binary companion has been confirmed \citep{ramsetal11}.

The distances are estimated from parallax measurements when a reliable measurement is available. For most of these stars, we use the re-calculated Hipparcos parallax from \citet{vanleeuw07}. When more precise estimates are available, i.e. from VLBI maser spot astrometry, these have been used. References are given in Table~\ref{sample}. To estimate distances for Mira variables without reliable parallax estimates, the period-luminosity relation of \citet{groewhit96} has been used for all stars for consistency, together with the spectral energy distribution (SED, Sect.~\ref{dustobs}). For semi-regular and irregular variables, and when the variability type has not been determined, a bolometric luminosity of $L_{\star}$\,=\,4000\,L$_{\odot}$ has been assumed (applies to four stars as indicated by a colon after the luminosity in Table~\ref{sample}). In the end, the $^{12}$CO/$^{13}$CO ratio is rather independent of the estimated distances for the stars. Uncertainties in the distance estimates will therefore not introduce biases when comparing the different samples to each other.

\begin{table}
\caption{The sample sources, separated according to spectral type, with variable type (V), period ($P$), luminosity ($L_{\star}$), parallax ($p$), and distance ($D$). }  
\begin{minipage}{9cm}
\label{sample}      
\centering              
\resizebox{\hsize}{!}{   
\begin{tabular}{l c c c c c}       
\hline\hline                
Source & V & $P$ & $L_{\star}$ & $p$ & $D$ \\    
	& & [d] & [L$_{\odot}$] & [mas]  & [pc] \\
\hline                  
{\em M-type stars:} & & & & & \\
\object{RX~Boo} & SRb & 162\phantom{$^{e}$} & \phantom{1}4000\phantom{:} & \phantom{1}7.31$\pm$0.5\footnote{\citet{kameetal12}} & \phantom{1}137 \\
\object{TX~Cam} & M & 557\phantom{$^{e}$} & \phantom{1}8600\phantom{:} & $\cdots$ & \phantom{1}380 \\
\object{R~Cas} & M & 431\phantom{$^{e}$} & \phantom{1}4000\phantom{:} & \phantom{1}5.67$\pm$2.0\footnote{\citet{vlemetal03}} & \phantom{1}176 \\
\object{R~Dor} & SRb & 338\phantom{$^{e}$} & \phantom{1}4000\phantom{:} & 18.31$\pm$1.0\footnote{\citet{vanleeuw07}} & \phantom{11}55 \\
\object{W~Hya} & SRa & 382\phantom{$^{e}$} & \phantom{1}6000\phantom{:} & 10.18$\pm$2.4$^{b}$ & \phantom{11}98 \\
\object{R~Leo} & M & 313\phantom{$^{e}$} & \phantom{1}2500\phantom{:} & 14.03$\pm$2.7$^{c}$ & \phantom{11}71\\
\object{GX~Mon} & M & 527\phantom{$^{e}$} & \phantom{1}8200\phantom{:} & $\cdots$ & \phantom{1}550 \\
\object{WX~Psc} & M & 660\phantom{$^{e}$} & 10300\phantom{:} & $\cdots$ & \phantom{1}700 \\
\object{RT~Vir} & SRb & 155\phantom{$^{e}$} & \phantom{1}4500\phantom{:} & \phantom{1}7.38$\pm$0.8\footnote{\citet{imaietal03}} & \phantom{1}226 \\
\object{SW~Vir} & SRb & 150\phantom{$^{e}$} & \phantom{1}4000\phantom{:} & \phantom{1}6.99$\pm$0.8$^{c}$ & \phantom{1}143\\
\object{IK~Tau} & M & 500\phantom{$^{e}$} & \phantom{1}7700\phantom{:} & $\cdots$ & \phantom{1}260 \\
\object{CIT~4} & M & 534\footnote{\citet{joneetal90}} & \phantom{1}4000\phantom{:} & $\cdots$ & \phantom{1}800 \\ 
\object{IRC+10365} & M & 500\phantom{$^{e}$} & \phantom{1}7700\phantom{:} & $\cdots$ & \phantom{1}650 \\
\object{IRC-10529} & M & 680 & 10600\phantom{:} & $\cdots$ & \phantom{1}620 \\
\object{IRC-30398} & M & 575 & \phantom{1}8900\phantom{:} & $\cdots$ & \phantom{1}550 \\
\object{IRC+40004} & M & 660 & 10300\phantom{:} & $\cdots$ & \phantom{1}600 \\
\object{IRC+50137} & M & 635 & \phantom{1}9900\phantom{:} & $\cdots$ & 1500 \\
\object{IRC+60169} & $\cdots$ & $\cdots$ & \phantom{1}4000: & $\cdots$ & \phantom{1}400 \\
\object{IRC+70066} & $\cdots$ & $\cdots$ & \phantom{1}4000: & $\cdots$ & \phantom{1}400 \\     
\noalign{\smallskip}
{\em S-type stars:} & & & & & \\
\noalign{\smallskip}
\object{R~And} & M & 409 & \phantom{1}6300\phantom{:} & $\cdots$ & \phantom{1}350 \\
\object{W~Aql} & M & 490 & \phantom{1}7600\phantom{:} & $\cdots$ & \phantom{1}300 \\
\object{TV~Aur} & SRb & 182 & \phantom{1}4000: & $\cdots$ & \phantom{1}400 \\
\object{AA~Cam} & Lb & $\cdots$ & \phantom{1}6000\phantom{:} & \phantom{1}1.28$\pm$0.7$^{c}$ & \phantom{1}780 \\
\object{S~Cas} & M & 611 & \phantom{1}8000\phantom{:} & $\cdots$ & \phantom{1}570 \\
\object{TT~Cen} & M & 462 & \phantom{1}6500\phantom{:} & $\cdots$ & 1180 \\
\object{T~Cet} & SRb & 159 & \phantom{1}4000\phantom{:} & \phantom{1}3.70$\pm$0.47$^{c}$ & \phantom{1}270 \\
\object{R~Cyg} & M & 514 & \phantom{1}8000\phantom{:} & $\cdots$ & \phantom{1}600 \\				
\object{$\chi$~Cyg} & M & 407 & \phantom{1}6500\phantom{:} & \phantom{1}5.53$\pm$1.10$^{c}$ & \phantom{1}181\\
\object{R~Gem} & M & 370 & \phantom{1}5700\phantom{:} & $\cdots$ & \phantom{1}650 \\
\object{ST~Her} & SRb & 148 & \phantom{1}4000\phantom{:} & \phantom{1}3.41$\pm$0.59$^{c}$ & \phantom{1}293\\
\object{Y~Lyn} & SRc & 110 & \phantom{1}4000\phantom{:} & \phantom{1}3.95$\pm$0.95$^{c}$ & \phantom{1}253 \\
\object{S~Lyr} & M & 438 & \phantom{1}6700\phantom{:} & $\cdots$ & 2000 \\
\object{RT~Sco} & M & 449 & \phantom{1}6900\phantom{:} & $\cdots$ & \phantom{1}400 \\
\object{T~Sgr} & M & 392 & \phantom{1}6000\phantom{:} & $\cdots$ & \phantom{1}700 \\
\object{DK~Vul} & SRa & 370 & \phantom{1}4000: & $\cdots$ & \phantom{1}750 \\
\object{EP~Vul} & Lb & $\cdots$ & \phantom{1}4000: & $\cdots$ & \phantom{1}510 \\
\noalign{\smallskip}
{\em Carbon stars:} & & & & & \\
\noalign{\smallskip}
\object{LP~And} & M & 614 & \phantom{1}9600\phantom{:} & $\cdots$ & \phantom{1}630 \\
\object{V~Aql} & SRb & 353 & \phantom{1}6500\phantom{:} & 2.76$\pm$0.69$^{c}$ & \phantom{1}362\\
\object{RV~Aqr} & M & 454 & \phantom{1}7000\phantom{:} & $\cdots$ & \phantom{1}550 \\
\object{UU~Aur} & SRb & 235 & \phantom{1}4000\phantom{:} & $\cdots$ & \phantom{1}240 \\
\object{X~Cnc} & SRb & 170 & \phantom{1}4500\phantom{:} & 2.92$\pm$0.78$^{c}$ & \phantom{1}342 \\
\object{Y~CVn} & SRb & 158 & \phantom{1}5800\phantom{:} & 3.12$\pm$0.34$^{c}$ & \phantom{1}321 \\
\object{V~Cyg} & M & 421 & \phantom{1}6000\phantom{:} & 2.73$\pm$1.58$^{c}$ & \phantom{1}366\\
\object{RY~Dra} & SRb & 173 & \phantom{1}4500\phantom{:} & 2.32$\pm$0.59$^{c}$ & \phantom{1}431 \\
\object{UX~Dra} & SRb & 168 & \phantom{1}4000\phantom{:} & 2.59$\pm$0.29$^{c}$ & \phantom{1}386 \\
\object{U~Hya} & SRb & 450 & \phantom{1}4000\phantom{:} & 4.80$\pm$0.23$^{c}$ & \phantom{1}208 \\
\object{CW~Leo} & M & 630 & \phantom{1}9800\phantom{:} & $\cdots$ & \phantom{1}120 \\
\object{R~Lep} & M & 432 & \phantom{1}5500\phantom{:} & 2.42$\pm$1.02$^{c}$ & \phantom{1}413 \\
\object{RW~LMi} & M & 640 & 10000\phantom{:} & $\cdots$ & \phantom{1}400 \\
\object{T~Lyr} & Lb & $\cdots$ & \phantom{1}9000\phantom{:} & 1.39$\pm$0.49$^{c}$ & \phantom{1}719 \\
\object{W~Ori} & SRb & 212 & \phantom{1}7000\phantom{:} & 2.65$\pm$0.95$^{c}$ & \phantom{1}377 \\
\object{V384~Per} & M & 535 & \phantom{1}8300\phantom{:} & $\cdots$ & \phantom{1}600 \\
\object{AQ~Sgr} & SRb & 200 & \phantom{1}3000\phantom{:} & 3.00$\pm$0.67$^{c}$ & \phantom{1}333 \\ 
\object{AFGL~3068} & M & 696 & 10900\phantom{:} & $\cdots$ & 1300 \\
\object{IRAS~15194-5115} & M & 575 & \phantom{1}8900\phantom{:} & $\cdots$ & \phantom{1}500 \\
\hline
\hline                                   
\end{tabular}}
\end{minipage}
\end{table}

\subsection{Data from the literature}

\subsubsection{Dust continuum emission}
\label{dustobs}

Spectral energy distributions of the stars are constructed from continuum observations from the literature. For all sample stars, J, H, and K band data from 2MASS and the IRAS fluxes are used. No new continuum data has been added since previous publications on this sample \citep{schoolof01,delgetal03,ramsetal08,ramsetal09} and more details can be found there.

\subsubsection{$^{12}$CO radio line observations}

The $^{12}$CO analysis is mainly based on previously published data. All references are given in Tables~\ref{12CO1}.1--4. For the S-type stars, a comparison between the IRAM 30\,m line intensities published in \citet{walletal11} and in \citet{ramsetal09} revealed a mistake in the scaling of the data in the latter publication. This has been corrected for all the IRAM 30\,m data used in the $^{12}$CO modelling of the S-type stars. Since the analysis in \citet{ramsetal09} is based on several different lines for each star, and since a good fit can be found for a rather broad range of input parameters \citep{ramsetal08}, the resultant physical parameters, i.e. $\dot{M}$ and $T_{\rm{kin}}(r)$, are only slightly affected by the re-scaling of the data. No transitions higher than $J$\,=\,6$\rightarrow$5 are included, and therefore a constant expansion velocity is assumed in the circumstellar model (Sect.~\ref{circmod}).

\subsubsection{$^{13}$CO radio line observations}

For the carbon stars, the $^{13}$CO data already published and analysed in \citet{schoolof00} is re-analysed in this paper, with some new data added for a handful of sources (see Table~\ref{13CO1}.3--4 for references). The re-analysis is motivated by substantial updates to the radiative transfer code since 2000, e.g. the inclusion of thermal dust emission based on the dust temperature structure and updated CO-H$_{2}$ collisional rates. For the M-type and S-type stars, the analysis is based on a large set of new data (see Sect.~\ref{newdata}). Additional data included in the analysis are from \citet[][M-type]{debeetal10,justetal12}, and the IRAM 30\,m data published in \citet[][S-type]{walletal11}. 

\subsection{New observations of $^{13}$CO radio line emission}

\label{newdata}
New observations of the $^{13}$CO $J$\,=\,$1\rightarrow0$ were performed at the Onsala 20\,m telescope (OSO), during 2007, of the brightest stars of all three chemical types in our sample. Furthermore, the $J$\,=\,$2\rightarrow1$ and $3\rightarrow2$ line emission was observed at APEX during several runs in 2009-2011, and at the JCMT during the same period. The relevant telescope data for the different observed transitions are given in Table~\ref{tele_table}.

\begin{table}
\caption{Telescope data relevant for the new observations of $^{13}$CO.}
\begin{minipage}{9cm}
\resizebox{\hsize}{!}{  
\label{tele_table}
$
\begin{array}{cccccccccccc}
\hline \hline
\noalign{\smallskip}
\multicolumn{1}{c}{{\mathrm{Transition}}} & &
\multicolumn{1}{c}{{\mathrm{Frequency}}} &  &
\multicolumn{1}{c}{{E_{\mathrm{up}}}} &  &
\multicolumn{1}{c}{{\mathrm{Telescope}}}  &&
\multicolumn{1}{c}{\eta_{\mathrm{mb}}}  && 
\multicolumn{1}{c}{\theta_{\mathrm{mb}}} \\ 
& & 
\multicolumn{1}{c}{{\mathrm{[GHz]}}} && 
\multicolumn{1}{c}{{\mathrm{[K]}}} &&  
&& &&
\multicolumn{1}{c}{[\arcsec]} \\
\noalign{\smallskip}
\hline
\noalign{\smallskip}
J=1\rightarrow0 && 110.201 && \phantom{0}5 && \mathrm{OSO} && 0.50 && 34 \\
J=2\rightarrow1 && 220.399 && 16 && \mathrm{APEX} && 0.75 && 28 \\  
  && && && \mathrm{JCMT} && 0.69 && 21 \\
J=3\rightarrow2 && 330.588 && 32 && \mathrm{APEX} && 0.73 && 19 \\
\noalign{\smallskip}
\hline \hline
\end{array}
$
}
\end{minipage}
\end{table}

Since the sources are not particularly extended, all observations were performed in dual beamswitch mode, where the source is placed alternately in the signal and in the reference beam, to attain flat baselines. Beam throws of 11{\arcmin}, 3{\arcmin}, and 2{\arcmin} at OSO, APEX, and JCMT, respectively, are used, which is sufficient to move off source. The pointing was checked regularly on stellar SiO masers (sometimes the source; OSO) and on strong CO and continuum sources (APEX, JCMT). Typically, the pointing was found to be consistent with the pointing model of the telescope to within $\approx$3\arcsec. The OSO and APEX receivers used for the observations are single sideband, while the JCMT observations were performed with a double sideband receiver.

Calibration to correct for the atmospheric attenuation, is performed at the telescope (OSO, JCMT) using the chopper-wheel method, and the spectra are collected in $T_{\rm{A}}^{\star}$-scale. APEX heterodyne data are calibrated regularly through a three-stage observation. In the first two stages, hot and cold load measurements are done to determine the receiver temperature. Finally a sky observation is made to determine the correction due to the atmospheric attenuation by adopting the model by \citet{pardetal01}. Conversion to main-beam-brightness-temperature scale ($T_{\rm{mb}}$) (for easier comparison to data collected at other sites) is done using $T_{\rm{mb}}=T_{\rm{A}}^{\star}$/$\eta_{\rm{mb}}$. $\eta_{\rm{mb}}$ is the adopted main-beam efficiency given in Table~\ref{tele_table}. The uncertainty in the calibrated spectra is estimated to be about $\pm20$\%, except when the signal-to-noise is very low and a slightly higher uncertainty ($\sim$\,30\%) is adopted. The upper-level energies of the transitions used in the analysis are also given in Table~\ref{tele_table}. With the $J\,=\,6\rightarrow5$ data from the literature, it ranges from 5 to 111\,K demonstrating that a large physical range of the CSE is probed by the analysis.

The data were reduced using XS\footnote{XS is a software package developed by P.~Bergman to reduce and analyse single-dish spectra. It is publicly available from: {\tt ftp://yggdrasil.oso.chalmers.se}} (OSO, APEX) and {\em SPLAT-VO} (JCMT). The individual scans were weighted with the system temperature and averaged. A low order (typically first) polynomial baseline was subtracted and the data were binned (typically to a velocity resolution of 1\,km\,s$^{-1}$) to improve the signal-to-noise ratio. The observed spectra are presented in Figs.~\ref{sp1}-\ref{sp3} and the peak and velocity-integrated line intensities are reported in Table~\ref{13CO1}.5--6 in $T_{\rm{mb}}$-scale.


\section{Radiative transfer modelling}
\label{s:radtrans}

\subsection{Modelling the circumstellar envelope}
\label{circmod}

The assumptions and details of the dust and $^{12}$CO radiative transfer modelling are described in previous publications \citep[see e.g.][and references therein]{schoetal13}. The CSE is assumed to be spherically symmetric and formed by a constant mass-loss rate. It is assumed to be expanding at a constant velocity, derived from fitting the $^{12}$CO line widths. The dust radiative transfer is performed using DUSTY\footnote{Ivezic, Z., Nenkova, M. \& Elitzur, M., 1999, User Manual for DUSTY, University of Kentucky Internal Report, accessible at: {\tt http://www.pa.uky.edu/$\sim$moshe/dusty}} and it provides the stellar temperature, the dust optical depth, and the radial dust temperature distribution.  Amorphous carbon grains \citep{suh00} and amorphous silicate grains \citep{justtiel92} are adopted for the carbon and the M- and S-type stars, respectively. The dust grains are assumed to be of the same size with a radius of 0.1\,$\mu$m, and have a density of 2\,g\,cm$^{-3}$ and 3\,g\,cm$^{-3}$ for the carbon and silicate grains, respectively. The dust temperature distribution is used as input to the $^{12}$CO radiative transfer modelling to calculate the radiation field due to thermal dust emission; however as discussed in Sect.~\ref{input}, the CO excitation is not strongly affected by the dust emission. Both the $^{12}$CO and $^{13}$CO radiative transfer modelling is performed with the non-LTE, non-local Monte Carlo code presented in \citet{schoolof01} and successfully benchmarked against other codes in \citet{zadeetal02}. 

\subsection{$^{12}$CO line modelling}
\label{12co}

Since the analysis is limited to $J$-levels below $J$=7, the expansion velocity is assumed to be constant across the region probed by the observations. With a constant mass-loss rate and expansion velocity, the density declines smoothly with $r^{-2}$. The abundance distribution of $^{12}$CO is based on the photochemical modelling of \citet{mamoetal88} adopting an initial photospheric abundance of $1\times10^{-3}$, $6\times10^{-4}$, and $2\times10^{-4}$ for carbon stars, S-type, and M-type stars, respectively. The kinetic temperature structure is calculated self-consistently by solving the energy balance equation. The gas is mainly heated through collisions with dust grains and cooled by line emission from CO (directly obtained from the excitation analysis) and H$_{2}$. Cooling due to the (adiabatic) expansion of the gas is also included. This results in a smoothly varying temperature distribution. 

The excitation analysis includes the first 41 rotational levels within the ground ($\nu$=0) and first ($\nu$=1) vibrationally excited states. The CO-H$_{2}$ collisional rates are taken from \citet{yangetal10}. An ortho-to-para ratio of H$_{2}$ of 3 was adopted when weighting together collisional rate coefficients for CO in collisions with ortho-H$_{2}$ and para-H$_{2}$. The collisional rate coefficients have been extrapolated to include energy levels up to $J$\,=\,41 and temperatures up to 3000\,K as described in \citet{schoetal05}.  Collisional excitation between the $\nu$=0 and 1 states, and within the $\nu$=1 state, can be neglected due to the fast radiative de-excitation from $\nu$=1 to 0. The molecular data files are available from the Leiden Atomic and Molecular Database (LAMBDA)\footnote{{\tt home.strw.leidenuniv.nl/$\sim$moldata/}}.

The mass-loss rate and the so-called $h$-parameter \citep[which determines the gas heating due to collisions with dust grains, see e.g.][for a definition]{schoolof01}, are free parameters in the $^{12}$CO modelling, which provides the physical parameters of the gas, i.e. the density and temperature structure. This is used as input to the $^{13}$CO model, where the $^{13}$CO abundance is the only free parameter.

\subsection{$^{13}$CO line modelling}

The basic physical parameters of the CSE, i.e. its density and temperature structure, are determined by the $^{12}$CO radiative transfer analysis and they are used as input to the model of the $^{13}$CO line emission. In this model the abundance distribution of the $^{13}$CO molecules (relative to H$_{2}$; $f(r)$) is given by
\begin{equation}
f(r) = f_{0}\, \mathrm{exp} \left[ -\mathrm{ln}\,2 \left( \frac{r}{r_{\rm{p}}} \right) ^{\alpha} \right],
\end{equation}
where $f_{0}$ is the initial photospheric abundance and $r_{\rm{p}}$ is the photodissociation radius (i.e. $f(r_{\rm{p}})$\,=\,$f_{0}/2$). The parameters $r_{\rm{p}}$ and $\alpha$ are assumed to be the same as for the $^{12}$CO envelope, based on the modelling of \citet{mamoetal88} and given by Eqs 9-11 in \citet{schoolof01}. The excitation properties of $^{13}$CO are included and derived in the same way as for $^{12}$CO (see Sect.~\ref{12co}). Due to the lower optical depth of the $^{13}$CO lines, the isotopologue is more readily photodissociated by interstellar radiation in the outer, cooler parts of the CSE. However, at temperatures below 35\,K, $^{12}$CO can produce $^{13}$CO from the chemical fractionation reaction $^{13}\mathrm{C}^{+}$+$^{12}\mathrm{CO}$ $\rightleftarrows$ $^{12}\mathrm{C}^{+}$+$^{13}\mathrm{CO}$, which somewhat compensates for the differential photodissociation. \citet{mamoetal88} concluded that the $^{12}$CO and $^{13}$CO abundance distributions always differ less than 20\% when tested over a large range in mass-loss rate. The parameter $f_{0}$ is varied until a satisfactory fit to the observed line profiles is found.  

\subsection{Finding the best-fit model}

For all three modelling steps involved in estimating the $^{12}$CO/$^{13}$CO ratio (i.e. the dust, $^{12}$CO, and ${13}$CO radiative transfer), the best-fit model is found by minimizing the total $\chi^{2}_{\mathrm{tot}}$ defined as
\begin{equation}
\label{chi2_sum}
\chi^2_{\mathrm{tot}} = \sum^N_{i=1} \left [ \frac{(I_{\mathrm{mod}}-I_{\mathrm{obs}})}{\sigma}\right ]^2, 
\end{equation}
where $I$ is the velocity-integrated line intensity (or flux density for the continuum measurements) of the model and observations, respectively, and $\sigma$ is the uncertainty in the measured value. This uncertainty is usually dominated by calibration uncertainties and around $\pm$20\%. In some cases, when the observed spectrum has a low signal-to-noise ratio, $\sigma$ is set to $\pm$30\% to take the added uncertainty into account. The summation is done over the $N$ independent observations. The reduced $\chi^2$ is given by
\begin{equation}
\chi^2_{\mathrm{red}}=\frac{\chi^2_{\mathrm{tot}}}{N-p},
\end{equation}
where $p$ is the number of parameters varied in the modelling. 


\section{Results}
\label{s:results}

\begin{table*}
\caption{Model results. See text for explanation.}
\label{results}
\centering
\resizebox{\hsize}{!}{   
\begin{tabular}{p{0.14\linewidth}cccccccccccccccccccr}
\hline \hline
\noalign{\smallskip}
& 
\multicolumn{8}{c}{SED modelling} & &
\multicolumn{5}{c}{$^{12}$CO modelling} & &
\multicolumn{3}{c}{$^{13}$CO modelling} & & \\
\noalign{\smallskip}
\cline{2-9}\cline{11-15}\cline{17-19}
\noalign{\smallskip}
\multicolumn{1}{c}{Source} &
\multicolumn{1}{c}{$D$} & 
\multicolumn{1}{c}{$L_{\star}$} &
\multicolumn{1}{c}{$T_{\star}$} &
\multicolumn{1}{c}{$\tau_{10}$} &
\multicolumn{1}{c}{$T_{\mathrm{d}}(r_{\mathrm{i}}$)} &
\multicolumn{1}{c}{$r_{\mathrm{i}}$} &
\multicolumn{1}{c}{$\chi^2_{\mathrm{red}}$} &
\multicolumn{1}{c}{$N$} & &
\multicolumn{1}{c}{$\dot{M}$} &
\multicolumn{1}{c}{$\upsilon_{\mathrm{e}}$} &
\multicolumn{1}{c}{$r_{\mathrm{p}}$} &
\multicolumn{1}{c}{$\chi^2_{\mathrm{red}}$} &
\multicolumn{1}{c}{$N$} & &
\multicolumn{1}{c}{$f_0$} & 
\multicolumn{1}{c}{$\chi^2_{\mathrm{red}}$} &
\multicolumn{1}{c}{$N$} &&
\multicolumn{1}{c}{$\frac{^{12}\mathrm{CO}}{^{13}\mathrm{CO}}$} \\
&
\multicolumn{1}{c}{[pc]}  &
\multicolumn{1}{c}{[L$_{\odot}$]} & 
\multicolumn{1}{c}{[K]} & &
\multicolumn{1}{c}{[K]} &
\multicolumn{1}{c}{[cm]} & & & &
\multicolumn{1}{c}{[M$_{\odot}$\,yr$^{-1}$]} &
\multicolumn{1}{c}{[km\,s$^{-1}$]} & \multicolumn{1}{c}{[cm]} & 
&&&
 & &  && \\
\noalign{\smallskip}
\hline
\noalign{\smallskip}
\em{M-type stars} & & &&&&&&&&&&&&&&&& && \\
\noalign{\smallskip}
RX Boo & \phantom{0}137 & \phantom{1}4000 & 2100 & 0.02\phantom{0} & \phantom{1}500 & 6.9$\times$10$^{14}$ & 0.4 & \phantom{1}7 && 7.0$\times$10$^{-7}$ & \phantom{1}9.5 & 3.0$\times$10$^{16}$ & 1.3 & 6  && 1.2$\times$10$^{-5}$ & 0.7 & 3 && 17 \\
TX Cam & \phantom{0}380 & \phantom{1}8600 & 2400 & 0.4\phantom{00} & \phantom{1}800 & 3.0$\times$10$^{14}$ & 0.6 & \phantom{1}8 && 5.5$\times$10$^{-6}$ & 18.5 & 7.7$\times$10$^{16}$ & 2.2 & 5 && 1.3$\times$10$^{-5}$  & 1.7 & 3 && 15\\
R Cas & \phantom{0}550 & \phantom{1}4000 & 1800 & 0.09\phantom{0} & \phantom{1}600 & 4.5$\times$10$^{14}$ & 0.8 & \phantom{1}7 && 9.0$\times$10$^{-7}$ & 11.5 & 3.3$\times$10$^{16}$ & 2.9 & 5 && 2.2$\times$10$^{-5}$ & $\cdots$ & 2 && \phantom{1}9 \\
R Dor & \phantom{00}55 & \phantom{1}4000 & 2100 & 0.05\phantom{0} & 1500 & 6.3$\times$10$^{13}$ & 0.6 & \phantom{1}7 && 1.6$\times$10$^{-7}$ & \phantom{1}6.0 & 1.6$\times$10$^{16}$ & 0.5 & 6 && 2.0$\times$10$^{-5}$ & 1.3 & 3 && 10 \\
W Hya & \phantom{00}98 & \phantom{1}6000 & 1800 & 0.08\phantom{0} & 1200 & 6.3$\times$10$^{13}$ & 0.5 & \phantom{1}7 && 1.5$\times$10$^{-7}$ & \phantom{1}7.0 & 1.5$\times$10$^{16}$ & 3.2 & 5 && 2.0$\times$10$^{-5}$ & $\cdots$ & 1 && 10\\
R Leo & \phantom{00}71 & \phantom{1}2500 & 2000 & 0.03\phantom{0} & 1200 & 1.3$\times$10$^{14}$ & 0.7 & \phantom{1}7 && 1.0$\times$10$^{-7}$ & \phantom{1}6.0 & 1.2$\times$10$^{16}$ & 0.6 & 3 && 3.2$\times$10$^{-5}$ & 2.5 & 3 && 6 \\
GX Mon & \phantom{0}650 & \phantom{1}8200 & 1800 & 0.5\phantom{00} & \phantom{1}500 & 6.8$\times$10$^{14}$ & 1.7 & \phantom{1}8 && 1.2$\times$10$^{-5}$ & 18.7 & 1.2$\times$10$^{17}$ & 0.7 & 4 && 1.8$\times$10$^{-5}$ & 2.5 & 3 && 11\\
WX Psc & \phantom{0}700 & 10300 & 1800 & 3.0\phantom{00} & \phantom{1}800 & 3.2$\times$10$^{14}$ & 2.7 & 11 && 4.0$\times$10$^{-5}$ & 19.3 & 2.5$\times$10$^{17}$ & 0.8 & 5 && 1.5$\times$10$^{-5}$ & 1.7 & 5 && 13\\
RT Vir & \phantom{0}226 & \phantom{1}4500 & 2000 & 0.09\phantom{0} & 1000 & 1.6$\times$10$^{14}$ & 1.0 & \phantom{1}7 && 4.5$\times$10$^{-7}$ & \phantom{1}7.8 & 2.5$\times$10$^{16}$ & 0.7 & 4 && 2.3$\times$10$^{-5}$ & 3.8 & 3 && \phantom{1}9 \\
SW Vir & \phantom{0}143 & \phantom{1}4000 & 2400 & 0.03\phantom{0} & \phantom{1}800 & 2.9$\times$10$^{14}$ & 0.6 & \phantom{1}7 && 5.0$\times$10$^{-7}$ & \phantom{1}7.5 & 2.7$\times$10$^{16}$ & 1.4 & 6 && 1.1$\times$10$^{-5}$ & $\cdots$ & 1 && 18 \\
IK Tau & \phantom{0}260 & \phantom{1}7700 & 2100 & 1.0\phantom{00} & 1000 & 1.8$\times$10$^{14}$ & 0.7 & 13 && 2.0$\times$10$^{-5}$ & 19.0 & 1.6$\times$10$^{17}$ & 0.5 & 5 && 2.1$\times$10$^{-5}$ & 3.3 & 3 && 10\\
CIT4 & \phantom{0}800 & \phantom{1}8300 & 2300 & 1.0\phantom{00} & \phantom{1}900 & 2.4$\times$10$^{14}$ & 0.3 & \phantom{1}7 && 1.4$\times$10$^{-5}$ & 19.0 & 1.3$\times$10$^{17}$ & $\cdots$ & 1 && 7.0$\times$10$^{-6}$ & $\cdots$ & 1 && 29\\
IRC+10365 & \phantom{0}650 & \phantom{1}7700 & 1800 & 0.9\phantom{00} & \phantom{1}800 & 2.7$\times$10$^{14}$ & 0.9 & \phantom{1}7 && 1.0$\times$10$^{-5}$ & 16.0 & 1.2$\times$10$^{17}$ & 0.9 & 3 && 1.6$\times$10$^{-5}$ & 3.3 & 3 && 13\\
IRC-10529 & \phantom{0}620 & 10600 & 1800 & 3.0\phantom{00} & \phantom{1}900 & 2.5$\times$10$^{14}$ & 3.7 & \phantom{1}9 && 1.0$\times$10$^{-5}$ & 14.0 &  1.2$\times$10$^{17}$ & 2.6 & 4 && 3.0$\times$10$^{-5}$ & $\cdots$ & 2 && \phantom{1}7\\
IRC-30398 & \phantom{0}550 & \phantom{1}8900 & 1800 & 0.5\phantom{00} & \phantom{1}800 & 2.6$\times$10$^{14}$ & 1.0 & \phantom{1}4 && 8.0$\times$10$^{-6}$ & 16.0 & 1.0$\times$10$^{17}$ & $\cdots$ & 1 && 1.6$\times$10$^{-5}$ & $\cdots$ & 1 && 13 \\
IRC+40004 & \phantom{0}600 & 10300 & 1800 & 0.2\phantom{00} & \phantom{1}500 & 6.5$\times$10$^{14}$ & 0.4 & \phantom{1}4 && 1.0$\times$10$^{-5}$ & 18.0 & 1.1$\times$10$^{17}$ & $\cdots$ & 2 && 1.0$\times$10$^{-5}$ & $\cdots$ & 1 && 20 \\
IRC+50137 & 1500 & \phantom{1}9900 & 1900 & 3.0\phantom{00} & \phantom{1}700 & 4.3$\times$10$^{14}$ & 0.7 & \phantom{1}7 && 3.0$\times$10$^{-5}$ & 18.5 & 2.1$\times$10$^{17}$ & $\cdots$ & 2 && 3.5$\times$10$^{-5}$ & $\cdots$ & 1 && \phantom{1}6\\
IRC+60169 & \phantom{1}400 & \phantom{1}4000 & 1800 & 0.3\phantom{00} & \phantom{1}500 & 6.6$\times$10$^{14}$ & 1.1 & \phantom{1}7 && 2.3$\times$10$^{-5}$ & 15.0 & 2.0$\times$10$^{17}$ & $\cdots$ & 2 && 7.0$\times$10$^{-6}$ & $\cdots$ & 1 && 29 \\
IRC+70066 & \phantom{0}400 & \phantom{1}4000 & 3000 & 1.0\phantom{00} & \phantom{1}500 & 1.4$\times$10$^{14}$ & 1.1 & \phantom{1}7 && 1.5$\times$10$^{-5}$ & 18.0 & 1.4$\times$10$^{17}$ & $\cdots$ & 2 && 3.0$\times$10$^{-6}$ & $\cdots$ & 1 && 66 \\
\noalign{\smallskip}
\em{S-type stars} & & &&&&&&&&&&&&&&&& && \\
\noalign{\smallskip}
R And & \phantom{0}350 & \phantom{1}6300 & 1900 & 0.05\phantom{0} & \phantom{1}600 & 4.6$\times$10$^{14}$ & 0.5 & \phantom{1}7 && 8.0$\times$10$^{-7}$ & \phantom{1}8.3 & 5.9$\times$10$^{16}$ & 0.7 & 5  && 2.5$\times$10$^{-5}$ & 1.2 & 3 && \phantom{1}24\\
W Aql & \phantom{0}300 & \phantom{1}7600 & 2400 & 0.5\phantom{00} & 1100 & 1.5$\times$10$^{14}$ & 0.3 & \phantom{1}7 && 2.7$\times$10$^{-6}$ & 17.2 & 9.0$\times$10$^{16}$ & 0.9 & 5  && 2.3$\times$10$^{-5}$ & 1.2 & 5 && \phantom{1}26\\
TV Aur & \phantom{0}400 & \phantom{1}4000 & 2700 & $\cdots$ & $\cdots$ & 1.0$\times$10$^{14}$ & 1.6 & \phantom{1}6 && 2.3$\times$10$^{-8}$ & \phantom{1}5.5 & 9.6$\times$10$^{15}$ & $\cdots$ & 2 && 2.7$\times$10$^{-4}$ & $\cdots$ & 2 && \phantom{0}$>$2 \\
AA Cam & \phantom{0}780 & \phantom{1}6000 & 3000 & $\cdots$ & $\cdots$ & 1.0$\times$10$^{14}$ & 2.9 & \phantom{1}6 && 5.0$\times$10$^{-8}$ & \phantom{1}5.0 & 1.5$\times$10$^{16}$ & 7.9 & 3 && 8.0$\times$10$^{-5}$ & $\cdots$ & 2 && \phantom{0}$>$8 \\
S Cas & \phantom{0}570 & \phantom{1}8000 & 1800 & 0.5\phantom{00} & 1100 & 3.4$\times$10$^{14}$ & 0.8 & \phantom{1}7 && 4.0$\times$10$^{-6}$ & 20.5 & 1.1$\times$10$^{17}$ & 0.2 & 4 && 8.5$\times$10$^{-6}$ & $\cdots$ & 2 && \phantom{1}71\\
TT Cen & 1180 & \phantom{1}6500 & 1900 & $\cdots$ & $\cdots$ & 2.6$\times$10$^{14}$ & 0.9 & \phantom{1}5 && 4.0$\times$10$^{-6}$ & 20.0 & 1.1$\times$10$^{17}$ & 2.0 & 3 && 3.0$\times$10$^{-5}$ & $\cdots$ & 1 && \phantom{1}20\\
T Cet & \phantom{0}270 & \phantom{1}5000 & 2400 & $\cdots$ & $\cdots$ & 1.4$\times$10$^{14}$ & 0.8 & \phantom{1}7 && 6.0$\times$10$^{-8}$ & \phantom{1}7.0 &1.5$\times$10$^{16}$ & 1.6 & 7 && 6.0$\times$10$^{-5}$ & 3.6 & 3 && 10\\
R Cyg & \phantom{0}600 & \phantom{1}8000 & 1900 & $\cdots$ & $\cdots$ & 1.8$\times$10$^{14}$ & 0.9 & \phantom{1}7 && 8.3$\times$10$^{-7}$ & \phantom{1}9.0 & 5.9$\times$10$^{16}$ & 1.4 & 5 && 2.3$\times$10$^{-5}$ & 3.9 & 3 && \phantom{1}26\\
$\chi$ Cyg & \phantom{0}180 & \phantom{1}6500 & 2200 & 0.2\phantom{00} & 1500 & 6.5$\times$10$^{13}$ & 0.9 & \phantom{1}7 && 6.0$\times$10$^{-7}$ & \phantom{1}8.5 & 4.9$\times$10$^{16}$ & 0.4 & 6 && 1.5$\times$10$^{-5}$ & 1.6 & 4 && \phantom{1}40\\
R Gem & \phantom{0}650 & \phantom{1}5700 & 3000 & $\cdots$ & $\cdots$ & 9.8$\times$10$^{13}$ & 1.6 & \phantom{1}6 && 3.5$\times$10$^{-7}$ & \phantom{1}4.5 & 4.7$\times$10$^{16}$ & 1.1 & 4 && 2.7$\times$10$^{-5}$ & $\cdots$ & 1 && \phantom{1}22\\
ST Her & \phantom{0}290 & \phantom{1}4000 & 2100 & 0.03\phantom{0} & \phantom{1}600 & 4.8$\times$10$^{14}$ & 0.8 & \phantom{1}7 && 1.3$\times$10$^{-7}$ & \phantom{1}8.5 & 2.1$\times$10$^{16}$ & 2.0 & 4 && 4.0$\times$10$^{-5}$ & $\cdots$ & 2 && \phantom{1}15\\
Y Lyn & \phantom{0}250 & \phantom{1}4000 & 2700 & $\cdots$ & $\cdots$ & 1.0$\times$10$^{14}$ & 0.8 & \phantom{1}7 && 1.8$\times$10$^{-7}$ & \phantom{1}7.5 & 2.6$\times$10$^{16}$ & 2.2 & 3 && 2.2$\times$10$^{-5}$ & 6.0 & 3 && \phantom{1}27\\
S Lyr & 2000 & \phantom{1}6700 & 2200 & 0.4\phantom{00} & \phantom{1}700 & 3.8$\times$10$^{14}$ & 1.2 & \phantom{1}7 && 3.5$\times$10$^{-6}$ & 13.0 & 1.2$\times$10$^{17}$ & $\cdots$ & 2 && 3.5$\times$10$^{-5}$ & $\cdots$ & 1 && \phantom{1}17\\
RT Sco & \phantom{0}400 & \phantom{1}6900 & 2100 & $\cdots$ & $\cdots$ & 1.9$\times$10$^{15}$ & 0.8 & \phantom{1}6 && 7.5$\times$10$^{-7}$ & 11.0 &  5.1$\times$10$^{16}$ & $\cdots$ & 2 && 2.0$\times$10$^{-5}$ & $\cdots$ & 1 && \phantom{1}30\\
T Sgr & \phantom{0}700 & \phantom{1}6000 & 2200 & $\cdots$ & $\cdots$ & 1.8$\times$10$^{14}$ & 0.7 & \phantom{1}6 && 1.7$\times$10$^{-7}$ & \phantom{1}7.5 &  2.5$\times$10$^{16}$ & 3.5 & 3 && 8.0$\times$10$^{-5}$ & $\cdots$ & 2 && \phantom{0}$>$8 \\
DK Vul & \phantom{0}750 & \phantom{1}4000 & 2900 & $\cdots$ & $\cdots$ & 8.9$\times$10$^{13}$ & 3.7 & \phantom{1}6 && 2.0$\times$10$^{-7}$ & \phantom{1}4.5 & 3.4$\times$10$^{16}$ & 2.5 & 3 && 2.2$\times$10$^{-5}$  & $\cdots$ & 1 && \phantom{1}27\\
EP Vul & \phantom{0}510 & \phantom{1}4000 & 2800 & $\cdots$ & $\cdots$ & 9.5$\times$10$^{13}$ & 1.6 & \phantom{1}7 && 2.3$\times$10$^{-7}$ & \phantom{1}6.0 & 3.2$\times$10$^{16}$ & 3.2 & 3 && 2.2$\times$10$^{-5}$ & $\cdots$ & 1 && \phantom{1}27 \\
\noalign{\smallskip}
\em{Carbon stars} & & &&&&&&&&&&&&&&&& && \\
\noalign{\smallskip}
LP And & \phantom{0}630 & \phantom{1}9600 & 1900 & 0.6\phantom{00} & 1100 & 1.8$\times$10$^{14}$ & 0.4 & 11 && 7.0$\times$10$^{-6}$ & 14.0 & 2.3$\times$10$^{17}$ & 0.3 & 5 && 1.8$\times$10$^{-5}$ & 3.2 & 3 && 56 \\
V Aql & \phantom{0}362 & \phantom{1}6500 & 2400 & $\cdots$ & $\cdots$ & 1.6$\times$10$^{14}$ & 1.8 & 7 && 1.5$\times$10$^{-7}$ & \phantom{1}8.5 & 2.9$\times$10$^{16}$ & 1.2 & 5 && 7.0$\times$10$^{-5}$ & $\cdots$ & 2 && $>$14 \\
RV Aqr & \phantom{0}550 & \phantom{1}7000 & 1900 & 0.2\phantom{00} & 1400 & 8.4$\times$10$^{13}$ & 0.3 & 7 && 2.0$\times$10$^{-6}$ & 16.0 & 1.0$\times$10$^{17}$ & 0.4 & 3 && 5.0$\times$10$^{-5}$ & $\cdots$ & 1 && 20 \\
UU Aur & \phantom{0}240 & \phantom{1}4000 & 2600 & $\cdots$ & $\cdots$ & 1.1$\times$10$^{14}$ & 1.1 & 7 && 6.0$\times$10$^{-7}$ & 11.0 & 5.7$\times$10$^{16}$ & 1.9 & 4 && 1.0$\times$10$^{-5}$ & $\cdots$ & 1 && 100 \\
X Cnc & \phantom{0}342 & \phantom{1}4500 & 2800 & $\cdots$ & $\cdots$ & 1.0$\times$10$^{14}$ & 1.3 & 7 && 7.0$\times$10$^{-8}$ & \phantom{1}7.0 & 2.0$\times$10$^{16}$ & 4.0 & 3 && 1.6$\times$10$^{-4}$ & $\cdots$ & 2 && $>$6 \\
Y CVn & \phantom{0}321 & \phantom{1}5800 & 2000 & $\cdots$ & $\cdots$ & 2.2$\times$10$^{14}$ & 1.0 & 7 && 1.5$\times$10$^{-7}$ & \phantom{1}8.0 & 2.9$\times$10$^{16}$ & 0.2 & 4 && 5.5$\times$10$^{-4}$ & 12.0 & 3 && 2 \\
V Cyg & \phantom{0}366 & \phantom{1}6000 & 2300 & 0.2\phantom{00} & 1400 & 9.4$\times$10$^{13}$ & 0.1 & 7 && 1.6$\times$10$^{-6}$ & 11.5 & 1.0$\times$10$^{17}$ & 1.9 & 5 && 2.6$\times$10$^{-5}$ & $\cdots$ & 2 && 38 \\
RY Dra & \phantom{0}431 & \phantom{1}4500 & 2300 & $\cdots$ & $\cdots$ & 1.5$\times$10$^{14}$ & 1.0 & 7 && 2.0$\times$10$^{-7}$ & 10.0 & 3.2$\times$10$^{16}$ & 5.0 & 3 && 4.0$\times$10$^{-4}$ & $\cdots$ & 1 && 2.5 \\
UX Dra & \phantom{0}386 & \phantom{1}4000 & 2500 & $\cdots$ & $\cdots$ & 1.2$\times$10$^{14}$ & 0.5 & 6 && 4.0$\times$10$^{-8}$ & \phantom{1}4.0 & 1.8$\times$10$^{16}$ & $\cdots$  & 2 && 3.5$\times$10$^{-4}$ & $\cdots$ & 1 && 3.0 \\
U Hya & \phantom{0}208 & \phantom{1}4000 & 2900 & $\cdots$ & $\cdots$ & 8.8$\times$10$^{13}$ & 0.4 & 6 && 1.2$\times$10$^{-7}$ & \phantom{1}6.5 & 2.8$\times$10$^{16}$ & 1.2 & 5 && 6.0$\times$10$^{-5}$ & $\cdots$ & 2 && 17 \\
CW Leo & \phantom{0}120 & \phantom{1}9800 & 2600 & 1.0\phantom{00} & 1300 & 1.6$\times$10$^{14}$ & 1.4 & 11 && 1.5$\times$10$^{-5}$ & 14.5 & 3.7$\times$10$^{17}$ & 0.1 & 4 && 1.4$\times$10$^{-5}$ & 1.6 & 7 && 71 \\
R Lep & \phantom{0}432 & \phantom{1}5500 & 2200 & 0.06\phantom{0} & 1000 & 1.8$\times$10$^{14}$ & 0.5 & 7 && 7.0$\times$10$^{-7}$ & 18.0 & 5.3$\times$10$^{16}$ & 0.7 & 5 && 4.5$\times$10$^{-5}$ & $\cdots$ & 2 && 22 \\
RW LMi & \phantom{0}400 & 10000 & 1800 & 0.5\phantom{00} & 1000 & 2.1$\times$10$^{14}$ & 1.5 & 11 &&  6.0$\times$10$^{-6}$ & 17.0 & 1.9$\times$10$^{17}$ & 1.6 & 4 && 2.2$\times$10$^{-5}$ & 2.0 & 4 && 45 \\
T Lyr & \phantom{0}719 & \phantom{1}9000 & 1900 & $\cdots$ & $\cdots$ & 2.8$\times$10$^{14}$ & 0.8 & 7 && 1.5$\times$10$^{-7}$ & 11.5 & 2.6$\times$10$^{16}$ & 11.3 & 3 && 1.1$\times$10$^{-4}$ & $\cdots$ & 1 && 9 \\
W Ori & \phantom{0}377 & \phantom{1}7000 & 2400 & $\cdots$ & $\cdots$ & 1.7$\times$10$^{14}$ & 1.0 & 7 && 1.4$\times$10$^{-7}$ & 11.0 & 2.6$\times$10$^{16}$ & 2.8 & 4 && 1.3$\times$10$^{-4}$ & $\cdots$ & 2 && $>$8 \\
V384 Per & \phantom{0}600 & \phantom{1}8300 & 1800 & 0.4\phantom{00} & 1500 & 8.0$\times$10$^{13}$ & 1.0 & 9 && 3.0$\times$10$^{-6}$ & 14.5 & 1.3$\times$10$^{17}$ & 3.8 & 4 && 2.3$\times$10$^{-5}$ & 7.1 & 3 && 43 \\
AQ Sgr & \phantom{0}333 & \phantom{1}3000 & 3000 & $\cdots$ & $\cdots$ & 7.1$\times$10$^{13}$ & 1.1 & 6 && 1.0$\times$10$^{-7}$  & 10.0 & 2.2$\times$10$^{16}$ & $\cdots$ & 2 && 1.5$\times$10$^{-4}$  & $\cdots$ & 2 && $>$7 \\
AFGL~3068 & 1300 & 10900 & 1800 & 5.0\phantom{00} & 1500 & 2.0$\times$10$^{14}$ & 1.9 & 11 && 2.5$\times$10$^{-5}$ & 14.0 & 5.2$\times$10$^{17}$ & 0.7 & 5 && 3.3$\times$10$^{-5}$ & 3.6 & 4 && 30 \\
IRAS~15194-5115 & \phantom{1}500 & \phantom{1}8900 & 2800 & 0.6\phantom{00} & 1300 & 1.5$\times$10$^{14}$ & 0.2 & 7 && 1.5$\times$10$^{-5}$ & 22.0 & 3.0$\times$10$^{17}$ & 3.3 & 4 && 1.0$\times$10$^{-4}$ & $\cdots$ & 2 && 10 \\
\hline
\hline
\end{tabular}
}
\end{table*}

\subsection{Dust continuum modelling}

The results of the dust continuum modelling: stellar temperature ($T_{\star}$), dust optical depth at 10\,$\mu$m ($\tau_{10}$), and the dust temperature at the inner radius ($T_{\mathrm{d}}(r_{\mathrm{i}})$) are given in Table~\ref{results}. The models give reasonable values that are consistent with our previous results. In general, the models give good fits to the observations with $\chi^2_{\mathrm{red}}$ of the order of 1. In cases where the dust optical depth was too low to be constrained ($\tau_{10}$\,$<$\,$0.01$), the modelling is only used to constrain $T_{\star}$. The inner radius, $r_{\mathrm{i}}$, is obtained from the estimated inner dust temperature. In cases where the dust properties could not be constrained, $r_{\mathrm{i}}$ is set to 5\,$R_{\star}$, where $R_{\star}$ is the stellar radius given by $T_{\star}$ and $L_{\star}$. As shown in \citet{ramsetal08}, the dust temperature at the inner radius is not very well constrained by the modelling; in general it is determined to within $\pm200$\,K. It is not a well-determined estimate of the dust-formation temperature, but should be regarded as the representative dust temperature just outside the dust formation zone. More reliable results could be attained by including a larger set of observations to better constrain the models. However, since the purpose is to investigate general trends across chemical/evolutionary types and mass-loss rates, and not to make the most accurate model for each individual star, similar constraints have been used for all stars in order not to introduce biases. Furthermore, we are here only interested in obtaining a reasonable estimate of the dust radiation field (for the molecular line excitation), rather than a more detailed study of the dust characteristics.
 
\subsection{Mass-loss rates from $^{12}$CO line modelling}

\begin{figure}
   \centering
   \includegraphics[width=8cm]{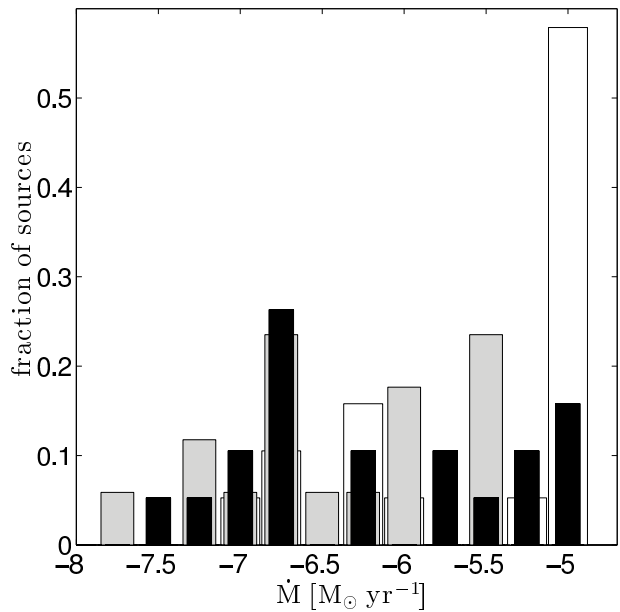}
      \caption{The mass-loss-rate distribution of the sample stars derived from the $^{12}$CO excitation analysis. The different spectral types are shown in the different colours: M-type stars in white, S-type stars in grey, carbon stars in black.}
         \label{mdotdist}
\end{figure}

The estimated mass-loss rates ($\dot{M}$) are given in Table~\ref{results} together with the expansion velocities ($\upsilon_{\mathrm{e}}$), the photodissociation radius ($r_{\mathrm{p}}$, also used in the $^{13}$CO modelling) and the $\chi^2_{\mathrm{red}}$ of the best-fit models. Figure~\ref{mdotdist} shows the $\dot{M}$-distribution for the different spectral types. The results are consistent with our previous results on the same sources. In general, the models produce good fits to the observed lines and the $\chi^{2}_{\mathrm{red}}$ are of the order $1-2$. For about 20\% of the models, the $\chi^{2}_{\mathrm{red}}$\,$>$\,3. This happens when one, or more of, the observed lines deviate significantly from the model results. This can be due to larger calibration uncertainties than assumed for individual lines, bad pointing, or even that the star is not well described by the adopted spherically symmetric model due to e.g. circumstellar asymmetries. The mass-loss rates given in Table~\ref{results} are the average stellar mass-loss rates during the creation of the CSE probed by the different lines used as constraints for the models. Even when $\chi^{2}_{\mathrm{red}}$ is rather large, the values given in Table~\ref{results} are still valid estimates of the mass-loss rate. For the S-type stars, the correction of the line intensities measured at the IRAM 30\,m telescope (see Sect.\ref{s:obs}) only marginally affected the results. This is because each estimate is based on a set of lines and not decisively dependent on the intensity of individual lines. 

As can be seen clearly in Fig.~\ref{mdotdist}, the three chemical types of AGB stars cover large, and similar, ranges in $\dot{M}$, about three orders of magnitude. For the M-type stars though, there is a larger number of higher mass-loss-rate objects, introduced by the difficulties in detecting the weak, low-abundance $^{13}$CO line emission.

\subsection{The $^{13}$CO abundance and the $^{12}$CO/$^{13}$CO ratio}

The estimated photospheric $^{13}$CO abundances ($f_{0}$), the $\chi^{2}_{\mathrm{red}}$ values of the models, the number of observed lines ($N$), and finally the $^{12}$CO/$^{13}$CO abundance ratio, are given in Table~\ref{results}. The observed lines are, in general, well fitted by the model. In some cases, the $\chi^{2}_{\mathrm{red}}$ values are rather large and again this is the consequence when it is not possible to find a model which is consistent with all lines. There is no correlation with high $\chi^{2}_{\mathrm{red}}$ values in the $^{12}$CO model (which would be the case if the difficulties were caused by circumstellar asymmetries), and therefore we conclude that the sometimes large $\chi^{2}_{\mathrm{red}}$ values are likely due to calibration uncertainties. 

\begin{figure}
   \centering
   \includegraphics[width=8cm]{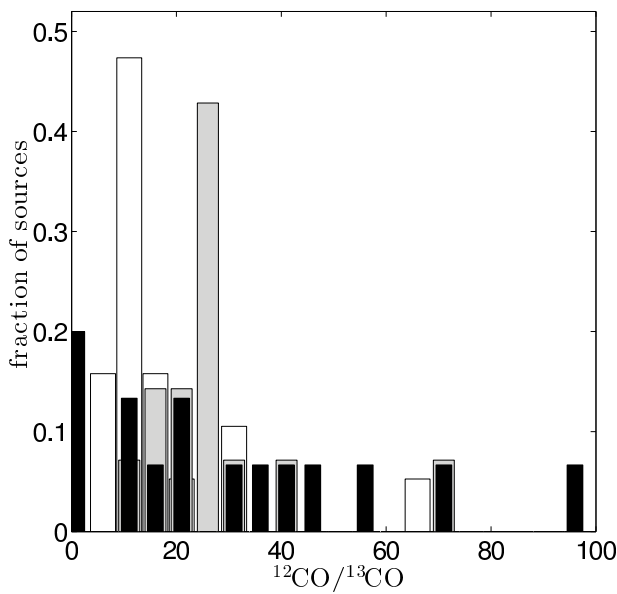}
      \caption{The $^{12}$CO/$^{13}$CO ratio distribution. The different spectral types are shown in the different colours: M-type stars in white, S-type stars in grey, carbon stars in black.}
         \label{abunddist}
   \end{figure}

\begin{figure}
   \centering
   \includegraphics[width=8cm]{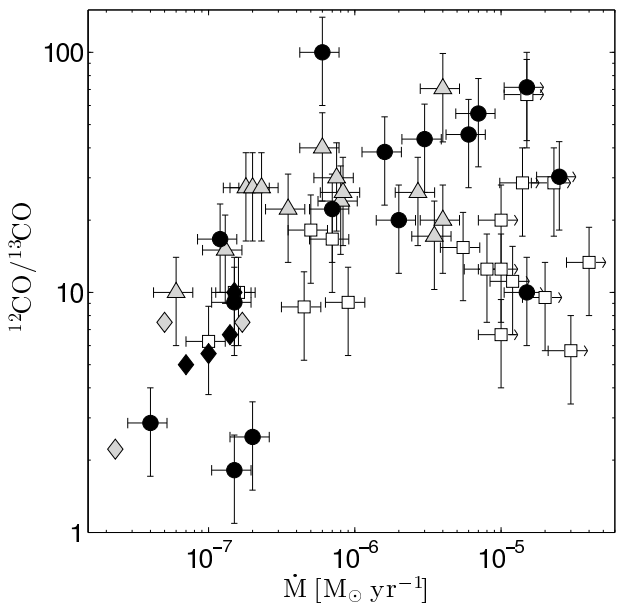}
      \caption{The $^{12}$CO/$^{13}$CO ratios with errors as a function of $\dot{M}$. The different spectral types are shown in the different colours: M-type stars are white squares, S-type stars are grey triangles, carbon stars are black circles. Lower limits are shown as diamonds in the corresponding colours. }
         \label{ratvsmdot}
   \end{figure}

In Fig.~\ref{abunddist} the distribution of $^{12}$CO/$^{13}$CO ratios for the sample is shown, and in Fig.~\ref{ratvsmdot} the $^{12}$CO/$^{13}$CO ratio is shown as a function of $\dot{M}$. The median value of the initial $^{13}$CO abundance ($f_0$) [excluding J-type carbon stars (Y CVn, RY Dra, UX Dra and T Lyr), and upper limit estimates] is $1.6\times10^{-5}$, $2.3\times10^{-5}$ and $3.0\times10^{-5}$ for the M-type, S-type, and carbon stars, respectively, corresponding to median $^{12}$CO/$^{13}$CO ratios of 13, 26, and 34. As apparent in Figs~\ref{abunddist} and \ref{ratvsmdot}, the carbon star ratios have a larger spread. This can be quantified by the ratio between the 90th and 10th percentile, i.e. the values below which 90\% and 10\% of the measurements are found. For the M- and S-type stars, the spread in the $^{13}$CO abundances is of the order of a few (4 and 3, respectively). For the carbon stars, the spread is close to a factor of 15. Whether the samples are different can be quantified (without making assumptions about the intrinsic distribution of the isotopologue ratios) by looking at the fraction of sources from one sample that have ratios above (or below) the median value of the other sample. For the carbon- (excluding J-type) and S-type-star samples, 83\% and 93\% of the sources, respectively, have ratios above the median value for the M-type sample. For the S-type sources, 85\% have ratios below the median value of the carbon star sample, showing that the samples most likely represent different populations, as also confirmed by the differences in spread. As seen in Fig.~\ref{ratvsmdot} (again disregarding ratios estimated from upper limits and the J-type carbon stars) we find no correlation between the $^{12}$CO/$^{13}$CO ratio and the mass-loss rate as expected if there is an increase in the mass-loss rate along the evolution on the AGB.

For the carbon stars, there is in general a good agreement between the estimates of \citet{schoolof00} and this work for the most reliable estimates (based on several detected lines). When the estimates differ substantially (a factor of two or more) they are mostly based on upper limits, or on only one $^{13}$CO line, and the difference cannot be attributed to the updates to the radiative transfer model since 2000. For the S-type stars, the ratios derived using full radiative transfer analysis (this work) are in general lower (of the order 20--30\%) than the values derived by \citet{walletal11} from a less detailed analysis.

\subsection{Uncertainty estimates}

The error bars in Fig.~\ref{ratvsmdot} are 1$\sigma$-errors. For $\dot{M}$, the errors are calculated by checking how much the $\dot{M}$ (given as an input parameter) can be varied while still giving a $\chi^{2}$-value within the 68\%-confidence limit. To estimate errors for the $^{12}$CO/$^{13}$CO abundance ratio, the $^{12}$CO abundance is varied in the $^{12}$CO model, assuming a constant $\dot{M}$ consistent with the best-fit model, giving the 1$\sigma$-error of its estimate. Similarly, the $^{13}$CO abundance is varied in the $^{13}$CO model to estimate the 1$\sigma$-error of its estimate. Finally the two errors are combined through normal error propagation to give the 1$\sigma$-error of the abundance ratio. The uncertainty of the isotopologue ratio is estimated to depend only very weakly on the mass-loss rate; however, as discussed in \citet{ramsetal08}, for very high mass-loss rates ($>$10$^{-5}$\,M$_{\odot}$\,yr$^{-1}$), it is difficult to constrain the mass-loss rate due to the saturation of the lines. This is indicated by arrows on the error bars in Fig.~\ref{ratvsmdot}. The estimated errors are of the same order as that estimated by \citet{khouetal14} from a detailed analysis of W~Hya.

\begin{table}[htbp]
   \caption{The change in the $^{13}$CO line intensities when varying different input parameters. See text for explanation.}
   \centering
   \resizebox{\hsize}{!}{    
   \begin{tabular}{lrcccc} 
      \hline \hline
      Parameter & Change & \multicolumn{4}{c}{$J$-transition} \\
      & & 1$\rightarrow$0 & 2$\rightarrow$1 & 3$\rightarrow$2 & 6$\rightarrow$5 \\
      \hline
      \noalign{\smallskip}
	{\bf R Dor} & & & & & \\
	\noalign{\smallskip}
      $r_{\rm{i}}$ & -33 \% & +3 \% & +6 \% & +5 \% & -3 \% \\
      				& +50 \% & -7 \% & -6 \% & -6 \% & +4 \% \\
	$L_{\star}$ & -33 \% & +13 \% & +13 \% & +8 \% & -9 \% \\
				& +50 \% & -13 \% & -13 \% & -11 \% & +9 \% \\
	$T_{\star}$ & +50 \% & +30 \% &  +25 \% & +15 \% & -17 \% \\
	$r_{\rm{p}}$ & -33 \% & -40 \% & -38 \% & -27 \% & -13 \% \\
				 & +50 \% & +50 \% & +40 \% & +20 \% & +8 \% \\
	no star		 &		   & +50 \% & +57 \% & +26 \% & -37 \% \\
	no dust		&		& +10 \% & +10 \% & +7 \% & -6 \% \\
	carbon dust   &		   & -30 \% & -31 \% & -28 \% & +16 \% \\
	\noalign{\smallskip}
	{\bf CW Leo} & & & & & \\
	\noalign{\smallskip}
	$r_{\rm{i}}$ & -33 \% & 0 \% & +3 \% & +3 \% & +1 \% \\
      				& +50 \% & -4 \% & -4 \% & -2 \% & -1 \% \\
	$L_{\star}$ & -33 \% & -1 \% & 0 \% & +2 \% & +5 \% \\
				& +50 \% & -1 \% & 0 \% & +1 \% & +1 \% \\
	$T_{\star}$ & +50 \% & -2 \% &  0 \% & +1 \% & +2 \% \\
	$r_{\rm{p}}$ & -33 \% & -2 \% & 0 \% & +1 \% & +1 \% \\
				 & +50 \% & -1 \% & 0 \% & +1 \% & +1 \% \\
	no star		 &		   & -1 \% & 0 \% & +1 \% & +1 \% \\
	no dust		&		& -25 \% & -28 \% & -23 \% & -27 \% \\
	silicate dust   &		   & -10 \% & -11 \% & -11 \% & -16 \% \\
	\hline \hline
   \end{tabular}}
   \label{diff}
\end{table}

\subsection{Dependence on input parameters}
\label{input}
To evaluate the sensitivity of the $^{13}$CO model results on the main derived or assumed input parameters we varied them by ${+50}$\% and ${-33}$\% to investigate the resulting effect on the line intensities. In Table~\ref{diff} the change (in percent) in the integrated line intensities of the $^{13}$CO($J$\,=\,$1-0$, $2-1$, $3-2$, and $6-5$) lines is shown for a low mass-loss-rate M-type star (R Dor; $\dot{M}$=1.6$\times$10$^{-7}$\,M$_\odot$\,yr$^{-1}$, $f_{0}$=2.0$\times$10$^{-5}$) and a high mass-loss-rate carbon star (CW Leo; $\dot{M}$=1.6$\times$10$^{-7}$\,M$_\odot$\,yr$^{-1}$, $f_{0}$=1.4$\times$10$^{-5}$) to cover the density range of the sample. The varied parameters are the inner radius ($r_{\rm{i}}$), the stellar luminosity ($L_{\star}$), the stellar temperature ($T_{\star}$), and the photodissociation radius ($r_{\rm{p}}$). Of course, the varied parameters are not independent of each other. Changing the stellar temperature or the luminosity will change the stellar radius; however, changing the stellar temperature will also change the peak of the stellar radiation field, independent of the total luminosity, and varying the luminosity can be done without affecting the shape of the energy distribution. The last three rows for each star in Table~\ref{diff} show the effect of changing the radiation field. First we assume that all radiation is emitted as thermal dust emission (i.e. no stellar radiation, labelled as "no star" in Table~\ref{diff}) and the energy distribution is determined by the dust temperature distribution. Then we remove the dust emission completely (labelled as "no dust" in Table~\ref{diff}). Finally the effect of changing the dust optical properties is tested by using carbon dust properties instead of those for silicate dust for R~Dor, and silicate dust properties for the CW~Leo model instead of carbon dust properties. The purpose is to test the sensitivity of the model to the assumed dust properties in particular, since this is the new addition to the model since \citet{schoolof00}.

When changing the stellar parameters, the line intensities vary within the calibration uncertainties. For these, and the other parameters, it is clear, and not surprising, that the low-density environment (R~Dor) is more sensitive compared to when the density is higher (CW~Leo). In the low-density case, changing the photodissociation radius has a significant effect, as expected when the gas is optically thin and well-excited in the entire envelope. For CW~Leo, the $^{13}$CO line intensities are not affected by changing the photodissociation radius or by not including a stellar radiation field likely since the molecules are collisionally excited within a region smaller than the photodissociation radius. However, turning off the dust emission results in a larger change indicating that the molecules are also to some small extent radiatively excited. For R~Dor, the line intensities are increased by 50\% for the lower transitions and decreased by almost 40\% for the $J$\,=\,6--5 transition when redistributing the spectral energy distribution to be determined only by the dust temperature. The intensities are less affected by changing the energy distribution by only changing the optical properties of the dust and there is also a minor effect when removing the dust radiation field entirely. This is indicative of the fact that the line intensities in the low-density gas is determined both by the radiation field from the star and from thermal dust emission. The dust radiation field also seems to influence the line intensities in the high-density case, since the intensities are more affected by changing the dust optical properties and by turning off the dust emission completely than by turning off the stellar radiation field. \citet{schoolof01} performed similar tests to evaluate the effect on the $^{12}$CO line intensities when varying stellar and model parameters for different optical depths (their Table~5). They found similar dependencies and we therefore believe the derived abundance ratio is rather insensitive to the uncertainties in these parameters. However, there are some differences (e.g. when varying the luminosity) which can result in additional uncertainties in the derived abundance ratio, in particular for the more sensitive, low-density sources.

The estimated isotopologue ratio is essentially independent of the assumed $^{12}$CO abundance since a change in the abundance will result in a corresponding change in the calculated mass-loss rate. Since the derived mass-loss rate is used as input when estimating the $^{13}$CO abundance, again, a corresponding change in the $^{13}$CO abundance will be the result. We tested this by running M-type models with $^{12}$CO/H$_{2}$=$1\times10^{-3}$ and carbon models with $^{12}$CO/H$_{2}$=$2\times10^{-4}$ for both the low- and high-density case.  


\section{Discussion}
\label{s:disc}

\subsection{Comparison with atmospheric estimates of the $^{12}$C/$^{13}$C ratio}

We have no reason to expect that the circumstellar isotopologue ratios that we present here are poorer estimates of the stellar isotope ratios than are e.g. those based on photospheric spectra (which are also based on molecular lines). The CO chemistry is relatively uncomplicated \citep{cher06}, and there is strong reason to expect that the isotopologue ratio in the CO gas that leaves the star is the same as the stellar isotope ratio, also from observations of different molecules \citep{milaetal09}. Processes in the CSE which can affect the circumstellar ratio are isotope-selective photodissociation and chemical fractionation, but they are expected to cancel out \citep{mamoetal88}. The different optical depths in the isotopologue lines, however, will need to be taken into account carefully through a radiative transfer analysis as performed by us. Nevertheless, a comparison with results obtained using photospheric probes is interesting.

$^{12}$C/$^{13}$C ratios have been estimated for some of the carbon stars in our sample using observations of atmospheric molecular absorption lines in the near-IR \citep{lambetal86} and in the optical \citep{ohnatsuj96}. These observations probe atmospheric gas closer to the star and the estimated ratios are based on a large set of lines ($\ge$200) formed in the stellar atmosphere (both CO and CN) where equilibrium chemistry prevails, and are therefore considered to be more directly representative of the stellar $^{12}$C/$^{13}$C ratio. However, the atmospheric values are very model dependent as made clear by the discrepancy between the results of \citet{lambetal86} and those of \citet{ohnatsuj96}, and the subsequent discussion \citep{gust97,delagust98,ohnatsuj98}. The optical and near-IR spectra are also (in the case of high-mass-loss-rate sources like CW Leo, severely) affected by the dust surrounding the star, while the circumstellar radio lines are less sensitive. 

\citet{schoolof00} showed that their $^{12}$CO/$^{13}$CO ratios were well correlated with the $^{12}$C/$^{13}$C ratios estimated by \citet{lambetal86} for the overlapping sources, but that they were generally lower. Nevertheless, stellar and circumstellar values agreed within the errors. The discrepancy between the Lambert et al.~values and the circumstellar values are somewhat accentuated by our analysis (for the overlapping sources the average value is a factor of two lower); however, the methods are very different and the results can be influenced by different systematic effects. The complexity of the isotope ratio estimates performed in the different studies (including Ohnaka \& Tsuji 1996) therefore precludes a more detailed comparison.

Given the progress made since the late 80:ies, it seems to us that there is reason to redo the analysis of the photospheric spectra using dynamical models to fit spectral observations covering the pulsation cycle (the previous analysis is based on hydrostatic model atmospheres and line-intensity fitting based on equivalent widths), in order to establish whether the $^{12}$C/$^{13}$C ratios of N-type carbon stars generally falls below 40 as suggested by our analysis, or above 40 as proposed by \citet{lambetal86}.

There is very good agreement between the atmospheric $^{12}$C/$^{13}$C ratios for M- and S-type stars \citep{smitlamb90,domiwall87} and our results; however there are only very few overlapping sources.

\subsection{The motivation for detailed radiative transfer modelling}

\begin{figure}
   \centering
   \includegraphics[width=8cm]{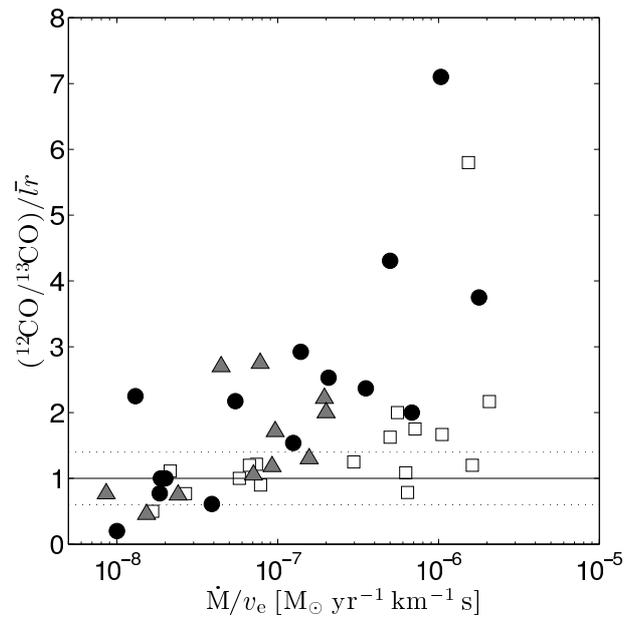}
      \caption{The ratio between the $^{12}$CO/$^{13}$CO ratios calculated by detailed radiative transfer and the average line-intensity ratio ($\bar{lr}$, see Table~\ref{tablineratios}). M-type stars are white squares, S-type stars are grey triangles and carbon stars are black dots. The solid line shows the one-to-one correlation with the dotted lines indicating the approximate errors from the detailed analysis. }
         \label{figlineratios}
   \end{figure}

There is no doubt that performing detailed radiative transfer modelling will result in a more accurate estimate of the $^{12}$CO/$^{13}$CO abundance ratio, compared to when using a simple line-intensity-ratio estimate. For carbon stars, \citet{schoolof00} showed that line ratios alone will underestimate the $^{12}$CO/$^{13}$CO ratio, mainly because the $^{12}$CO emission is optically thick and the amount of $^{12}$CO is underestimated. Since this is the first detailed analysis that also includes M- and S-type stars, it is important to investigate if the line-intensity ratios give a good estimate of the $^{12}$CO/$^{13}$CO ratio for these stars as is often assumed. If the line-intensity ratio still gives a reasonable estimate within the (rather large) estimated errors of the radiative transfer model, not much will be gained by performing a detailed analysis. Table~\ref{tablineratios} lists the line-intensity ratios of the measured lines (corrected for antenna size) for all transitions observed in both isotopologues together with the estimated value from the detailed analysis and the observed average line-intensity ratio ($\bar{lr}$). Figure~\ref{figlineratios} shows the ratio between the model value and the observed average line-intensity ratio as a function of the circumstellar density (as measured by $\dot{M}/\upsilon_{\rm{e}}$).

\begin{table}[htbp]
   \centering
   \caption{Comparison between the $^{12}$CO/$^{13}$CO results from the abundance analysis modelling and line intensity ratios. The second column repeats the results from the detailed radiative transfer (RT) modelling, also given in Table~\ref{results}. The following four columns give the $^{12}$CO/$^{13}$CO line-intensity ratios corrected for antenna size for the different observed rotational transitions. The last column gives the average of the observed line-intensity ratios from columns 3--6.} 
   \resizebox{\hsize}{!}{    
   \begin{tabular}{lcccccc} 
   \hline \hline
   Source & RT model & \multicolumn{4}{l}{Line ratios, $^{12}$CO/$^{13}$CO} & Average \\
    &  & 1$\rightarrow$0 & 2$\rightarrow$1 & 3$\rightarrow$2 & 6$\rightarrow$5 & $\bar{lr}$ \\
   \hline
      \noalign{\smallskip}
	{\em M-type stars} & & & & & &\\
	\noalign{\smallskip}
	RX Boo & \phantom{1}17 & & 18 & 10 & & 14\\
	TX Cam & \phantom{1}15 & & 19 & \phantom{1}7 & 10 & 12 \\
	R Cas & \phantom{11}9 & & 10 & & 10 & 10\\
	R Dor & \phantom{1}10 & & 20 & 13 & \phantom{1}6 & 13\\
	W Hya & \phantom{1}10 & & & \phantom{1}9 & & \phantom{1}9\\
	R Leo & \phantom{11}6 & & 15 & \phantom{1}9 & & 12 \\
	GX Mon & \phantom{1}11 & 21 & \phantom{1}6 & & & 14 \\
	WX Psc & \phantom{1}13 & 11 & \phantom{1}5 & \phantom{1}3 & \phantom{1}4 & \phantom{1}6 \\
	RT Vir & \phantom{11}9 & & 11 & \phantom{1}7 & & \phantom{1}9 \\
	SW Vir & \phantom{1}18 & & 15 & & & 15 \\
	IK Tau & \phantom{1}10 & & \phantom{1}8 & \phantom{1}6 & \phantom{1}5 & \phantom{1}6\\
	IRC+10365 & \phantom{1}13 & 17 & \phantom{1}6 & & & 12 \\
	IRC-10529 & \phantom{11}7 & & \phantom{1}5 & \phantom{1}2 & & \phantom{1}4\\
	IRC-30398 & \phantom{1}13 & & \phantom{1}8 & & & \phantom{1}8 \\
	IRC+40004 & \phantom{1}20 & & 10 & & & 10 \\
	IRC+50137 & \phantom{11}6 & & \phantom{1}5 & & & \phantom{1}5 \\
	IRC+60169 & \phantom{1}29 & & \phantom{1}5 & & & \phantom{1}5 \\
	\noalign{\smallskip}
	{\em S-type stars} & & & & & &\\
	\noalign{\smallskip}
	R And & \phantom{1}24 & 17 & 11 & & & 14 \\
	W Aql & \phantom{1}26 & 27 & 18 & 16 & & 20 \\
	S Cas & \phantom{1}71 & 34 & 29 & & & 32 \\
	TT Cen & \phantom{1}20 & & 10 & & & 10 \\
	T Cet & \phantom{1}10 & & 12 & 14 & & 13 \\
	R Cyg & \phantom{1}26 & 24 & 19 & & & 22 \\
	$\chi$ Cyg & \phantom{1}40 & 53 & 36 & 25 & & 38 \\
	R Gem & \phantom{1}22 & & \phantom{1}8 & & & \phantom{1}8 \\ 
	ST Her & \phantom{1}15 & 35 & 31 & & & 33 \\
	Y Lyn & \phantom{1}27 & 35 & 37 & & & 36 \\
	DK Vul & \phantom{1}27 & & 10 & & & 10 \\ 
	\noalign{\smallskip}
	{\em Carbon stars} & & & & & &\\
	\noalign{\smallskip}
	LP And & \phantom{1}56 & 18 & 10 & 12 & & 13 \\
	RV Aqr & \phantom{1}20 & 13 & & & & 13 \\
	UU Aur & 100 & 46 & & & & 46 \\
	Y CVn & \phantom{11}2 & \phantom{1}1 & & \phantom{1}3 & & \phantom{1}2 \\
	V Cyg & \phantom{1}38 & 12 & & 13 & & 13 \\
	RY Dra & \phantom{11}3 & & & \phantom{1}3 & & \phantom{1}3 \\
	UX Dra & \phantom{11}3 & 15 & & & & 15 \\
	U Hya & \phantom{1}17 & 27 & 17 & & & 22 \\
	CW Leo & \phantom{1}71 & 17 & \phantom{1}7 & \phantom{1}7 & & 10 \\
	R Lep & \phantom{1}22 & & 36 & & & 36 \\
	RW LMi & \phantom{1}45 & 31 & \phantom{1}9 & 18 & & 19 \\
	T Lyr & \phantom{11}9 & & & \phantom{1}4 & & \phantom{1}4 \\
	V384 Per & \phantom{1}43 & 27 & 14 & \phantom{1}9 & & 17 \\
	AFGL 3068 & \phantom{1}30 & 15 & \phantom{1}6 & \phantom{1}3 & & \phantom{1}8 \\
	IRAS 15194-5115 & \phantom{1}10 & \phantom{1}4 & \phantom{1}5 & & & \phantom{1}5 \\
	\hline \hline
   \end{tabular}}
   \label{tablineratios}
\end{table}

There is a large scatter among the line-intensity ratios for individual sources, indicating that this type of analysis is very sensitive to which transitions that have been observed. This is also, to some extent, true for the detailed analysis. From Fig.~\ref{figlineratios} it is clear that for low-density CSEs of M- and S-type stars, a simple line-intensity-ratio estimate of the isotopologue ratio gives a reasonably good estimate. However, at higher density, the isotopologue ratio is also underestimated for M- and S-type stars. For the S-type stars, this becomes a problem already at $\dot{M}/\upsilon_{\rm{e}}\geq5\times10^{-8}$\,M$_\odot$\,yr$^{-1}$\,km$^{-1}$\,s, which includes the majority of the S-type stars of our sample. We conclude that it is appropriate to perform detailed radiative transfer analysis already at lower densities and abundances, and that it is necessary at higher densities, also for M-type stars, in order to estimate reliable isotopologue ratios.

\subsection{Evolutionary models}

The aim of this work is not to derive exact values for the $^{12}$C/$^{13}$C ratios of the individual stars, but rather to gather enough data for a sufficiently large sample to be able to search for trends in order to draw some more general conclusions. There is always a trade-off between the sample size and the level of detail and accuracy of the derived abundances. The drawback when using nearby stars is the large distance uncertainties which can possibly introduce errors in the final abundances. The largest advantage is the reliability of the observational data. By using the most reliable distance estimates available for our sources and by using the same methods for the different chemical types, we are confident that we have dealt with this problem appropriately. Even though the uncertain distances might have introduced errors in the $^{13}$CO abundances, the abundance ratio is much less sensitive and the relative trends are very reliable. 

Evolutionary models aim to describe full evolutionary sequences and not the conditions of a star at a specific instance in time, as observations do. It should, however, be possible to compare the trends seen among the different chemical types in our sample with the expectations for thermally pulsing AGB stars (it is likely that all stars in our sample has reached the TP-AGB given their well-developed CSEs) to evaluate the evolutionary state of the stars, and to provide observational constraints for the models. From the models, the amount of $^{12}$C at the surface should increase as the star evolves \citep[see e.g.][and references therein]{lattboot97} leading to an increase in the overall carbon abundance (eventually creating carbon stars if the process is efficient enough) and in the $^{12}$C/$^{13}$C ratio. Known processes that can disrupt this general trend are cool bottom processing (CBP) and hot bottom burning (HBB). CBP refers to material being mixed down (e.g. by rotation) from the bottom of the convective envelope to close to the H-burning shell (where it is processed) and up again, changing the composition of the higher layers. CBP can occur in low-mass AGB stars and has been suggested as an explanation for the low $^{18}$O/$^{16}$O and $^{12}$C/$^{13}$C ratios observed in AGB stars, and in particular in RGB stars \citep[e.g.][]{harretal87,wassetal95,bootsack99,tsui07}. HBB refers to when the temperature at the bottom of the convective envelope is high enough for nuclear burning to take place. This only occurs in massive AGB stars ($\gtrsim$4\,M$_{\odot}$). HBB will also decrease the $^{12}$C/$^{13}$C ratio as $^{12}$C is destroyed (and converted into $^{14}$N) and some small amounts of $^{13}$C is produced; however, the effect on the overall carbon budget is such that it prevents the formation of carbon stars \citep[e.g.][]{bootetal93}.

The stars of our sample seem to follow the general trend expected from evolutionary models, the M-type stars have the lowest $^{12}$C/$^{13}$C ratio and the carbon stars the highest  (see Fig.~\ref{abunddist}). The S-type stars fall in between as expected if they are transition objects. Notably, the carbon stars show a large spread in the isotopologue ratio (also when excluding the J-type stars discussed below), which is as expected from the evolution on the AGB. The spread in the isotope ratio for an M-type AGB star is restricted to the difference in its initial value after the RGB and about twice this (assuming that C/O\,$\approx$\,0.6 when it arrives on the AGB). For a carbon star there is, in principle, no upper limit to the isotope ratio as more and more $^{12}$C is dredged up as the star evolves. There are also a number of low-mass-loss-rate carbon stars that show $^{12}$C/$^{13}$C ratios as low as 3 (classified as J-type). \citet{bootetal93} show that there should be a narrow region for which HBB produces enough $^{13}$C to decrease the isotope ratio down to a few, while there is still enough $^{12}$C to make it a carbon star. This should correspond to stars with $-6.4$\,$<$\,$M_{\rm{bol}}$\,$<$\,$-6.3$. None of the carbon stars in our sample have such high luminosities, and they are likely not massive enough for HBB to occur. From a study of galactic J-type stars, Abia \& Isern (ApJ 536, 438, 2000) concluded that these are most likely low-mass stars, less evolved than the N-type carbon stars. Proposed explanations for the low isotope ratios are an extra mixing process on the early AGB, mixing at the He-core flash, or evolution in a binary system (see below). We are in the process of collecting data on the oxygen isotopologues, using circumstellar CO lines, of our sample to investigate this further. 

Another aspect that will significantly alter the abundance evolution of a star is whether it is in a binary system \citep[as shown by the work of Izzard and collaborators, e.g.][]{izza04}. Binary evolution affecting the $^{12}$C/$^{13}$C ratio in the case of J-type stars has been discussed by \citet{abiaiser00} and \citet{sengetal13}, but no firm conclusions have been drawn. Unfortunately, the binary fraction of our sample is not known. None of the stars in our sample have been detected in UV or X-rays through serendipitous observations or all-sky surveys \citep{ramsetal12}, which could be indicative of a hot companion. However, in particular for the carbon stars, it is difficult for high-energy radiation from a binary companion to escape the thick AGB CSE and the observations at hand provide rather poor constraints. Targeted observations of e.g. the J-type carbon stars would better constrain the likelihood for binarity, and give better grounds for evaluating the process behind the low $^{12}$C/$^{13}$C ratios. A hot companion could possibly influence the circumstellar isotopologue ratio through differential photodissociation, making it less representative of the stellar $^{12}$C/$^{13}$C ratio, as suggested by \citet{vlemetal13} for the case of R~Scl.

\subsection{Comparison with $^{12}$C/$^{13}$C ratios for post-AGB stars}

\begin{figure}
   \centering
   \includegraphics[width=8cm]{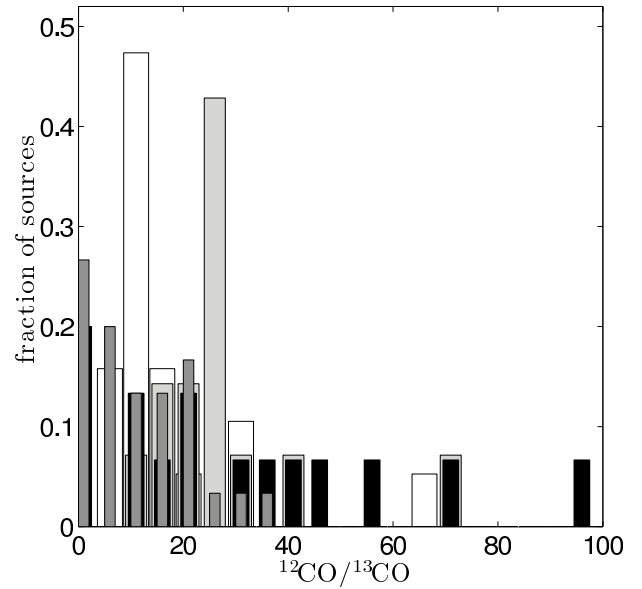}
      \caption{The $^{12}$CO/$^{13}$CO ratio distribution together with estimates for post-AGB stars. The different spectral types are shown in the different colours: M-type stars in white, S-type stars in light grey, carbon stars in black. The post-AGB stars are shown in dark grey.}
         \label{histpn}
   \end{figure}

In Fig.~\ref{histpn} we show our results together with estimates of the $^{12}$C/$^{13}$C ratio from circumstellar CO line emission for the post-AGB stars and PNe presented by \citet{palletal00}, \citet{balsetal02}, and \citet{sancsaha12}.
 
The post-AGB ratios are estimated from line-intensity ratios under the assumption that the emission is optically thin. It is beyond the scope of this paper to evaluate the validity of that assumption, but if not valid, the post-AGB values give lower limits for the isotopologue ratio. From the sample of \citet{sancsaha12}, we only include post-AGB objects. Most of the stars of their sample are classified as M-type. It is clear that the post-AGB isotopologue ratios are more in line with what we find for M-type AGB stars than for the carbon stars. This might indicate that they have evolved directly from M-type stars with low initial (when leaving the RGB) isotopic ratios. It could also indicate that PNe with rich molecular envelopes are formed mainly from more massive stars (for which HBB prevents a transformation to a carbon star), or alternatively that binarity may alter the isotopologue ratios. However, whether a simple line-intensity-ratio estimate is indeed representative of the stellar $^{12}$C/$^{13}$C ratio (as determined by the evolution of the star) would have to be investigated further before any firm conclusions can be drawn. 

\section{Conclusions}
We have estimated $^{12}$CO/$^{13}$CO ratios for a sample of AGB stars, 19 M-type, 17 S-type, and 19 carbon stars. The median ratio is 13 for the M-type stars, 26 for the S-type stars, and 34 for the carbon stars, and we find no correlation with mass-loss rate. While the results for the M- and S-type stars are rather well limited in range, the results for the carbon stars show a large spread from an abundance ratio of a few, for the J-type carbon stars, up to close to 100 for the highest ratios measured. We arrive at the following conclusions:

\begin{itemize}
\item{The median $^{12}$CO/$^{13}$CO ratio increases from M- to S-type to carbon stars. This is expected if the spectral types constitute an evolutionary sequence, but this is the first time this has been shown observationally for a sample including a significant number of M- and S-type stars. We want to emphasize that the results for the S-type stars are statistically different from the M-type stars providing further support to a scenario where the S-type AGB stars are transition objects.}
\item{The spread in the $^{13}$CO abundance, quantified by the ratio between the 90th and 10th percentile, is 4, 3, and 15 for the M-type, S-type and carbon stars, respectively. The larger spread in the isotopologue ratios found for the carbon stars can be explained as a consequence of the evolution on the AGB as a limited amount of $^{12}$C can be dredged-up before the star becomes a carbon star.}
\item{Circumstellar $^{12}$CO/$^{13}$CO ratios estimated from detailed radiative transfer analysis are representative of the stellar $^{12}$C/$^{13}$C ratio.}
\item{Detailed radiative transfer is appropriate already at lower densities and abundances in order to estimate accurate abundance ratios, also for M- and S-type stars. At higher abundances and densities ($\dot{M}/v_{\rm{e}}$$>$3$\times$10$^{-7}$\,M$_{\odot}$\,yr$^{-1}$\,km$^{-1}$\,s for M-type stars), it is necessary for a reliable estimate.}
\item{The isotopologue-ratio distribution for post-AGB stars is similar to that derived for the M-type AGB stars which might indicate that they have evolved directly from M-type stars with low initial (when leaving the RGB) ratios, or that molecular envelopes are more common around massive post-AGB stars, or alternatively that binarity may alter the isotopologue ratios. However, further investigation is necessary in order to confirm the correlation.}
\end{itemize}

\begin{acknowledgements}   
This publication is based on data acquired with the Atacama Pathfinder Experiment (APEX) telescope, the IRAM 30\,m telescope, the James Clerk Maxwell Telescope (JCMT), the Swedish-ESO Submillimeter Telescope (SEST), and the Onsala 20\,m telescope. APEX is a collaboration between the Max-Planck-Institut fur Radioastronomie, the European Southern Observatory (ESO), and the Swedish National Facility for Radio Astronomy, Onsala Space Observatory (OSO). IRAM is supported by INSU/CNRS (France), MPG (Germany) and IGN (Spain). The JCMT is operated by the Joint Astronomy Centre on behalf of the Science and Technology Facilities Council of the United Kingdom, the Netherlands Organisation for Scientific Research, and the National Research Council of Canada. The Onsala 20\,m telescope is operated by OSO. The SEST was operated jointly by ESO and OSO. \\

SR is partly funded by the Sch\"onberg fellowship at Uppsala University. HO acknowledges support from the Swedish Research Council. \\

Finally, the authors would like to thank Andr\'es P\'erez S\'anchez (Argelander-Institut f\"ur Astronomie, Bonn, Germany) for performing the observations at the JCMT.   
\end{acknowledgements}

\bibliographystyle{aa}
\bibliography{dabib}

\clearpage
\appendix

\section{Observational data}

\begin{table}[h]
\label{12CO1}
\caption{$^{12}$CO observations used to constrain the physical parameters.}   
\begin{minipage}{9cm}   
\centering        
\resizebox{\hsize}{!}{    
\begin{tabular}{l c c c c c l}       
\hline\hline                
Source & Code\footnote{The first letter of the telescope name followed by the $J$-numbers of the observed transition.} & D & $\theta$ & T$_{\rm{mb}}$ & I$_{\rm{obs}}$  &  Ref.\\    
	&  & [pc] & [''] & [K] & [K km s$^{-1}$] &  \\
\hline
\noalign{\smallskip}
{\em M-type stars:} & & & & & \\
\noalign{\smallskip}
RX Boo & O10 & 137 & 33 & 0.8\phantom{0}  & \phantom{1}14.0  & K\&O99 \\
	& I10 & & 21 & 1.1\phantom{0} & \phantom{1}28.0  & Netal98 \\ 
	& I21 & & 11 & 3.4\phantom{0} & 149.0  & Netal98 \\
	& J32 & & 14 & 3.5\phantom{0} & \phantom{1}53.8  & K\&O99 \\
	& J43 & & 13 & 2.6\phantom{0} & \phantom{1}37.4  & Tetal06 \\
	& C65 & & 10 & 2.5\phantom{0} & \phantom{1}34.8  & Tetal06 \\
	& J65 & & 8 & 0.43 & \phantom{11}4.1  & Detal10 \\
TX Cam & O10 & 380 & 33 & 0.8\phantom{0} & \phantom{1}20.0  & Retal08 \\
	& J21 & & 21 & 2.2\phantom{0} & \phantom{1}61.0  & Retal08 \\
	& J32 & & 14 & 2.6\phantom{0} & \phantom{1}71.0 & Retal08 \\
	& J43 & & 11 & 5.5\phantom{0} & 149.0  & Retal08 \\
	& H65 & & 31 & - & \phantom{1}14.2  & Jetal12 \\
R Cas & O10 & 176 & 33 & 0.5\phantom{0} & \phantom{11}8.4 &  Detal03 \\
	& J21 & & 21 & 1.8\phantom{0} & \phantom{1}32.1  & Detal03 \\ 
	& J32 & & 14 & 5.5\phantom{0} & 100.2  & Detal03 \\
	& J43 & & 11 & 5.4\phantom{0} & \phantom{1}89.6  & Detal03 \\
	& H65 & & 31 & - & \phantom{1}14.7 & Jetal12 \\
R Dor & S10 & \phantom{1}55 & 45 & 0.4\phantom{0} & \phantom{11}5.0 &  K\&O99 \\
	& A21 & & 27 & 4.0\phantom{0} & \phantom{1}39.2 &  R\&O14 \\
	& A32 & & 18 & 5.7\phantom{0} & \phantom{1}63.0 & R\&O14 \\
	& H65 & & 31 & - & \phantom{1}14.9 & Jetal12 \\
W Hya & S10 & \phantom{1}98 & 45 & 0.04 & \phantom{11}0.8 & Jetal05 \\
	& J21 & & 21 & 1.4\phantom{0} & \phantom{1}18.7 &  Jetal05 \\
	& J32 & & 14 & 2.8\phantom{0} &  \phantom{1}40.2 &  Jetal05 \\
	& J43 & & 11 & 3.1\phantom{0} &  \phantom{1}44.6 &  Jetal05 \\
	& H65 & & 31 & - & \phantom{11}9.6 &  Jetal12 \\
R Leo & O10 & \phantom{1}71 & 33 & 0.3\phantom{0} & \phantom{11}2.4 &  Detal03 \\
	& J21 & & 21 & 1.3\phantom{0} & \phantom{1}15.0 &  Detal03 \\
	& J32 & & 14 & 3.9\phantom{0} & \phantom{1}41.6 &  Detal03 \\ 
GX Mon & O10 & 550 & 33 & 0.9\phantom{0} & \phantom{1}31.0 &  Retal08 \\ 
	& J21 & & 21 & 2.0\phantom{0} & \phantom{1}61.0 &  Retal08 \\
	& J32 & & 14 & 2.5\phantom{0} & \phantom{1}80.0 &  Retal08 \\
	& J43 & & 11 & 2.4\phantom{0} & \phantom{1}79.0  & Retal08 \\
WX Psc & O10 & 700 & 33 & 1.3\phantom{0} & \phantom{1}44.0 & Retal08 \\
	& J21 & & 21 & 2.8\phantom{0} & \phantom{1}81.0 &  Retal08 \\ 
	& J32 & & 14 & 2.4\phantom{0} & \phantom{1}70.0 &  Retal08 \\ 
	& J43 & & 11 & 3.1\phantom{0} & \phantom{1}84.0 &  Retal08 \\ 
	& H65 & & 31 & - & \phantom{11}4.4 & Jetal12 \\ 
IK Tau & O10 & 500 & 33 & 1.3\phantom{0} & \phantom{1}46.0 &  Retal08 \\
	& J21 & & 21 & 3.2\phantom{0} & \phantom{1}95.0 & Retal08 \\
	& J32 & & 14 & 4.2\phantom{0} & 125.0 &  Retal08 \\
	& J43 & & 11 & 4.4\phantom{0} & 130.0 &  Retal08 \\
	& H65 & & 31 & - & \phantom{1}11.6 &  Jetal12 \\
RT Vir & O10 & 226 & 33 & 0.4\phantom{0} & \phantom{11}5.2 &  K\&O99 \\
	& J21 & & 21 & 1.0\phantom{0} & \phantom{1}14.1 &  K\&O99 \\  
	& J32 & & 14 & 1.4\phantom{0} & \phantom{1}19.0 &  K\&O99 \\
	& J43 & & 11 & 1.4\phantom{0} & \phantom{1}16.1 &  K\&O99 \\
\noalign{\smallskip}
\hline\hline   
\end{tabular}}
\tablebib{K\&O99-\citet{kersolof99}; Netal98-\citet{nerietal98}; Tetal06-\citet{teysetal06}; Detal10-\citet{debeetal10}; Retal08-\citet{ramsetal08}; Jetal12-\citet{justetal12}; Detal03-\citet{delgetal03}; R\&O14-This work; Jetal05-\citet{justetal05}}
\end{minipage}
\end{table}

\begin{table}
\label{12CO2}
\caption{$^{12}$CO observations used to constrain the physical parameters.}      
\centering        
\resizebox{\hsize}{!}{    
\begin{tabular}{l c c c c c c l}       
\hline\hline                
Source & Code & D & $\theta$ & T$_{\rm{mb}}$ & I$_{\rm{obs}}$ &  Ref.\\    
	&  & [pc] & [''] & [K] & [K km s$^{-1}$] & \\
\hline
\noalign{\smallskip}
{\em M-type stars:} & & & & & \\
\noalign{\smallskip}
SW Vir & S10 & 143 & 45 & 0.3\phantom{0} & \phantom{11}4.3 &  K\&O99 \\
	& O10 & & 33 & 0.7\phantom{0} & \phantom{1}10.2 &  K\&O99 \\
	& S21 & & 23 & 1.3\phantom{0} & \phantom{1}18.4 & K\&O99 \\
	& J21 & & 21 & 2.0\phantom{0} & \phantom{1}26.9 &  K\&O99 \\
	& J32 & & 14 & 3.0\phantom{0} & \phantom{1}40.2 &  K\&O99 \\
	& J43 & & 11 & 2.5\phantom{0} & \phantom{1}36.0 &  K\&O99 \\
CIT4 & O10 & 800 & 33 & - & \phantom{1}15: &  Oetal98 \\
IRC+10365 & O10 & 650 & 33 & 0.8\phantom{0} & \phantom{1}21.5 &  Detal03 \\
	& J21 & & 21 & 1.5\phantom{0} & \phantom{1}42.0 &  Detal03 \\
	& J32 & & 14 & 2.4\phantom{0} & \phantom{1}59.1 &  R\&O14 \\ 
IRC-10529 & O10 & 620 & 33 & 0.7\phantom{0}  & \phantom{1}16.0 &  Retal08 \\
	& J21 & & 21 & 2.2\phantom{0}  & \phantom{1}45.0 &  Retal08 \\
	& J32 & & 14 & 1.4\phantom{0}  & \phantom{1}27.0 &  Retal08 \\
	& J43 & & 11 & 1.1\phantom{0} & \phantom{1}22.0 &  Retal08 \\
IRC-30398 & J21 & 550 & 21 & 1.8\phantom{0} & \phantom{1}44.6 &  Detal03 \\
IRC+40004 & O10 & 600 & 33 & 0.8\phantom{0} & \phantom{1}24.8 &  Detal03 \\
	& J21 & & 21 & 1.4\phantom{0} & \phantom{1}41.5 &  Detal03 \\
IRC+50137 & J21 & 1500 & 21 & 1.4\phantom{0} & \phantom{1}36.8 &  Detal03 \\
	& J32 & & 14 & 1.3\phantom{0} & \phantom{1}35.7 & Detal03 \\
IRC+60169 & J21 & 400 & 21 & 1.3\phantom{0} & \phantom{1}26.8 &  JCMT-A \\
	& J32 & & 14 & 1.3\phantom{0} & \phantom{1}31.4 &  JCMT-A \\
IRC+70066 & J21 & 400 & 21 & 0.85 & \phantom{1}25.5 &  JCMT-A \\
	& J32 & & 14 & 1.4\phantom{0} & \phantom{1}39.4 &  JCMT-A \\
\noalign{\smallskip}
{\em S-type stars:} & & & & & & \\
\noalign{\smallskip}
R And & N10 & 350 & 55 & 0.4\phantom{0}  & \phantom{11}5.7 & B\&L94 \\
	& I10 & & 21 & 1.8\phantom{0} & \phantom{1}25.8 &  Retal09 \\
	& I21 & & 11 & 5.0\phantom{0} & \phantom{1}70.5 &  Retal09 \\
	& J21 & & 21 & 2.5\phantom{0} & \phantom{1}32.0 & JCMT-A \\
	& J32 & & 14 & 3.3\phantom{0} & \phantom{1}43.0 &  JCMT-A \\
W Aql & S10 & 300 & 45 & 1.3\phantom{0} & \phantom{1}28.4 &  Netal92 \\
	& I10 & & 21 & 3.9\phantom{0} & 112.7 &  Wetal11 \\
	& I21 & & 11 & 7.0\phantom{0} & 203.3 &  Wetal11 \\
	& C21 & & 33 & 2.1\phantom{0} & \phantom{1}54.8 &  Ketal98 \\
	& C32 & & 25 & 3.5\phantom{0} & \phantom{1}93.2 &  Ketal98 \\
	& A32 & & 18 & 5.0\phantom{0} & 136.2 &  Retal09 \\
	& J43 & & 11 & 7.4\phantom{0} & 199.5 &  JCMT-A \\
TV Aur & I10 & 400 & 21 & 0.04 & \phantom{11}0.3 &  Wetal11 \\
	& I21 & & 11 & 0.1\phantom{0} & \phantom{11}0.9 & Wetal11 \\
AA Cam & I10 & 780 & 21 & 0.09 & \phantom{11}0.5 &  Wetal11 \\
	& I21 & & 11 & 0.17 & \phantom{11}1.1 &  Wetal11 \\
	& J32 & & 14 & 0.17 & \phantom{11}0.8 &  Retal09 \\
S Cas & O10 & 570 & 33 & 0.4\phantom{0} & \phantom{1}13.8 & Retal09 \\
	& I10 & & 21 & 0.8\phantom{0} & \phantom{1}27.2 &  Wetal11 \\
	& I21 & & 11 & 2.0\phantom{0} & \phantom{1}68.5 &  Wetal11 \\
	& J32 & & 14 & 1.1\phantom{0} & \phantom{1}31.0 &  Retal09 \\
TT Cen & S10 & 1080 & 45 & \phantom{1}0.07 & \phantom{11}2.7 &  S\&L95 \\
	& S21 & & 23 & \phantom{1}0.25 & \phantom{11}9.4 & S\&L95 \\
	& A32 & & 18 & \phantom{1}0.5\phantom{0} & \phantom{1}13.9 &  Retal09 \\
T Cet & I10 & \phantom{1}270 & 21 & \phantom{1}0.2\phantom{0} & \phantom{11}1.6 &  Wetal11 \\
	& I21 & & 11 & \phantom{1}0.3\phantom{0} & \phantom{11}2.2 &  Wetal11 \\
	& S21 & & 23 & \phantom{1}0.3\phantom{0} & \phantom{11}2.0 &  Retal09 \\
	& J21 & & 21 & \phantom{1}0.4\phantom{0} & \phantom{11}3.7 &  Retal09 \\
	& A32 & & 18 & \phantom{1}0.7\phantom{0} & \phantom{11}6.1 &  Retal09 \\
	& J32 & & 14 & \phantom{1}0.8\phantom{0} & \phantom{11}7.2 & Retal09 \\
	& J43 & & 11 & \phantom{1}1.0\phantom{0} & \phantom{11}9.5 &  Retal09 \\
	& J65 & & \phantom{1}8 & \phantom{1}1.1\phantom{0} & \phantom{1}10.0 &  Retal09 \\
R Cyg & N10 & \phantom{1}600 & 55 & \phantom{1}0.15 & \phantom{11}2.5 & Retal09 \\
	& O10 & & 33 & \phantom{1}0.3\phantom{0} &  \phantom{11}4.4 &  Retal09 \\
	& I10 & & 21 & \phantom{1}0.7\phantom{0} &  \phantom{1}12.1 &  Retal09 \\
	& I21 & & 11 & \phantom{1}2.2\phantom{0} &  \phantom{1}36.5 &  Retal09 \\
	& C32 & & 20 & \phantom{1}2.3\phantom{0} &  \phantom{1}14.6 &  Retal09 \\
\hline\hline   
\end{tabular}}
\tablebib{K\&O99-\citet{kersolof99}; Oetal98-\citet{olofetal98}; Detal03-\citet{delgetal03}; R\&O14-This work; Retal08-\citet{ramsetal08}; JCMT-A-The JCMT archive; B\&L94-\citet{bieglatt94}; Retal09-\citet{ramsetal09}; Netal92-\citet{nymaetal92}; Wetal11-\citet{walletal11}; Ketal98-\citet{knapetal98}; S\&L95-\citet{sahaliec95}}
\end{table}

\begin{table}
\label{12CO3}
\caption{$^{12}$CO observations used to constrain the physical parameters.}      
\centering        
\resizebox{\hsize}{!}{    
\begin{tabular}{l c c c c c c l}       
\hline\hline                
Source & Code & D & $\theta$ & T$_{\rm{mb}}$ & I$_{\rm{obs}}$ &  Ref.\\    
	&  & [pc] & [''] & [K] & [K km s$^{-1}$] & \\
\hline
\noalign{\smallskip}
{\em S-type stars:} & & & & & & \\
\noalign{\smallskip}
$\chi$ Cyg & O10 & \phantom{1}180 & 33 & \phantom{1}1.8\phantom{0} & \phantom{1}27.2 &  Retal09 \\
	& I10 & & 21 & \phantom{1}4.4\phantom{0} & \phantom{1}52.5 &  Retal09 \\ 
	& I21 & & 11 & 10.3\phantom{0} & 138.2 &  Retal09 \\
	& J21 & & 21 & \phantom{1}4.2\phantom{0} & \phantom{1}60.2 &  Retal09 \\
	& J32 & & 14 & \phantom{1}8.3\phantom{0} & 119.5 &  Retal09 \\
	& J43 & & 11 & \phantom{1}9.6\phantom{0} & 134.6 &  Retal09 \\
R Gem & O10 & \phantom{1}650 & 33 & \phantom{1}0.3\phantom{0} & \phantom{11}2.4 &  Retal09 \\
	& I10 & & 21 & \phantom{1}1.0\phantom{0} & \phantom{11}7.2 &  Retal09 \\
	& I21 & & 11 & \phantom{1}2.7\phantom{0} & \phantom{1}17.7 &  Retal09 \\
	& A32 & & 18 & \phantom{1}1.0\phantom{0} & \phantom{11}6.5 &  Retal09 \\
ST Her & O10 & \phantom{1}290 & 33 & \phantom{1}0.2\phantom{0}  & \phantom{11}1.9 &  Retal09 \\
	& I10 & & 21 & \phantom{1}0.4\phantom{0} & \phantom{11}5.3 &  Wetal11 \\
	& I21 & & 11 & \phantom{1}1.5\phantom{0} & \phantom{1}18.4 &  Wetal11 \\
	& C32 & & 20 & \phantom{1}0.3\phantom{0} & \phantom{11}5.1 &  Ketal98 \\
Y Lyn & O10 & \phantom{1}250 & 33 & \phantom{1}0.3\phantom{0}  & \phantom{11}4.1 &  Retal09 \\
	& I10 & & 21 & \phantom{1}0.8\phantom{0}  & \phantom{1}10.6 &  Retal09 \\
	& I21 & & 11 & \phantom{1}2.3\phantom{0}  & \phantom{1}29.3 &  Retal09 \\
S Lyr & O10 & 2000 & 33 & \phantom{1}0.1\phantom{0} & \phantom{11}1.9 &  Retal09 \\
	& A32 & & 18 & \phantom{1}0.3\phantom{0} & \phantom{11}6.9 &  Retal09 \\
RT Sco & S10 & \phantom{1}400 & 45 & \phantom{1}0.2\phantom{0} & \phantom{11}4.4 &  S\&L95 \\
	& A32 & & 18 & \phantom{1}1.8\phantom{0} & \phantom{1}32.6 &  Retal09 \\
T Sgr & I10 & \phantom{1}700 & 21 & \phantom{1}0.2\phantom{0} & \phantom{11}1.2 &  Retal09 \\ 
	& I21 & & 11 & \phantom{1}0.4\phantom{0} & \phantom{11}6.1 &  Retal09 \\
	& A32 & & 18 & \phantom{1}0.2\phantom{0} & \phantom{11}2.5 &  Retal09 \\ 
DK Vul & I10 & & 21 & \phantom{1}0.6\phantom{0} & \phantom{11}2.2 &  Retal09 \\
	& I21 & & 11 & \phantom{1}1.7\phantom{0} & \phantom{11}7.1 &  Retal09 \\
	& A32 & & 18 & \phantom{1}0.7\phantom{0} & \phantom{11}3.6 &  Retal09 \\
EP Vul & N10 & \phantom{1}510 & 55 & \phantom{1}0.12 & \phantom{11}0.9 &  B\&L94 \\
	& S10 & & 45 & \phantom{1}0.1\phantom{0} & \phantom{11}0.8 &  S\&L95 \\
	& A32 & & 18 & \phantom{1}0.35 & \phantom{11}3.1 &  Retal09 \\
\noalign{\smallskip}
{\em Carbon stars:} & & & & & & \\
\noalign{\smallskip}
LP And & O10 & 630 & 33 & 3.0\phantom{0} & \phantom{1}63.0 & Retal08 \\
	& J21 & & 21 & 5.2\phantom{0} & 104.0 & Retal08 \\
	& J32 & & 14 & 6.5\phantom{0} & 124.0 & Retal08 \\
	& J43 & & 11 & 7.6\phantom{0} & 143.0 & Retal08 \\
	& J65 & & \phantom{0}8 & 7.0\phantom{0} & 139.0 & Retal08 \\
V Aql & S10 & 362 & 45 & 0.2\phantom{0} & \phantom{11}2.8 & S\&O01 \\
	& O10 & & 33 & 0.35 & \phantom{11}3.2 & S\&O01 \\
	& S21 & & 23 & 0.65 & \phantom{11}8.2 & S\&O01 \\
	& J21 & & 21 & 0.75 & \phantom{11}9.0 & S\&O01 \\
	& J32 & & 14 & 1.0\phantom{0} & \phantom{1}11.2 & S\&O01 \\
RV Aqr & S10 & 550 & 45 & 0.31 & \phantom{11}7.5 & S\&O01 \\
	& O21 & & 23 & 0.84 & \phantom{1}18.1 & S\&O01 \\
	& S32 & & 16 & 0.83 & \phantom{1}18.6 & S\&O01 \\
UU Aur & O10 & 240 & 33 & 0.45 & \phantom{11}7.9 & S\&O01 \\
	& I10 & & 21 & 0.88 & \phantom{1}18.8 & S\&O01 \\
	& J21 & & 21 & 0.9\phantom{1} & \phantom{1}16.8 & S\&O01 \\
	& I21 & & 11 & 2.12 & \phantom{1}39.0 & S\&O01 \\
X Cnc & N10 & 342 & 55 & 0.06 & \phantom{11}0.7 & S\&O01 \\
	& O10 & & 33 & 0.16 & \phantom{11}1.7 & S\&O01 \\
	& I21 & & 11 & 0.98 & \phantom{1}11.1 & S\&O01 \\
Y CVn & O10 & 321 & 33 & 0.33 & \phantom{11}4.5 & S\&O01 \\
	& I10 & & 21 & 0.75 & \phantom{1}10.3 & S\&O01 \\
	& J21 & & 21 & 0.95 & \phantom{1}11.9 & S\&O01 \\
	& J32 & & 14 & 2.2\phantom{1} & \phantom{1}20.9 & S\&O01 \\
V Cyg & O10 & 366 & 33 & 1.37 & \phantom{1}27.5 & S\&O01 \\
	& N21 & & 27 & 2.9\phantom{1} & \phantom{1}50.1 & S\&O01 \\
	& J21 & & 21 & 3.9\phantom{1} & \phantom{1}69.6 & S\&O01 \\
	& J32 & & 14 & 5.2\phantom{1} & \phantom{1}88.9 & S\&O01 \\
	& J43 & & 11 & 7.5\phantom{0} & 123.4 & S\&O01 \\
	\hline\hline   
\end{tabular}}
\tablebib{Retal09-\citet{ramsetal09}; Wetal11-\citet{walletal11}; Ketal98-\citet{knapetal98}; S\&L95-\citet{sahaliec95}; B\&L94-\citet{bieglatt94}; Retal08-\citet{ramsetal08}; S\&O01-\citet[][and references therein]{schoolof01}}
\end{table}

\begin{table}
\label{12CO4}
\caption{$^{12}$CO observations used to constrain the physical parameters.}      
\centering        
\resizebox{\hsize}{!}{    
\begin{tabular}{l c c c c c c l}       
\hline\hline                
Source & Code & D & $\theta$ & T$_{\rm{mb}}$ & I$_{\rm{obs}}$ &  Ref.\\    
	&  & [pc] & [''] & [K] & [K km s$^{-1}$] & \\
\hline
\noalign{\smallskip}
{\em Carbon stars:} & & & & & & \\
\noalign{\smallskip}
RY Dra & O10 & 431 & 33 & 0.12 & \phantom{11}2.4 & S\&O01 \\
	& I21 & & 11 & 0.56 & \phantom{1}21.6 & S\&O01 \\
	& J32 & & 14 & 0.9\phantom{1} & \phantom{1}18.9 & S\&O01 \\
UX Dra & O10 & 386 & 33 & 0.4\phantom{1} & \phantom{11}2.1 & S\&O01 \\
	& I21 & & 11 & 1.1\phantom{1} & \phantom{11}6.8 & S\&O01 \\
U Hya & S10 & \phantom{1}208 & 45 & \phantom{1}0.46 & \phantom{111}5.4 & S\&O01 \\
	& S21 & & 23 & \phantom{1}1.4\phantom{1} & \phantom{11}13.8 & S\&O01 \\
	& J21 & & 21 & \phantom{1}2.0\phantom{1} & \phantom{11}20.2 & S\&O01 \\
	& I21 & & 11 & \phantom{1}4.8\phantom{1} & \phantom{11}48.8 & S\&O01 \\
	& A32 & & 18 & \phantom{1}2.3\phantom{0} & \phantom{11}24.3 &  R\&O14 \\
CW Leo & O10 & \phantom{1}120 & 33 & 15.3\phantom{1} & \phantom{1}386\phantom{.1} & Retal08 \\
	& J21 & & 21 & 30.7\phantom{1} & \phantom{1}689\phantom{.1} & Retal08 \\
	& J32 & & 14 & 48.7\phantom{1} & 1070\phantom{.1} & Retal08 \\
	& J43 & & 11 & 56.8\phantom{1} & 1230\phantom{.1} & Retal08 \\
R Lep & S10 & \phantom{1}432 & 45 & \phantom{1}0.21 & \phantom{111}6.2 & S\&O01 \\
	& I10 & & 21 & \phantom{1}0.92 & \phantom{11}31.7 & S\&O01 \\
	& S21 & & 23 & \phantom{1}0.80 & \phantom{11}18.1 & S\&O01 \\
	& I21 & & 11 & \phantom{1}1.91 & \phantom{11}56.4 & S\&O01 \\
	& J32 & & 14 & \phantom{1}0.9\phantom{0} & \phantom{11}27.9 &  S\&O01 \\
RW LMi & O10 & \phantom{1}400 & 33 & \phantom{1}3.6\phantom{1} & \phantom{1}108\phantom{.1} & Retal08 \\
	& J21 & & 21 & \phantom{1}4.7\phantom{1} & \phantom{1}123\phantom{.1} & Retal08 \\
	& J32 & & 14 & \phantom{1}9.6\phantom{1} & \phantom{1}244\phantom{.1} & Retal08 \\
	& J43 & & 11 & 10.0\phantom{1} & \phantom{1}246\phantom{.1} & Retal08 \\
T Lyr & O10 & \phantom{1}719 & 33 & \phantom{1}0.03 & \phantom{111}0.7 & S\&O01 \\
	& I21 & & 11 & \phantom{1}0.25 & \phantom{111}5.6 & S\&O01 \\
	& J32 & & 14 & \phantom{1}0.3\phantom{1} & \phantom{111}6.4 & S\&O01 \\
W Ori & S10 & \phantom{1}377 & 45 & \phantom{1}0.06 & \phantom{111}1.2 & S\&O01 \\
	& I10 & & 21 & \phantom{1}0.27 & \phantom{111}5.0 & S\&O01 \\
	& S21 & & 23 & \phantom{1}0.27 & \phantom{111}4.9 & S\&O01 \\
	& I21 & & 11 & \phantom{1}1.01 & \phantom{11}19.0 & S\&O01 \\
V384 Per & O10 & \phantom{1}600 & 33 & \phantom{1}1.8\phantom{1} & \phantom{11}35\phantom{.1} & Retal08 \\
	& J21 & & 21 & \phantom{1}2.8\phantom{1} & \phantom{11}63\phantom{.1} & Retal08 \\
	& J32 & & 14 & \phantom{1}1.9\phantom{1} & \phantom{11}40\phantom{.1} & Retal08 \\
	& J43 & & 11 & \phantom{1}3.9\phantom{1} & \phantom{11}79\phantom{.1} & Retal08 \\
AQ Sgr & S10 & \phantom{1}333 & 45 & \phantom{1}0.09 & \phantom{111}1.4 & S\&O01 \\
	& S21 & & 23 & \phantom{1}0.30 & \phantom{111}4.1 & S\&O01 \\
AFGL 3068 & O10 & 1300 & 33 & \phantom{1}2.3\phantom{1} & \phantom{11}48\phantom{.1} & Retal08 \\
	& J21 & & 21 & \phantom{1}3.9\phantom{1} & \phantom{11}72\phantom{.1} & Retal08 \\
	& J32 & & 14 & \phantom{1}4.2\phantom{1} & \phantom{11}73\phantom{.1} & Retal08 \\
	& J43 & & 11 & \phantom{1}5.3\phantom{1} & \phantom{11}93\phantom{.1} & Retal08 \\
	& J65 & & \phantom{1}8 & \phantom{1}3.8\phantom{1} & \phantom{11}55\phantom{.1} & Retal08 \\
IRAS 15194-5115 & S10 & \phantom{1}500 & 45 & \phantom{1}1.5\phantom{1} & \phantom{11}60.7 & Netal93 \\
	& S21 & & 23 & \phantom{1}4.3\phantom{1} & \phantom{1}150.9 & Netal93 \\
	& S32 & & 14 & \phantom{1}4.1\phantom{1} & \phantom{1}130.4 & Retal99 \\
	& A32 & & 18 & \phantom{1}5.3\phantom{1} & \phantom{1}163.0 &  R\&O14 \\
	\hline\hline   
\end{tabular}}
\tablebib{S\&O01-\citet[][and references therein]{schoolof01}; R\&O14-This work; Retal08-\citet{ramsetal08}; Netal93-\citet{nymaetal93}; Retal99-\citet{rydeetal99}}
\end{table}

\begin{table}
\label{13CO1}
\caption{$^{13}$CO observations used to constrain the $^{12}$CO/$^{13}$CO ratio.}      
\centering        
\resizebox{\hsize}{!}{    
\begin{tabular}{l c c c c c l}       
\hline\hline                
Source & Code & D & $\theta$ & T$_{\rm{mb}}$ & I$_{\rm{obs}}$ &  Ref.\\    
	&  & [pc] & [''] & [K] & [K km s$^{-1}$]  & \\
\hline
\noalign{\smallskip}
{\em M-type stars:} & & & & & & \\
\noalign{\smallskip}
RX Boo & J21 & \phantom{1}137 & 22 & 0.17 & \phantom{1}2.3 & R\&O14 \\
	& A21 & & 28 & 0.15 & \phantom{1}2.1 & R\&O14 \\ 
	& J32 & & 15 & 0.14 & \phantom{1}5.5 & Detal10 \\
TX Cam & J21 & \phantom{1}380 & 22 & 0.13 & \phantom{1}3.3 & R\&O14 \\
	& J32 & & 15 & 0.29 & 10.9 & Detal10 \\
	& H65 & & 31 & - & \phantom{1}1.4 & Jetal12 \\
R Cas & J21 & \phantom{1}176 & 22 & 0.19 & \phantom{1}3.1 & R\&O14 \\
	& H65 & & 31 & - & \phantom{1}1.5 & Jetal12 \\
R Dor & A21 & \phantom{11}55 & 28 & 0.2\phantom{1} & \phantom{1}2.0 & R\&O14 \\
	& A32 & & 19 & 0.5\phantom{1} & \phantom{1}4.7 & R\&O14 \\
	& H65 & & 31 & - & \phantom{1}2.7 & Jetal12 \\
W Hya & H65 & \phantom{11}98 & 31 & - & \phantom{1}1.1 & Jetal12 \\
R Leo & J21 & \phantom{11}71 & 22 & 0.14 & \phantom{1}1.0 & R\&O14 \\
	& A21 & & 28 & 0.12 & \phantom{1}1.3 & R\&O14 \\
	& A32 & & 19 & 0.22 & \phantom{1}2.7 & R\&O14 \\
GX Mon & O10 & \phantom{1}550 & 33 & 0.05 & \phantom{1}1.5 & R\&O14 \\
	& J21 & & 22 & 0.27 & 10.0 & R\&O14 \\
	& A21 & & 28 & 0.23 & \phantom{1}8.5 & R\&O14 \\
WX Psc & O10 & \phantom{1}700 & 33 & 0.11 & \phantom{1}4.1 & R\&O14 \\
	& J21 & & 22 & 0.4\phantom{1} & 15.5 & R\&O14 \\
	& A21 & & 28 & 0.31 & 10.9 & R\&O14 \\
	& A32 & & 19 & 0.32 & 10.8 & R\&O14 \\
	& H65 & & 31 & - & \phantom{1}1.1 & Jetal12 \\
RT Vir & J21 & \phantom{1}226 & 22 & 0.1\phantom{1} & \phantom{1}1.3 & R\&O14 \\
	& A21 & & 28 & 0.11 & \phantom{1}1.2 & R\&O14 \\ 
	& A32 & & 19 & 0.12 & \phantom{1}1.7 & R\&O14 \\
SW Vir & A21 & \phantom{1}143 & 28 & 0.1\phantom{1} & \phantom{1}1.1 & R\&O14 \\
IK Tau & J21 & \phantom{1}500 & 22 & 0.3\phantom{1} & 12.4 & R\&O14 \\
	& A32 & & 19 & 0.38 & 12.9 & R\&O14 \\
	& H65 & & 31 & - & \phantom{1}2.6 & Jetal12 \\
CIT4 & J21 & \phantom{1}800 & 22 & 0.08 & \phantom{1}2.6 & R\&O14 \\
IRC+10365 & O10 & \phantom{1}650 & 33 & 0.05 & \phantom{1}1.3 & R\&O14 \\
	& J21 & & 22 & 0.2\phantom{1} & \phantom{1}7.3 & R\&O14 \\
	& A21 & & 28 & 0.2\phantom{1} & \phantom{1}6.0 & R\&O14 \\
IRC-10529 & A21 & \phantom{1}620 & 28 & 0.4\phantom{1} & \phantom{1}9.7 & R\&O14 \\
	& A32 & & 19 & 0.3\phantom{1} & \phantom{1}7.2 & R\&O14 \\ 
IRC-30398 & A21 & \phantom{1}550 & 28 & 0.12 & \phantom{1}3.5 & R\&O14 \\
IRC+40004 & J21 & \phantom{1}600 & 22 & 0.12 & \phantom{1}4.0 & R\&O14 \\
IRC+50137 & J21 & 1500 & 22 & 0.2\phantom{1} & \phantom{1}7.0 & R\&O14 \\
IRC+60169 & J21 & & 21 & 0.15 & \phantom{1}5.6 & R\&O14 \\
IRC+70066 & O10 & \phantom{1}400 & 34 & 0.05 & \phantom{1}1.1 & R\&O14 \\
\noalign{\smallskip}
{\em S-type stars:} & & & & & \\
\noalign{\smallskip}
R And & I10 & \phantom{1}350 & 21 & 0.08 & \phantom{1}1.5 & Wetal11 \\
	& I21 & & 11 & 0.25 & \phantom{1}6.4  & Wetal11 \\
	& J21 & & 21 & 0.18 & \phantom{1}2.4 & R\&O14 \\
W Aql & I10 & \phantom{1}300 & 21 & 0.2\phantom{1} & \phantom{1}4.1 & Wetal11 \\
	& I21 & & 11 & 0.4\phantom{1} & 11.1 & Wetal11 \\
	& J21 & & 22 & 0.23 & \phantom{1}6.2 & R\&O14 \\
	& A21 & & 28 & 0.24 & \phantom{1}7.1 & R\&O14 \\
	& A32 & & 19 & 0.32 & \phantom{1}8.7 & R\&O14 \\
TV Aur & I10 & \phantom{1}400 & 21 & - & $<$0.17 & Wetal11 \\
	& I21 & & 11 & - & $<$0.25 & Wetal11 \\
AA Cam & I10 & \phantom{1}780 & 21 & - & $<$0.1\phantom{1} & Wetal11 \\
	& I21 & & 11 & - & $<$0.09 & Wetal11 \\
S Cas & I10 & \phantom{1}570 & 21 & 0.03 & \phantom{1}0.8 & Wetal11 \\
	& I21 & & 11 & 0.07 & \phantom{1}2.4 & Wetal11 \\
TT Cen & A21 & 1080 & 28 & 0.02 & \phantom{1}0.7 & R\&O14 \\
T Cet & J21 & \phantom{1}270 & 21 & - & \phantom{1}0.3 & Detal10 \\
	& J32 & & 15 & - & \phantom{1}0.5 & Detal10 \\
	& A32 & & 19 & 0.05 & \phantom{1}0.3 & R\&O14 \\
R Cyg & I10 & \phantom{1}600 & 21 & 0.04 & \phantom{1}0.5 & Wetal11 \\
	& I21 & & 11 & 0.1\phantom{0} & \phantom{1}1.9 & Wetal11 \\
	& J21 & & 21 & 0.08 & \phantom{1}1.3 & R\&O14 \\
\hline \hline            
\end{tabular}}
\tablebib{R\&O14-This work; Detal10-\citet{debeetal10}; Jetal12-\citet{justetal12}; Wetal11-\citet{walletal11}}
\end{table}

\begin{table}
\label{13CO2}
\caption{$^{13}$CO observations used to constrain the $^{12}$CO/$^{13}$CO ratio.}      
\centering        
\resizebox{\hsize}{!}{    
\begin{tabular}{l c c c c c l}       
\hline\hline                
Source & Code & D & $\theta$ & T$_{\rm{mb}}$ & I$_{\rm{obs}}$ &  Ref.\\    
	&  & [pc] & [''] & [K] & [K km s$^{-1}$]  & \\
\hline
\noalign{\smallskip}
{\em S-type stars:} & & & & & \\
\noalign{\smallskip}
$\chi$ Cyg & O10 & \phantom{1}180 & 33 & 0.07 & \phantom{1}0.8 & R\&O14 \\
	& I10 & & 21 & 0.1\phantom{1} & \phantom{1}1.0 & Wetal11 \\
	& I21 & & 11 & 0.3\phantom{1} & \phantom{1}3.8 & Wetal11 \\
	& J32 & & 21 & - & \phantom{1}4.7 & Detal10 \\
R Gem & J21 & \phantom{1}650 & 22 & 0.08 & \phantom{1}0.6\phantom{1} & R\&O14 \\
ST Her & I10 & \phantom{1}290 & 21 & 0.01 & \phantom{1}0.15 & Wetal11 \\
	& I21 & & 11 & 0.04 & \phantom{1}0.6\phantom{1} & Wetal11 \\
Y Lyn & I10 & \phantom{1}250 & 21 & 0.02 & \phantom{1}0.3\phantom{1} & Wetal11 \\
	& I21 & & 11 & 0.06 & \phantom{1}0.8\phantom{1} & Wetal11 \\
	& J21 & & 22 & 0.04 & \phantom{1}0.7\phantom{1} & R\&O14 \\
S Lyr & J21 & 2000 & 22 & 0.04 & \phantom{1}0.6\phantom{1} & R\&O14 \\
RT Sco & A21 & \phantom{1}400 & 28 & 0.06 & \phantom{1}0.6\phantom{1} & R\&O14 \\
T Sgr & I10 & \phantom{1}700 & 21 & -& $<$0.2\phantom{1} & Wetal11 \\
	& I21 & & 11 & - & $<$0.5\phantom{1} & Wetal11 \\
DK Vul & J21 & & 22 & 0.05 & \phantom{1}0.2\phantom{1} & R\&O14 \\ 
EP Vul & J21 & & 22 & 0.03 & \phantom{1}0.3\phantom{1} & R\&O14 \\
\noalign{\smallskip}
{\em Carbon stars:} & & & & & \\
\noalign{\smallskip}
LP And & O10 & \phantom{1}630 & 33 & 0.16 & \phantom{1}3.6\phantom{1} & R\&O14 \\
	& J21 &  & 22 & 0.43 & 10.2\phantom{1} & R\&O14 \\
	& J32 &  & 15 & - & 10.0\phantom{1} & S\&O00 \\
V Aql & S10 & \phantom{1}362 & 45 & - & $<$0.23 & S\&O00 \\
      & S21 & & 24 & - & $<$0.28 & S\&O00 \\
RV Aqr & S10 & \phantom{1}550 & 45 & 0.02 & \phantom{1}0.6\phantom{1} & S\&O00 \\
UU Aur & O10 & \phantom{1}240 & 33 & 0.02 & \phantom{1}0.17 & S\&O00 \\
X Cnc & S10 & \phantom{1}342 & 45 & - & $<$0.28 & S\&O00 \\
      & S21 & & 24 & - & $<$0.27 & S\&O00 \\
Y CVn & N10 & \phantom{1}321 & 55 & - & \phantom{1}1.0\phantom{1} & S\&O00 \\
      & O10 & & 33 & 0.26 & \phantom{1}3.4\phantom{1} & S\&O00 \\
      & J32 & & 15 & - & \phantom{1}7.9\phantom{1} & S\&O00 \\
V Cyg & O10 & \phantom{1}366 & 33 & 0.10 & \phantom{1}2.3\phantom{1} & S\&O00 \\
      & J32 & & 15 & - & \phantom{1}6.7\phantom{1} & S\&O00 \\
RY Dra & J32 & \phantom{1}431 & 15 & - & \phantom{1}5.8\phantom{1} & S\&O00 \\
UX Dra & O10 & \phantom{1}386 & 33 & 0.02 & \phantom{1}0.14 & S\&O00  \\
U Hya & S10 & \phantom{1}208 & 45 & 0.02 & \phantom{1}0.2\phantom{1} & S\&O00 \\
	& S21 & & 24 & 0.09 & \phantom{1}0.8\phantom{1} & S\&O00 \\
CW Leo & N10 & \phantom{1}321 & 55 & - & 16.9\phantom{1} & S\&O00 \\
	& S10 & & 45 & 1.44 & 24.3\phantom{1} & S\&O00 \\
	& O10 & & 33 & 1.5\phantom{1} & 22.1\phantom{1} & R\&O14 \\
 	& S21 & & 24 & - & 64.9\phantom{1} & S\&O00 \\
	& J21 & & 22 & 5.0\phantom{1} & 91.7\phantom{1} & R\&O14 \\
	& A21 & & 28 & 4.8\phantom{1} & 91.6\phantom{1} & R\&O14 \\
	& J32 & & 15 & - & 157.3\phantom{1} & S\&O00 \\
R Lep & S10 & \phantom{1}432 & 45 & - & $<$0.2\phantom{1} & S\&O00 \\
	& S21 & & 24 & 0.02 & \phantom{1}0.5\phantom{1} & S\&O00 \\
RW LMi & N10 & \phantom{1}400 & 55 & - & \phantom{1}2.4\phantom{1} & S\&O00 \\ 
	& O10 & & 33 & 0.2\phantom{1} & \phantom{1}3.5\phantom{1} & R\&O14 \\
	& J21 & & 22 & 0.6\phantom{1} & 13.7\phantom{1} & R\&O14 \\
	& J32 & & 15 & - & 13.8\phantom{1} & S\&O00 \\
T Lyr & J32 & \phantom{1}719 & 15 & - & \phantom{1}1.8\phantom{1} & S\&O00 \\
W Ori & S10 & \phantom{1}377 & 45 & - & $<$0.18 & S\&O00 \\
	& S21 & & 24 & - & $<$0.25 & S\&O00 \\
V384 Per & O10 & \phantom{1}600 & 33 & 0.05 & \phantom{1}1.3\phantom{1} & R\&O14 \\
	& J21 & & 22 & 0.2\phantom{1} & \phantom{1}4.6\phantom{1} & R\&O14 \\
	& J32 & & 15 & - & \phantom{1}4.4\phantom{1} & S\&O00 \\
AQ Sgr & S10 & \phantom{1}333 & 45 & - & $<$0.12 & S\&O00 \\
	& S21 & & 23 & - & $<$0.27 & S\&O00 \\
AFGL 3068 & O10 & 1300 & 33 & 0.12 & \phantom{1}3.3\phantom{1} & R\&O14 \\
	& J21 & & 22 & 0.55 & 12.8\phantom{1} & R\&O14 \\
	& A21 & & 28 & 0.6\phantom{1} & 13.2\phantom{1} & R\&O14 \\
	& A32 & & 19 & 0.6\phantom{1} & 14.2\phantom{1} & R\&O14 \\
IRAS15194-5115 & S10 & \phantom{1}500 & 45 & - & 14.1\phantom{1} & S\&O00 \\
	& S21 & & 24 & - & 30.5\phantom{1} & S\&O00 \\
\hline               
\end{tabular}}
\tablebib{R\&O14-This work; Wetal11-\citet{walletal11}; S\&O00-\citet{schoolof00}}
\end{table}

\begin{figure*}[h]
\center
   \includegraphics[width=3.5cm]{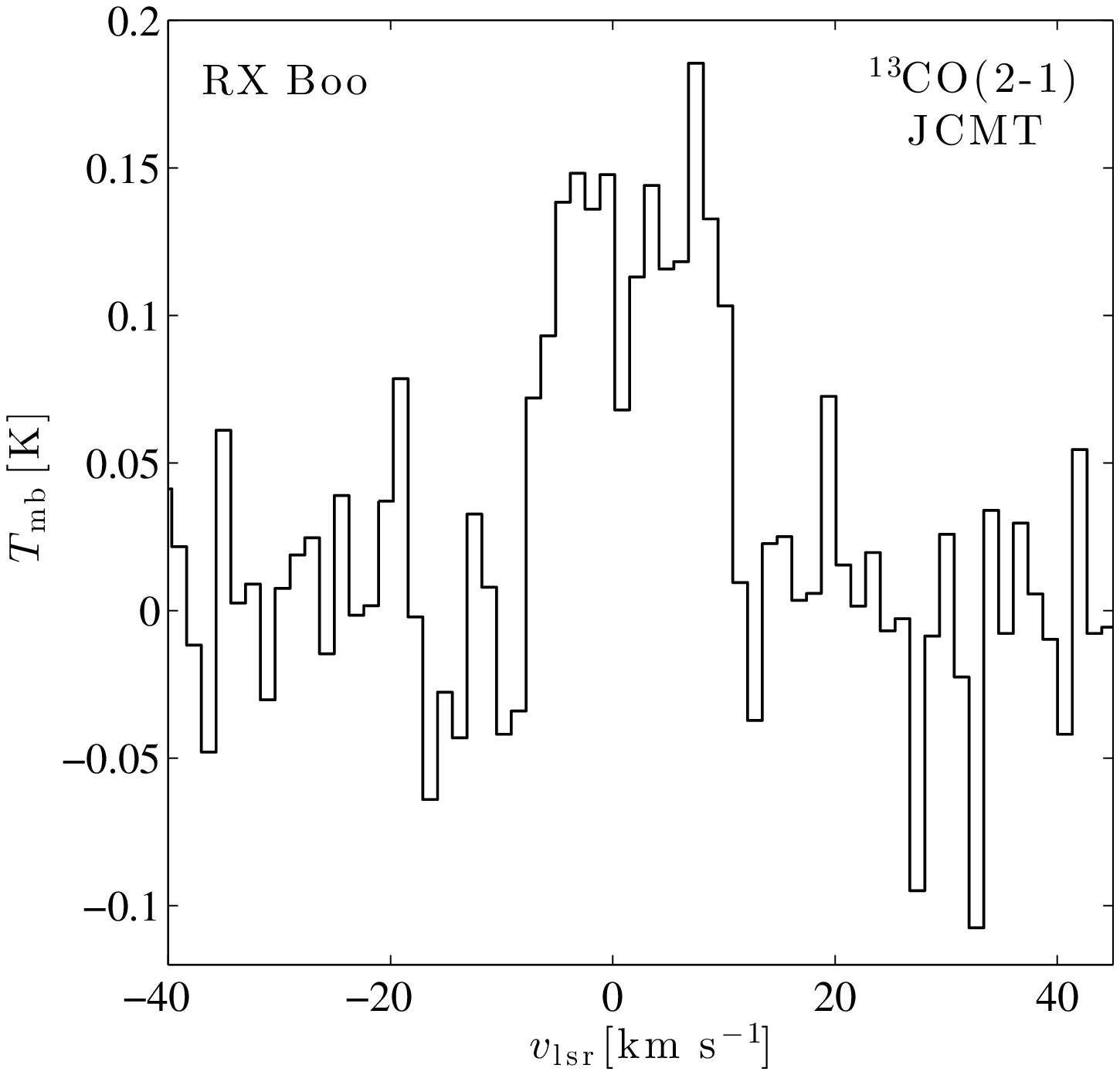} 
   \includegraphics[width=3.5cm]{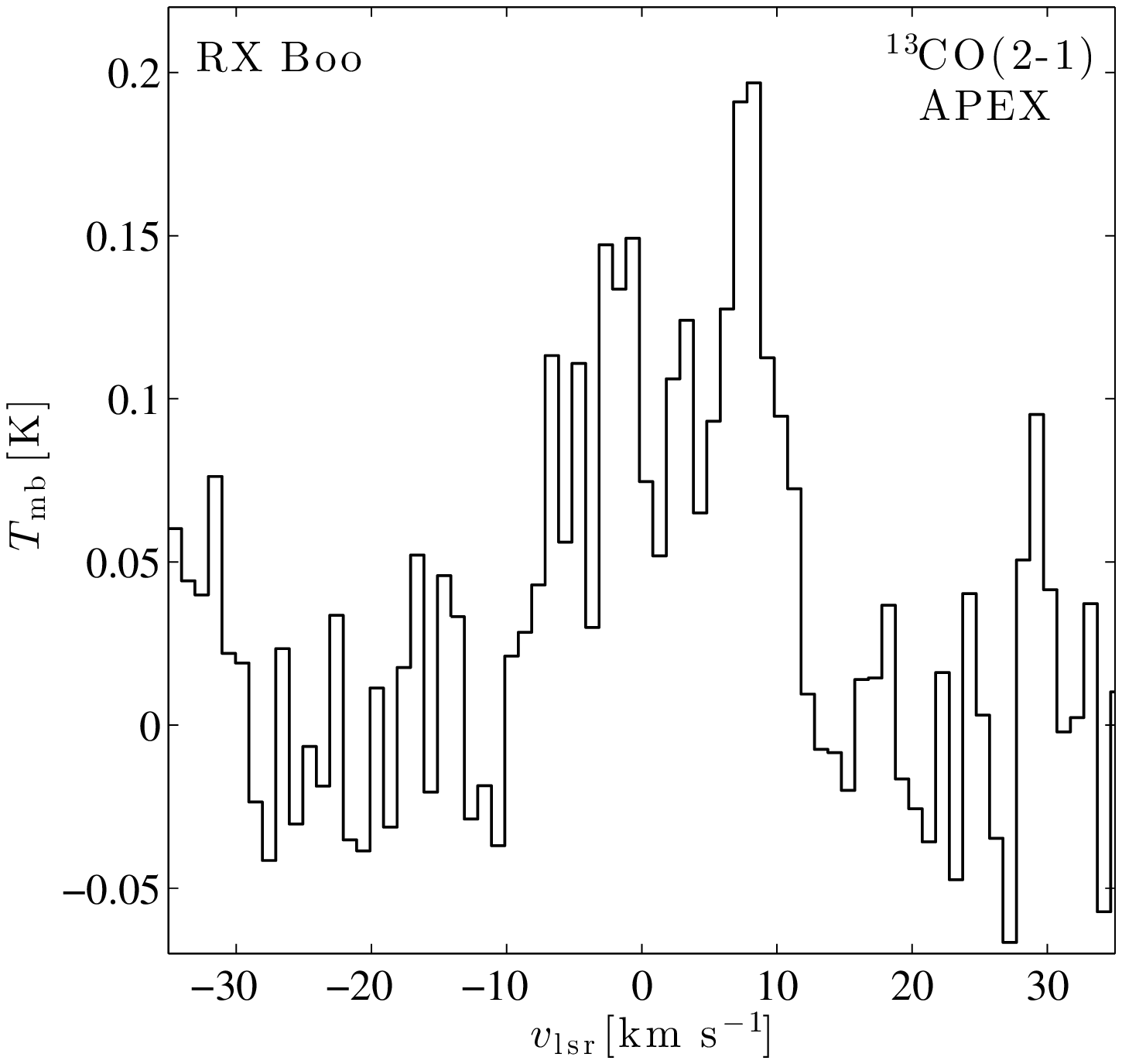}
   \includegraphics[width=3.5cm]{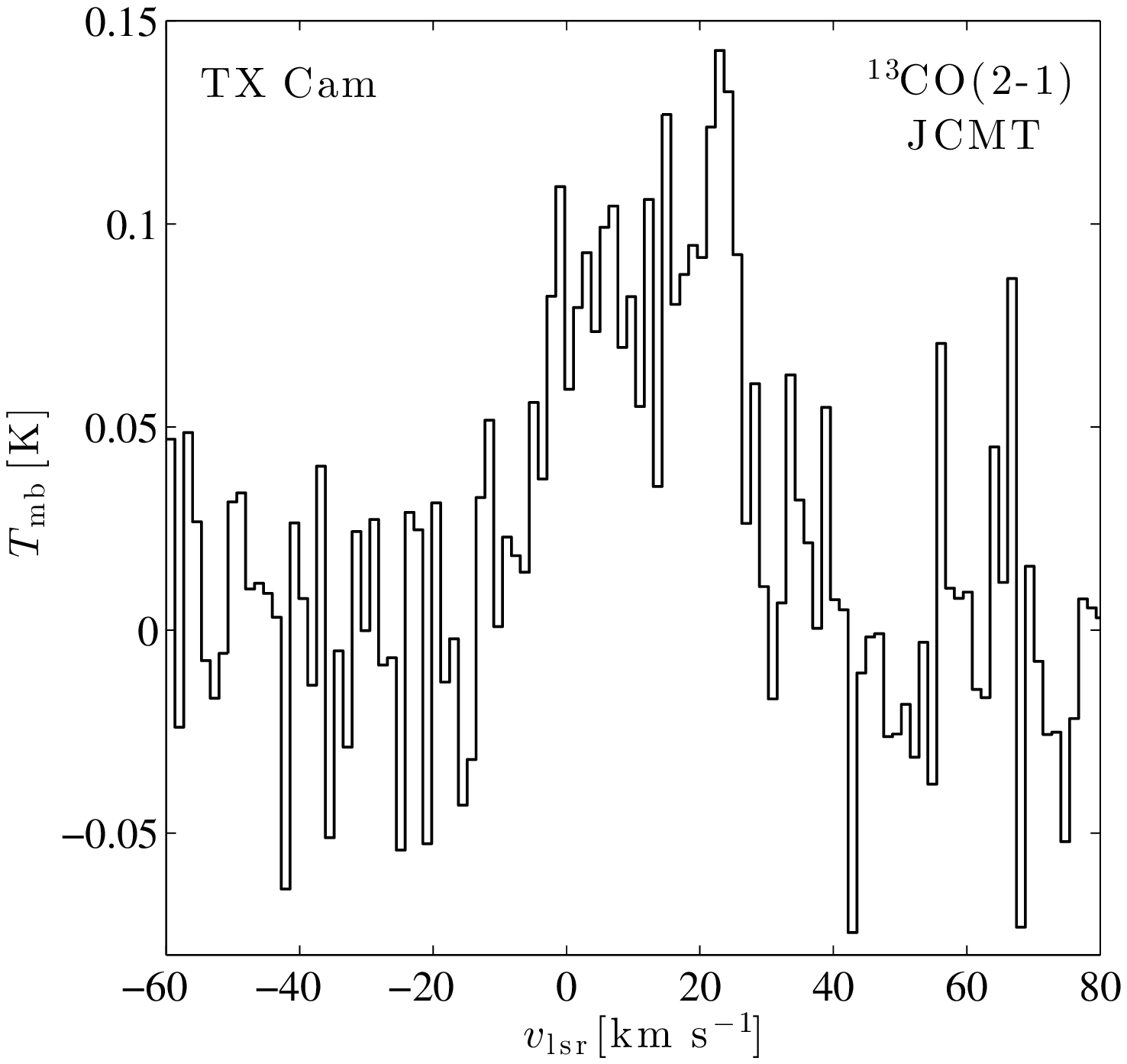}
   \includegraphics[width=3.5cm]{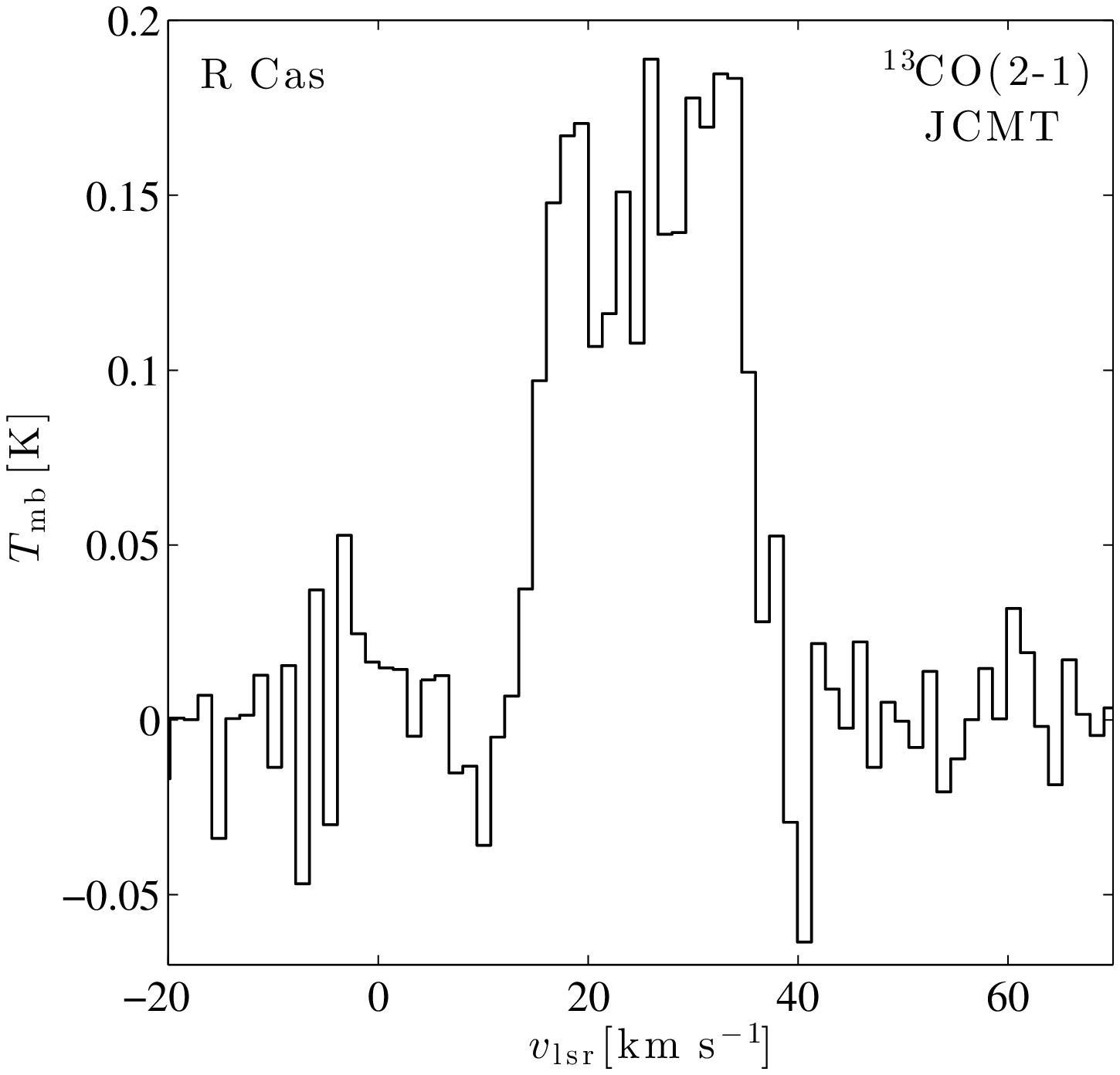}
   \includegraphics[width=3.5cm]{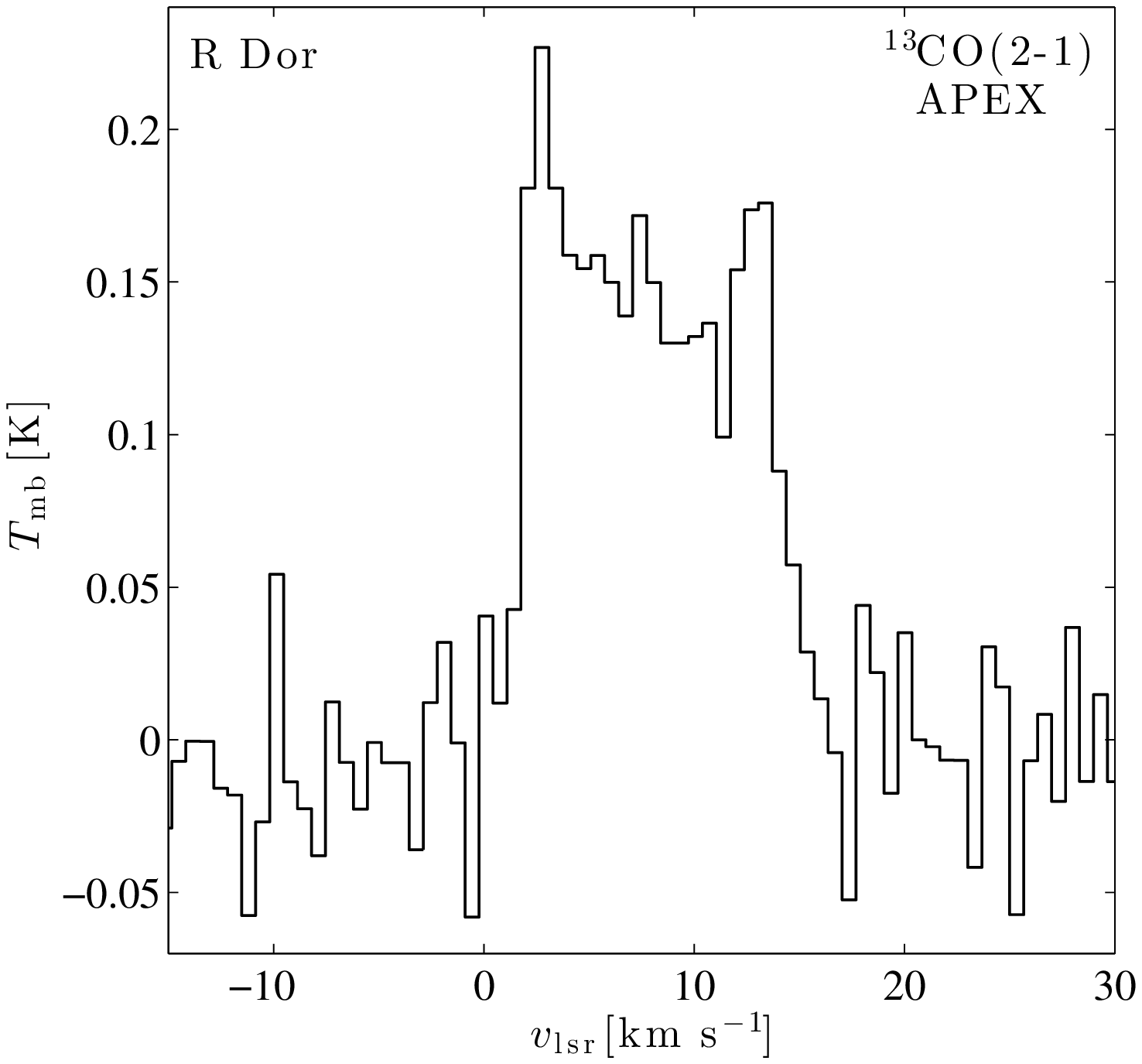}
   \includegraphics[width=3.5cm]{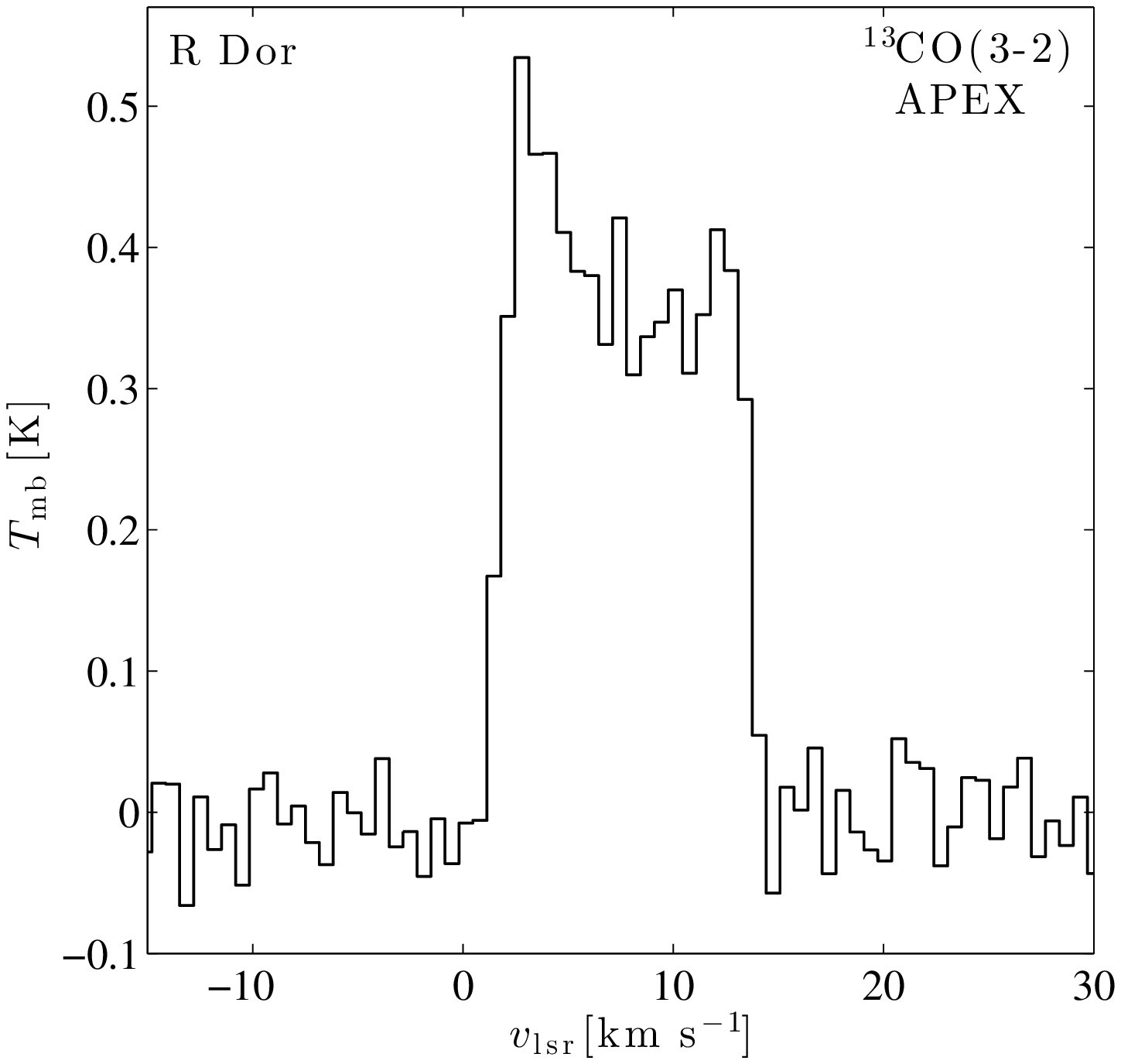} 
   \includegraphics[width=3.5cm]{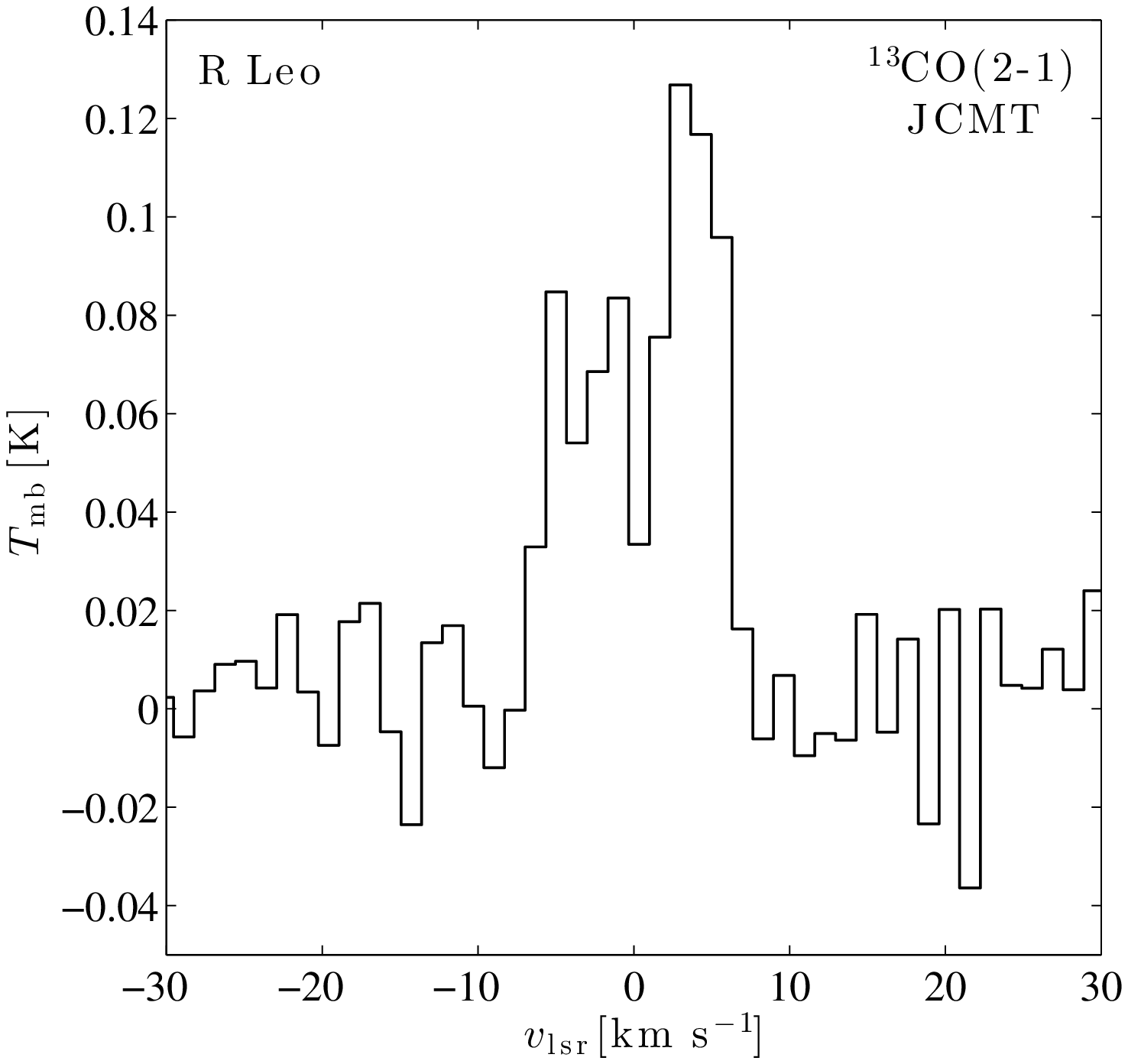}
   \includegraphics[width=3.5cm]{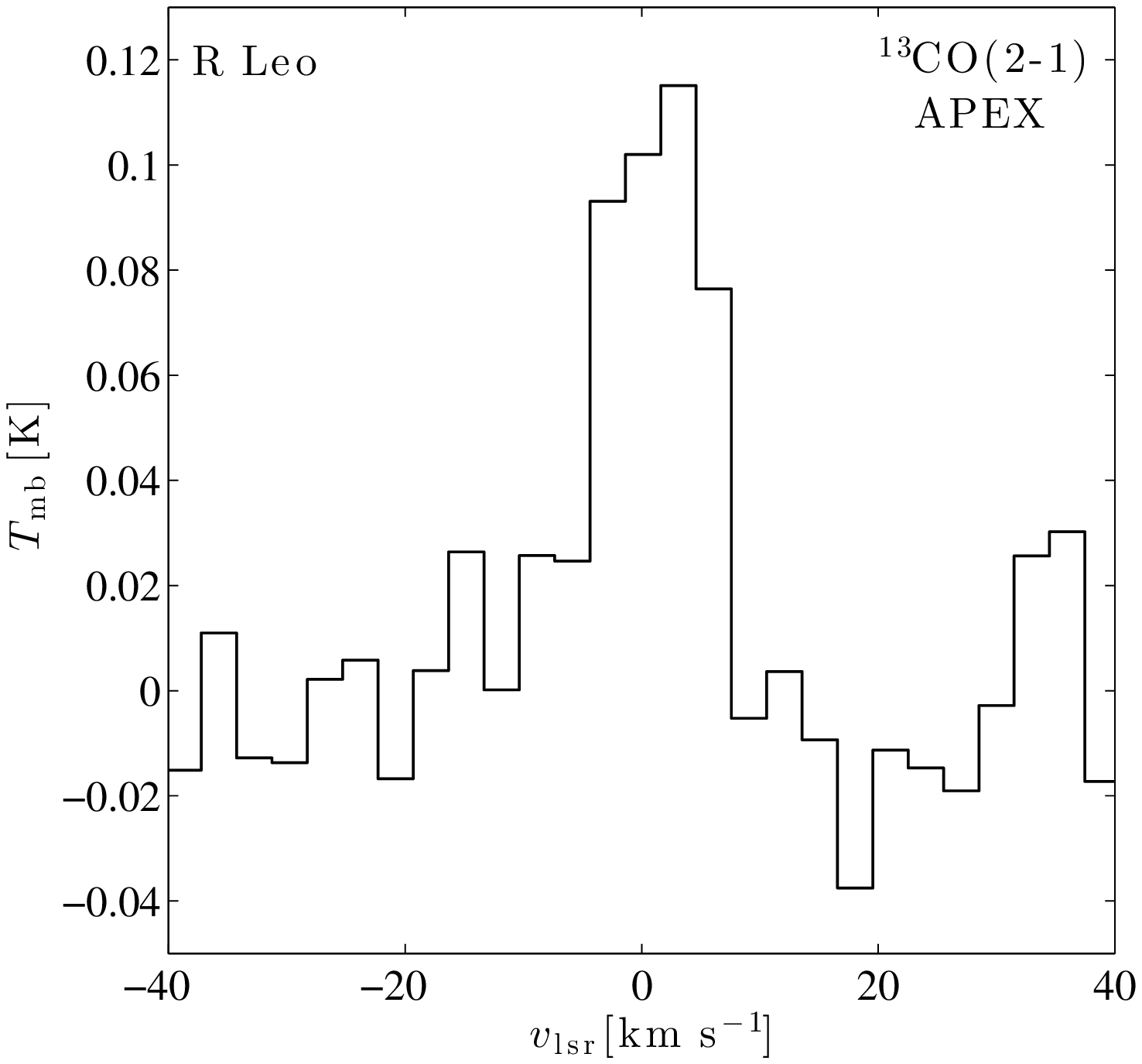}
   \includegraphics[width=3.5cm]{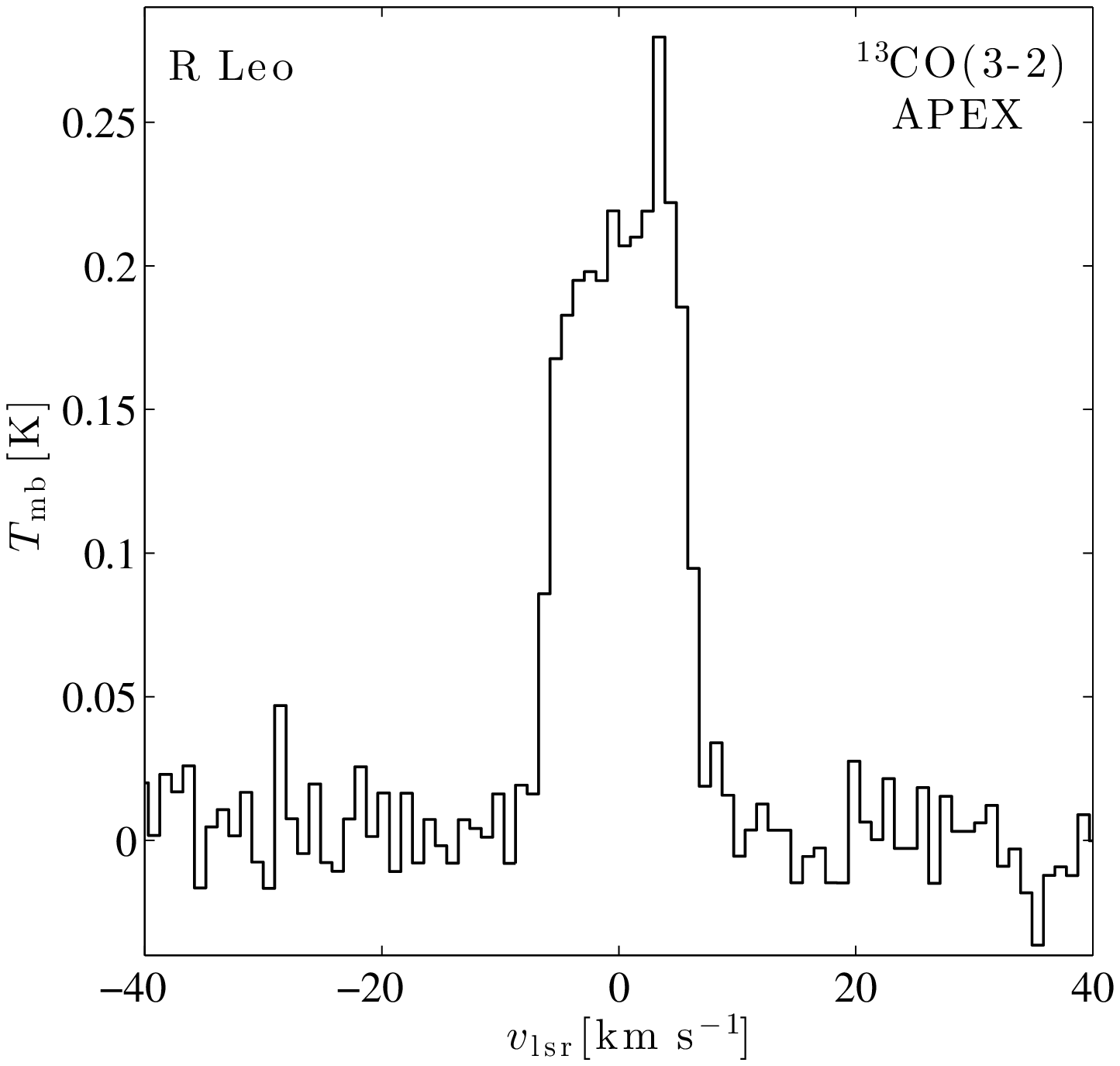}
   \includegraphics[width=3.5cm]{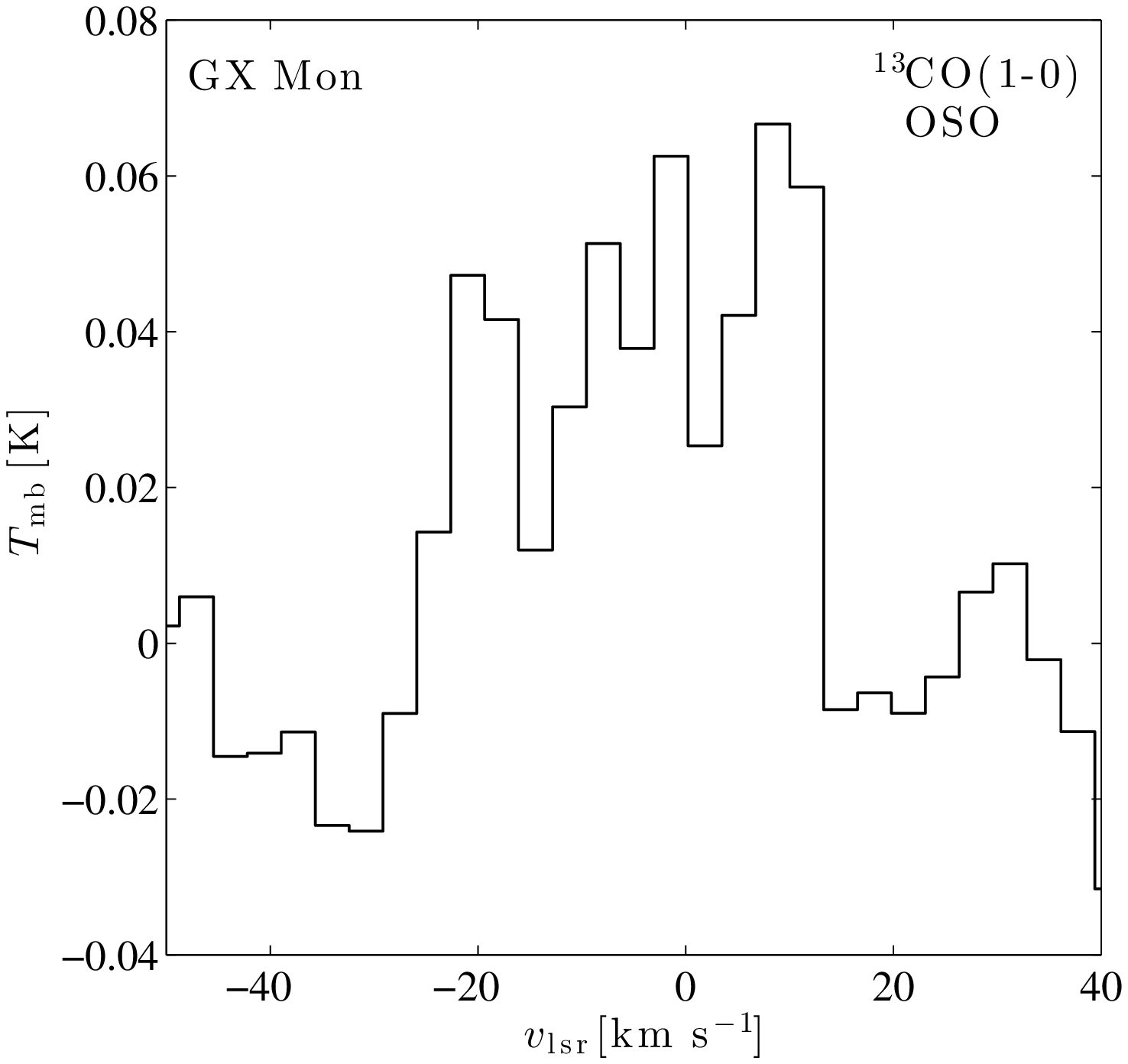}
   \includegraphics[width=3.5cm]{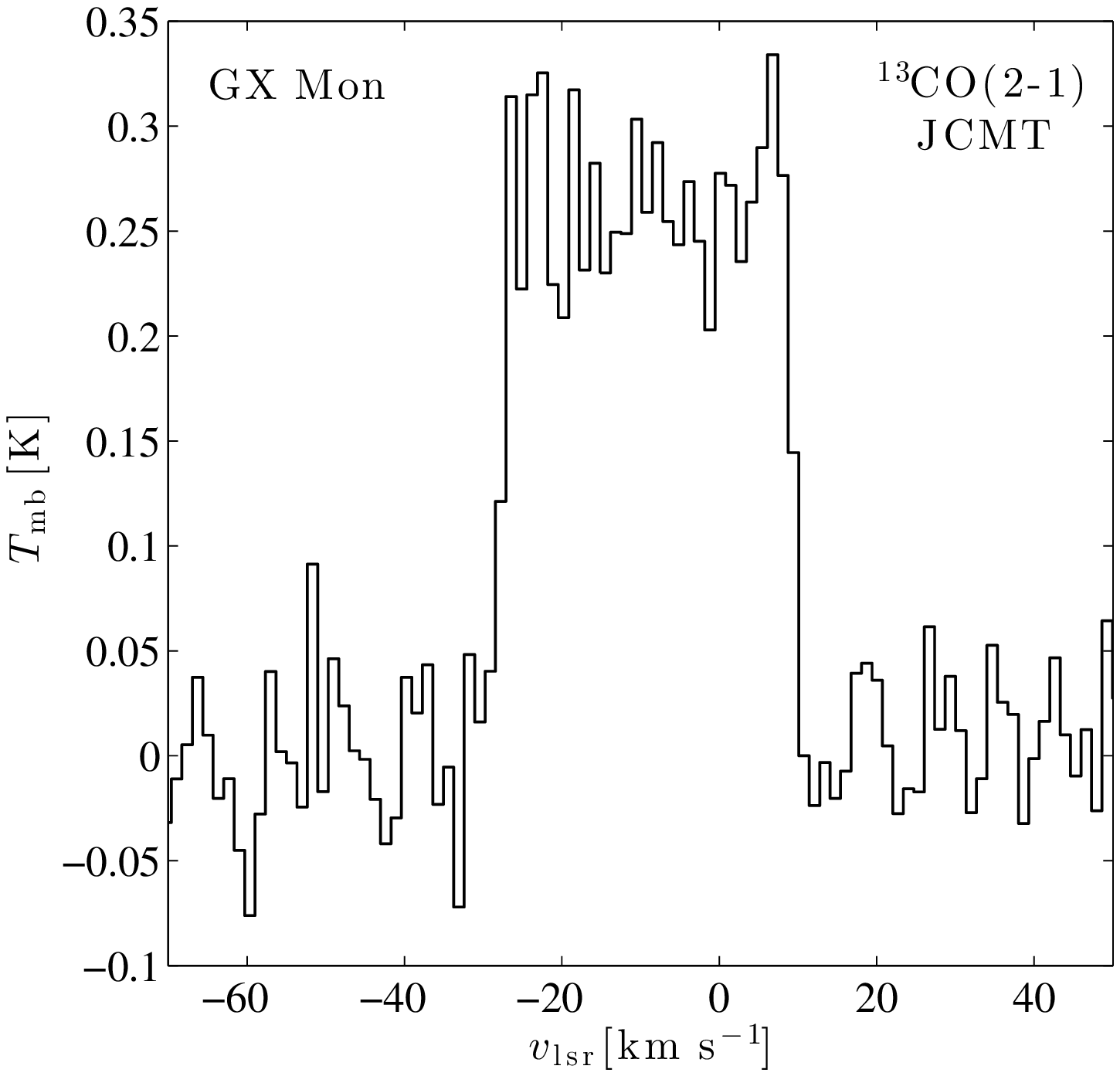} 
   \includegraphics[width=3.5cm]{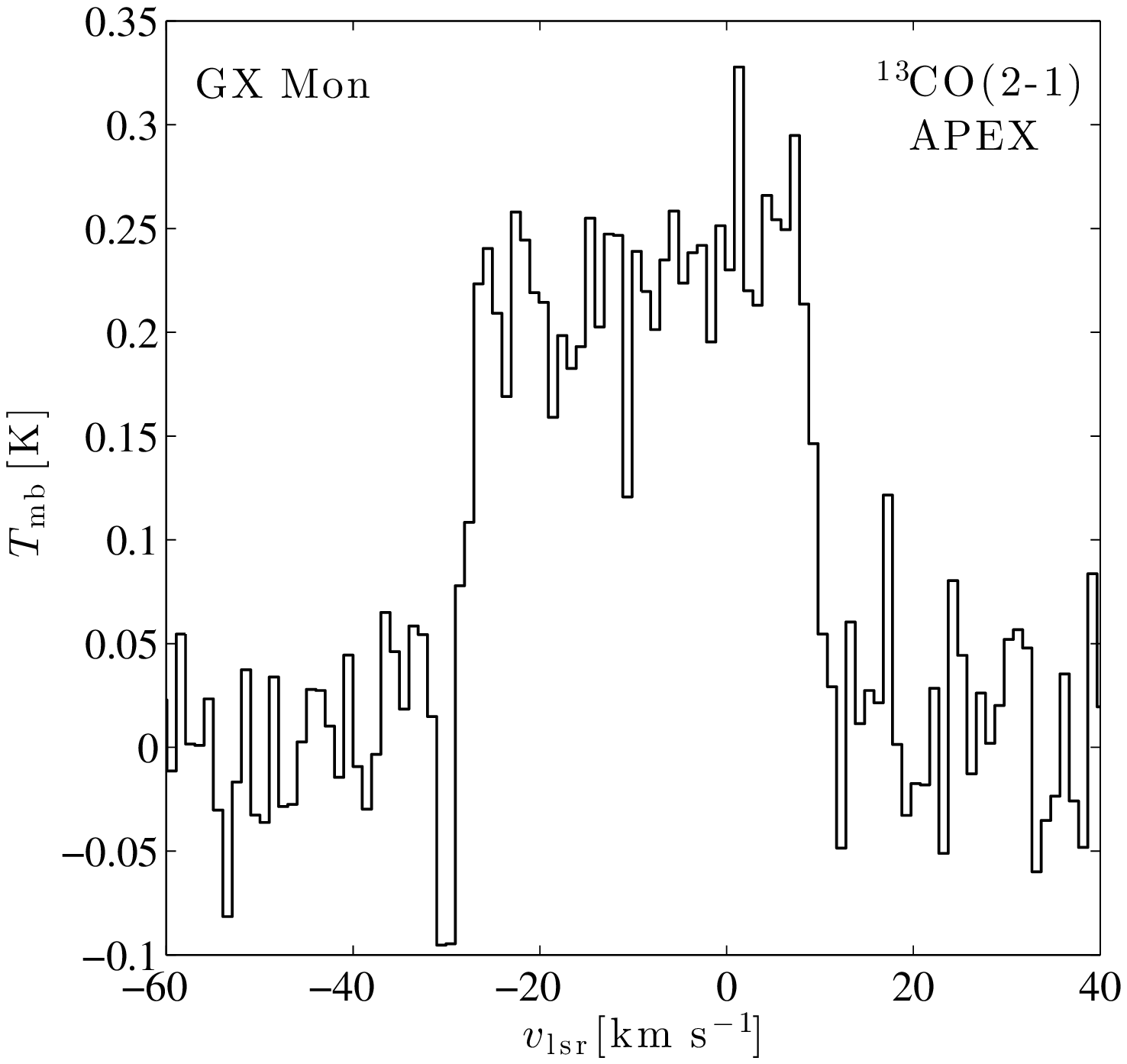}
   \includegraphics[width=3.5cm]{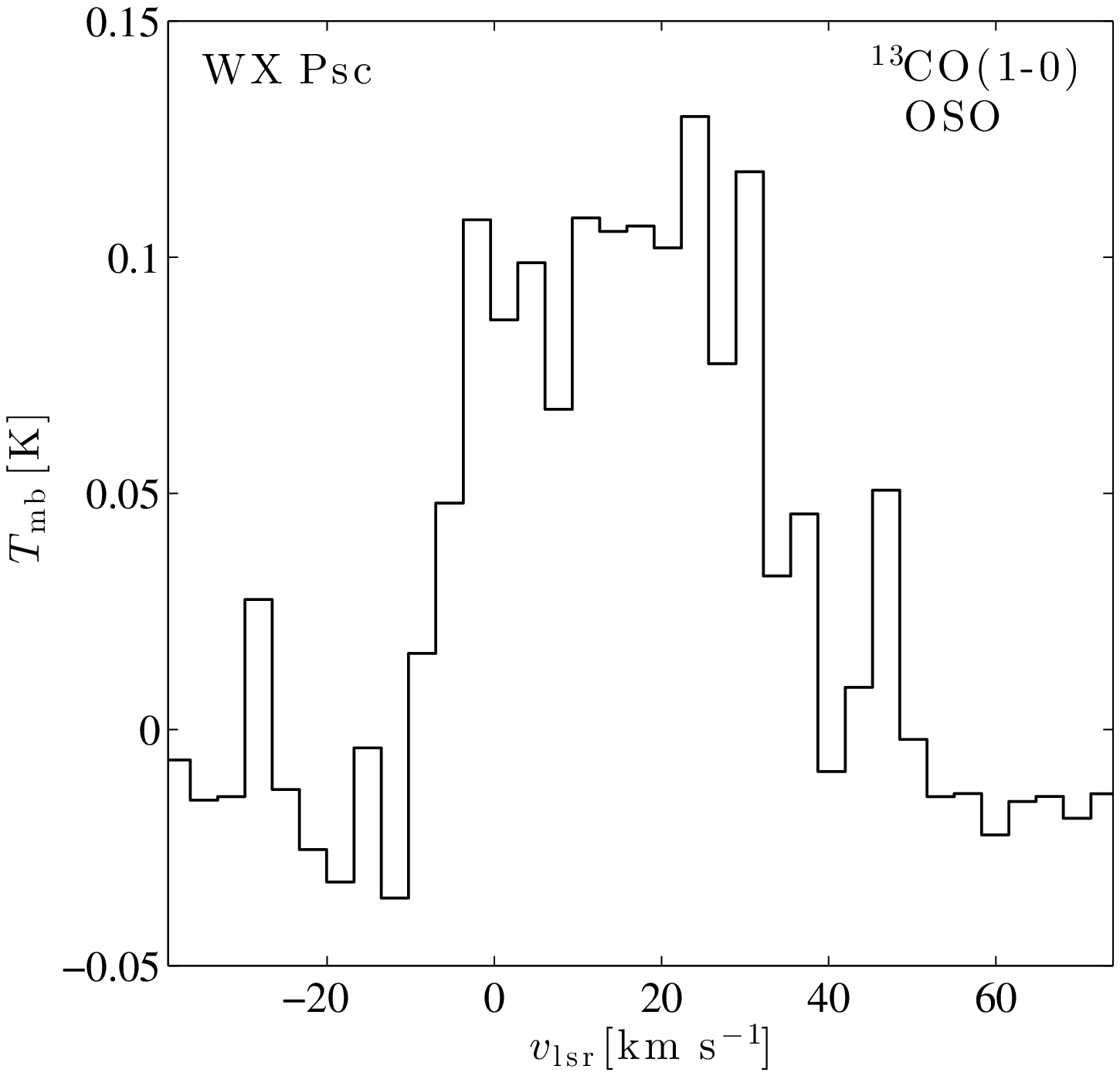}
   \includegraphics[width=3.5cm]{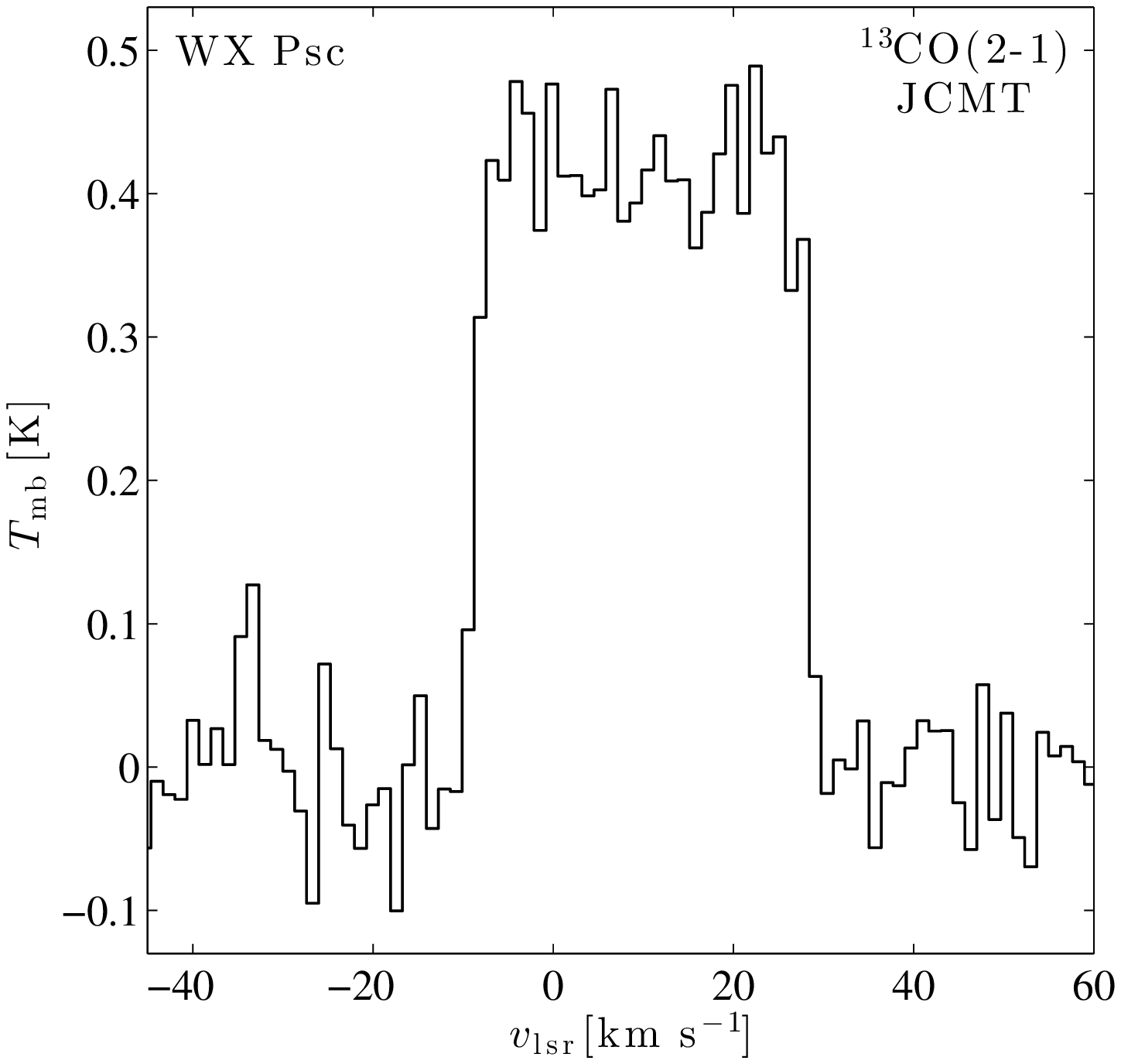}
   \includegraphics[width=3.5cm]{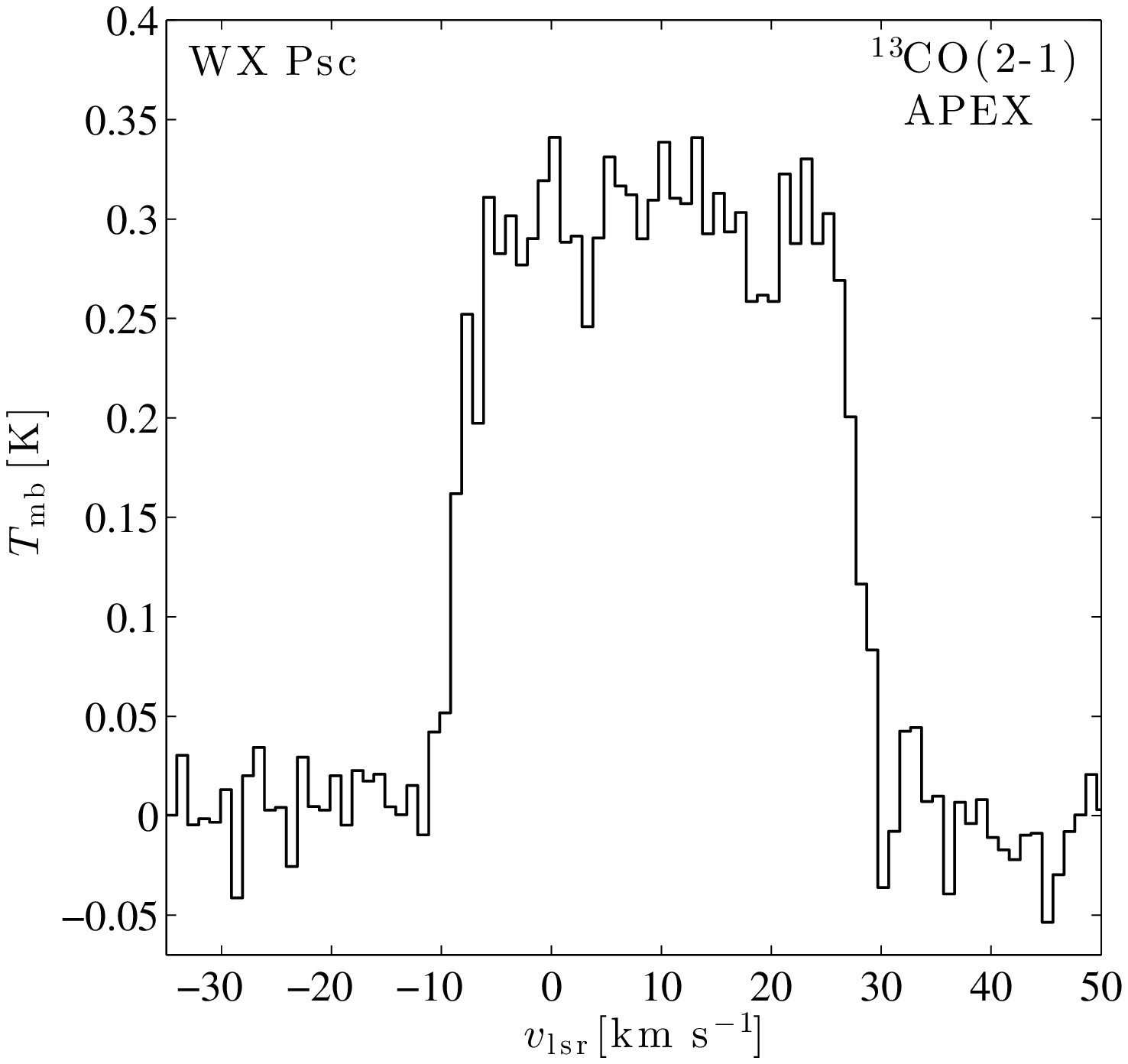}
   \includegraphics[width=3.5cm]{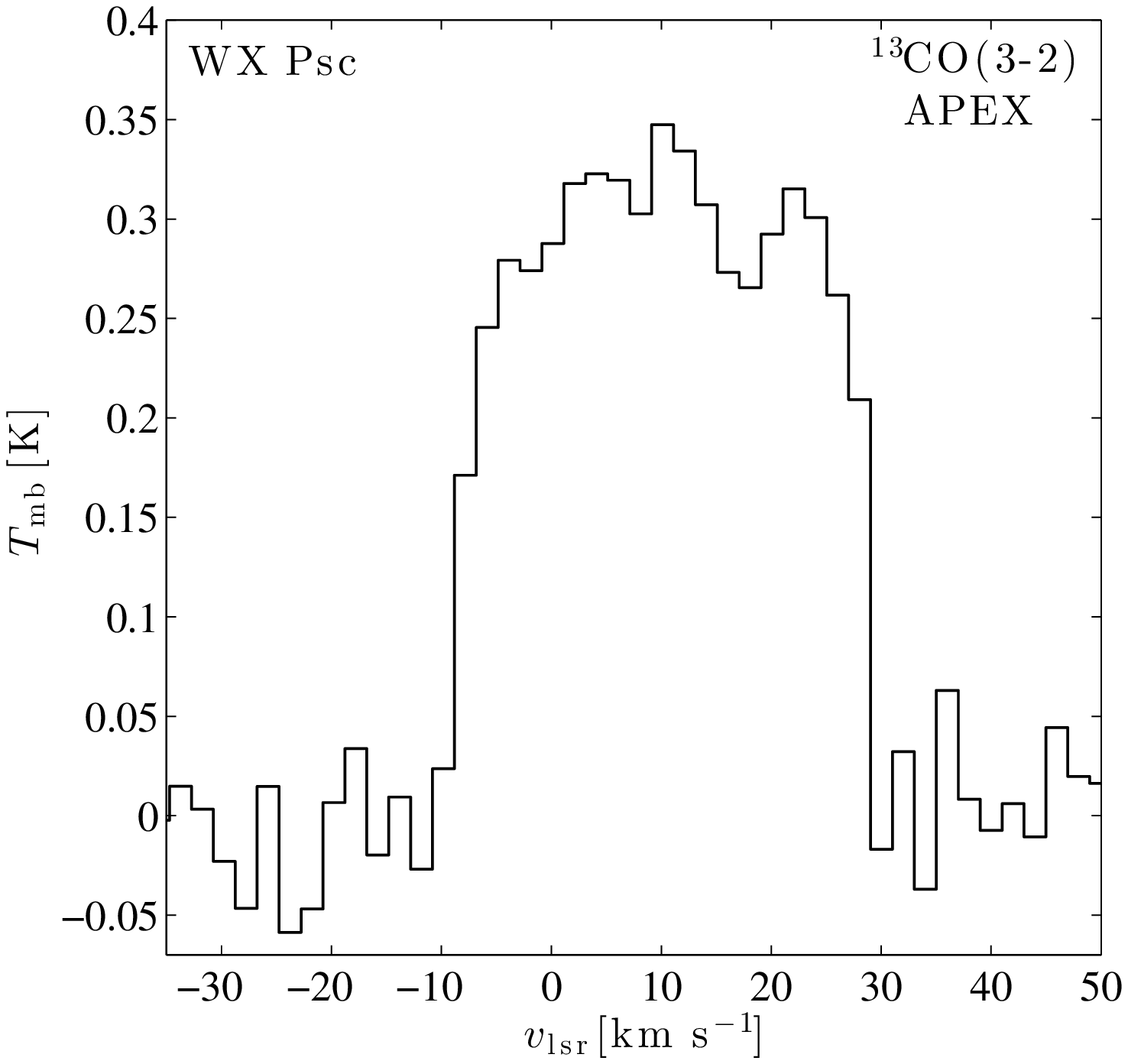} 
   \includegraphics[width=3.5cm]{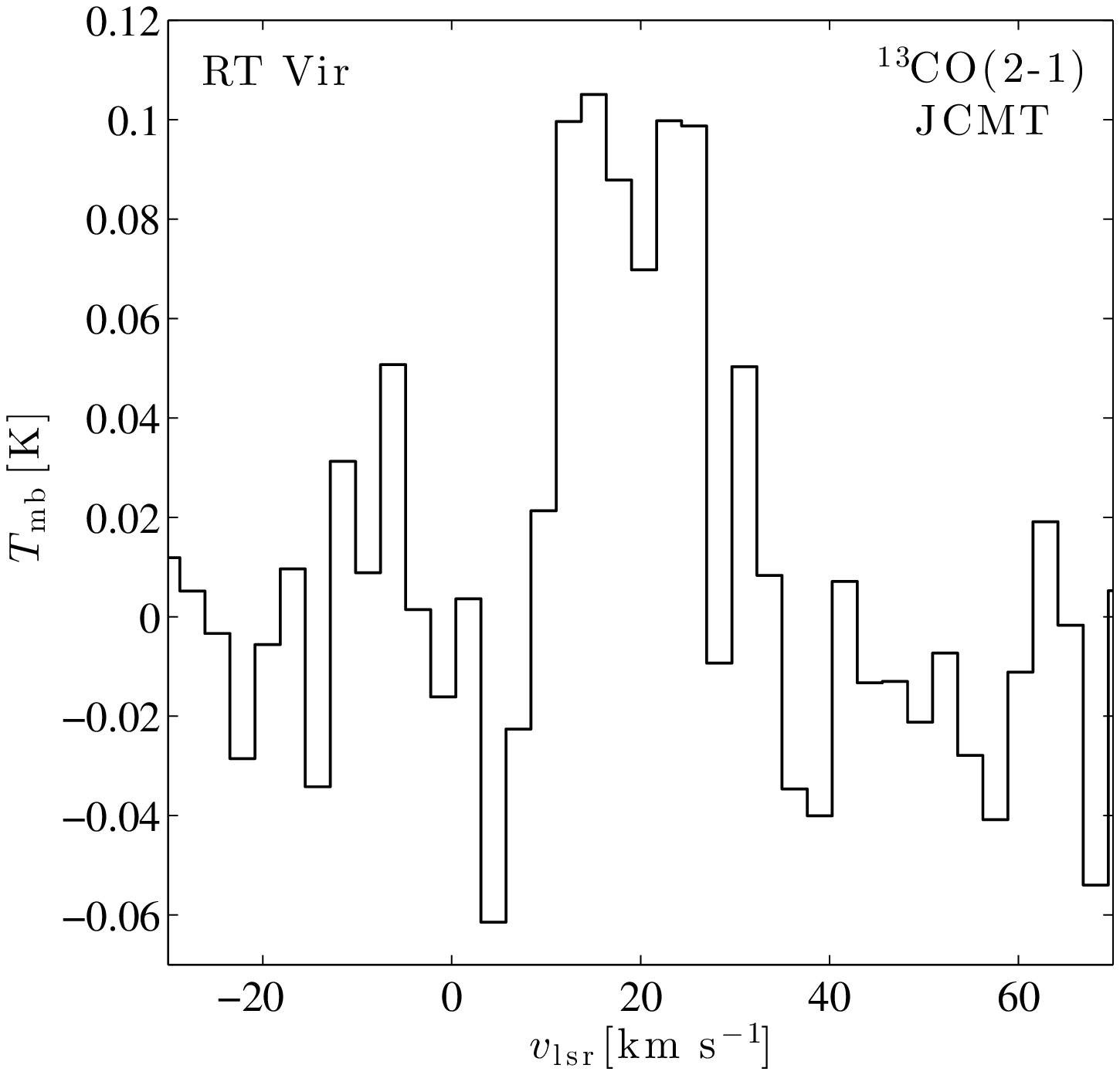}
   \includegraphics[width=3.5cm]{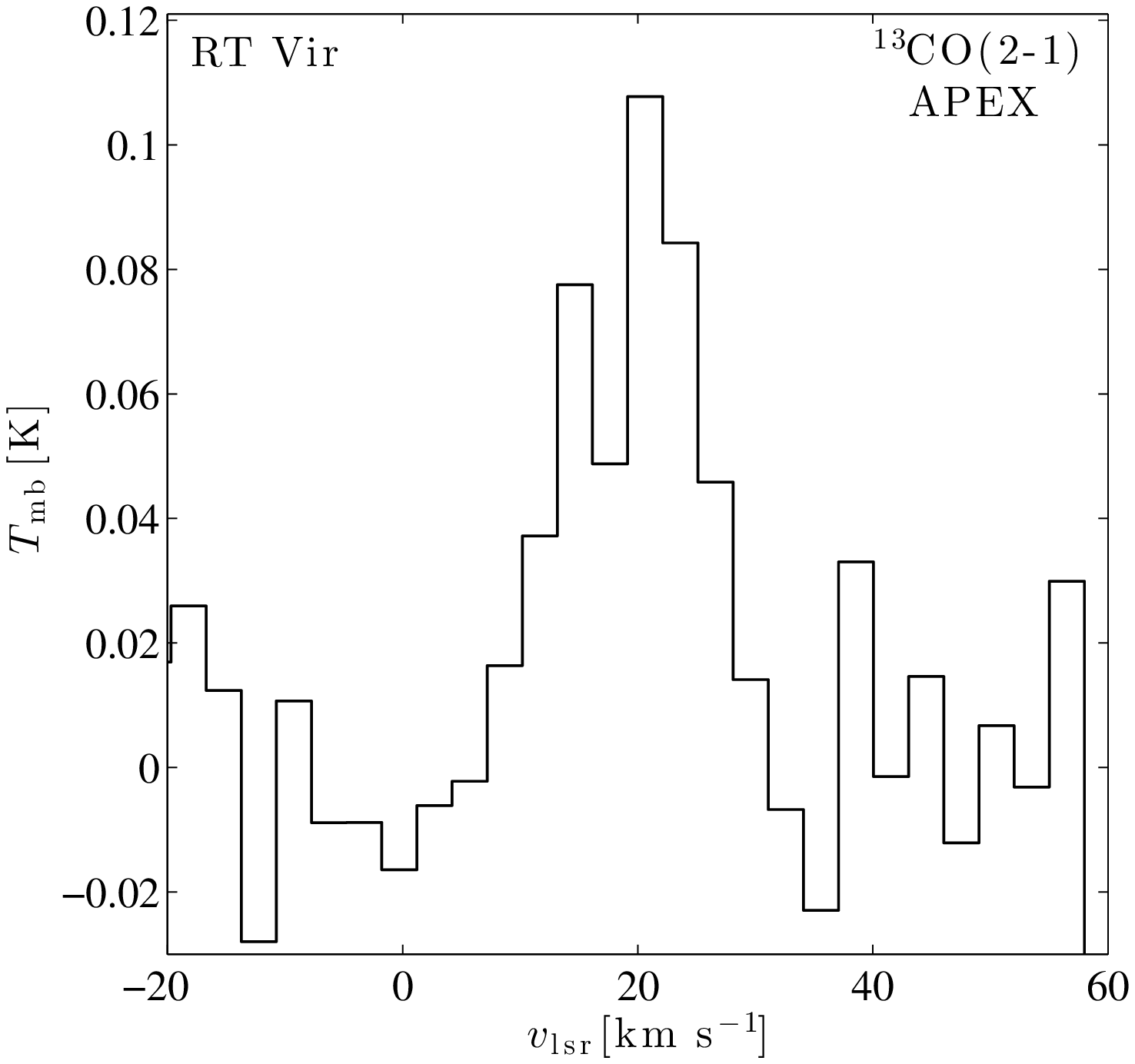}
   \includegraphics[width=3.5cm]{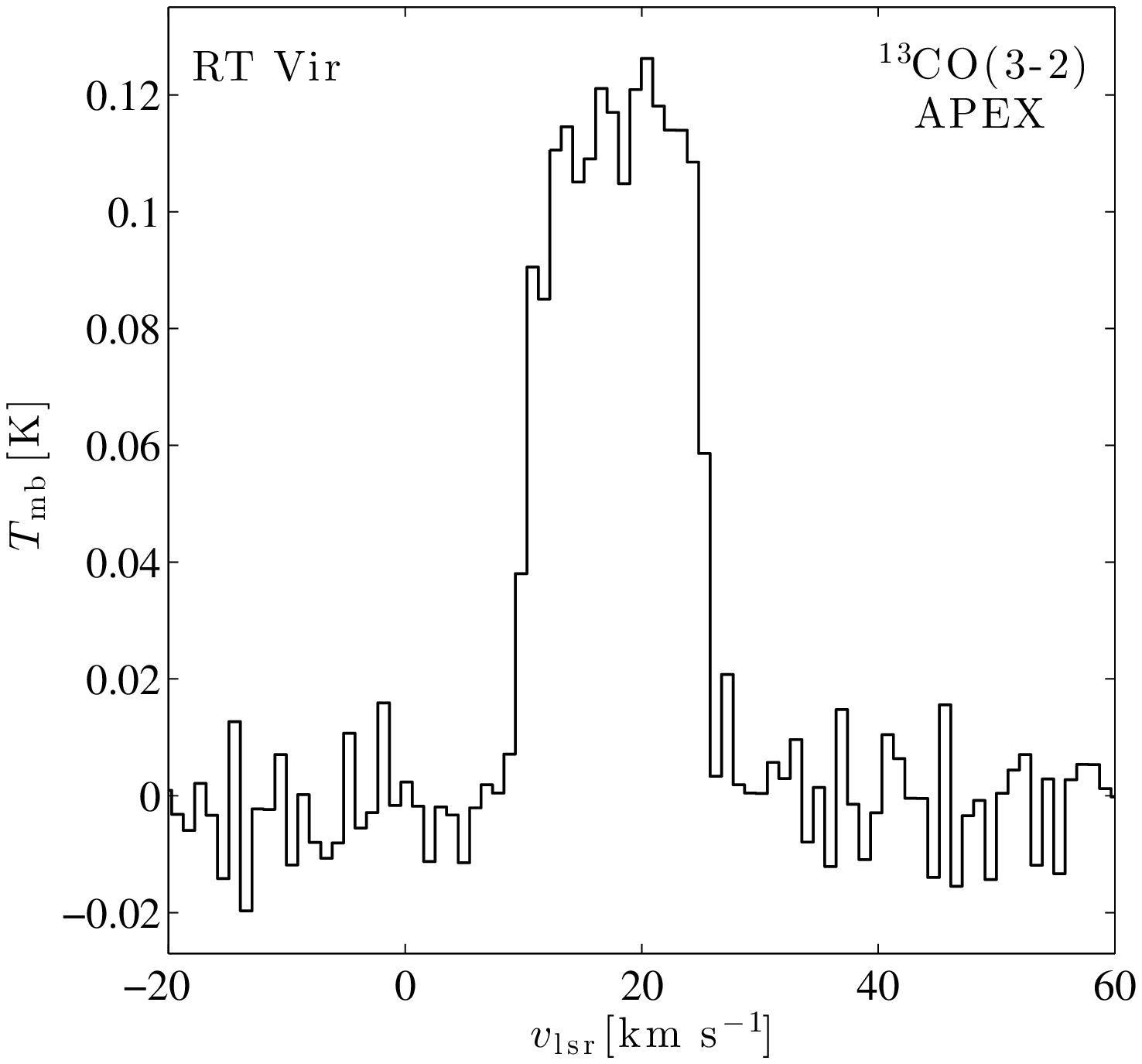}
   \includegraphics[width=3.5cm]{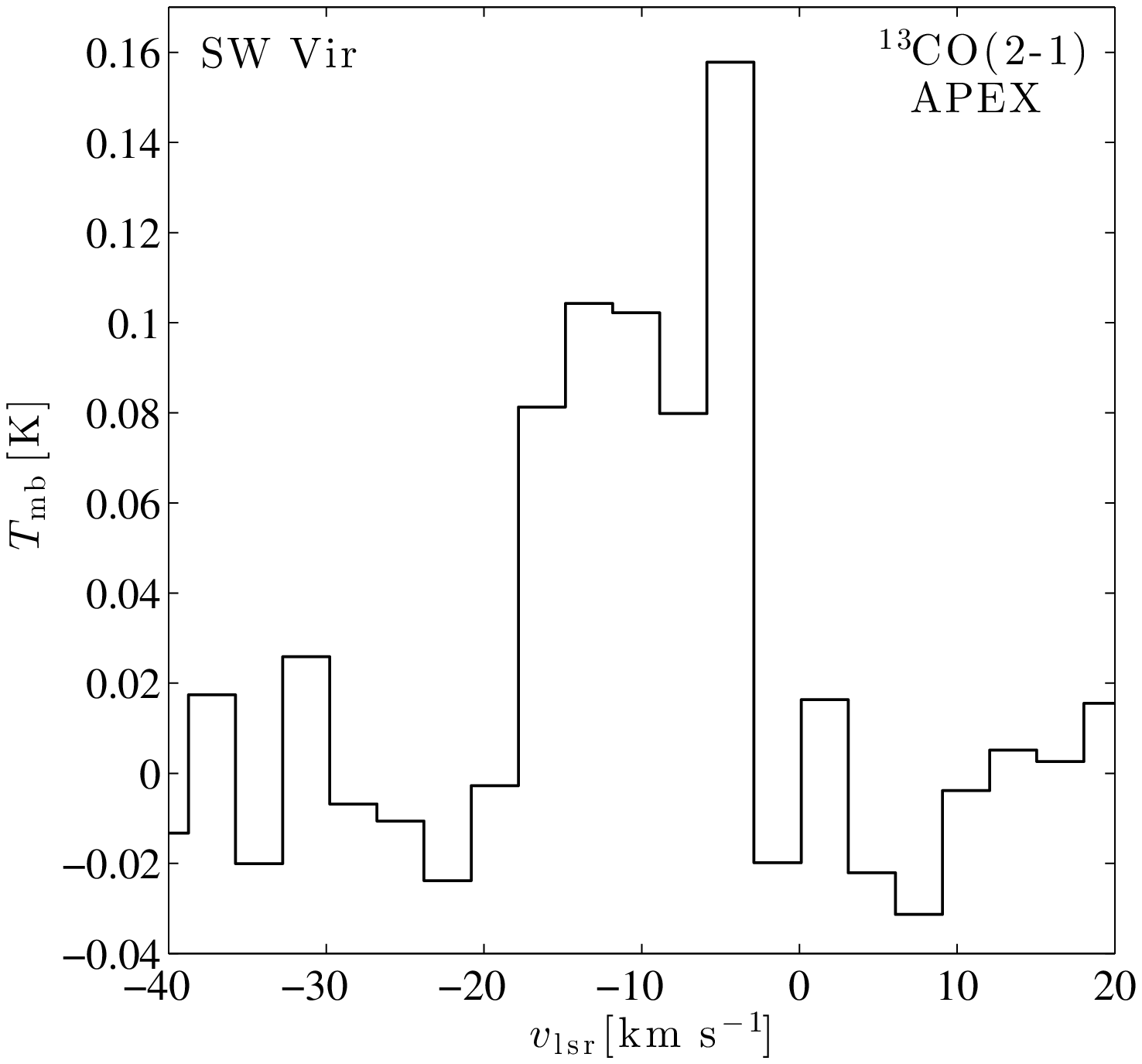}
   \includegraphics[width=3.5cm]{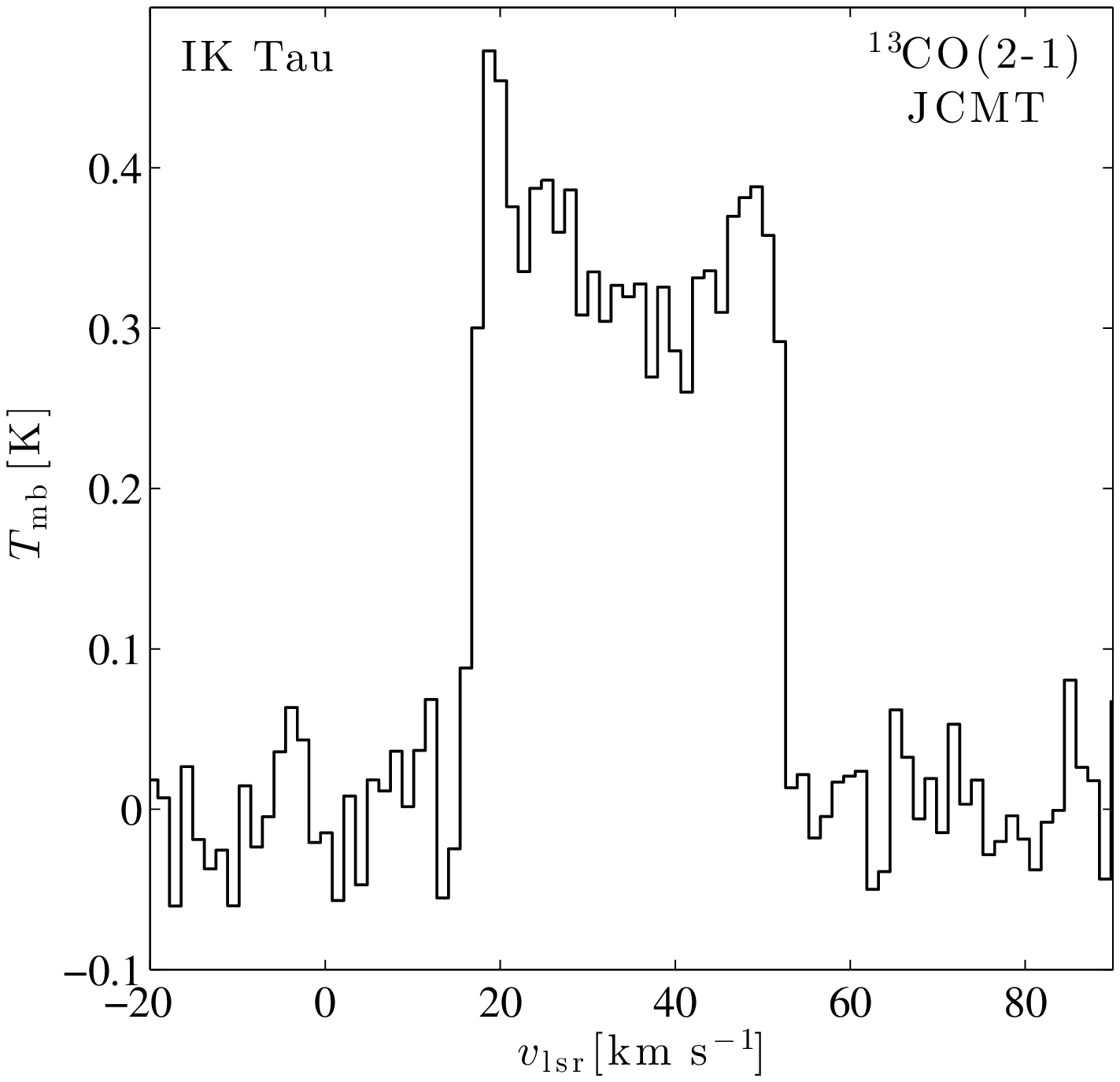} 
   \includegraphics[width=3.5cm]{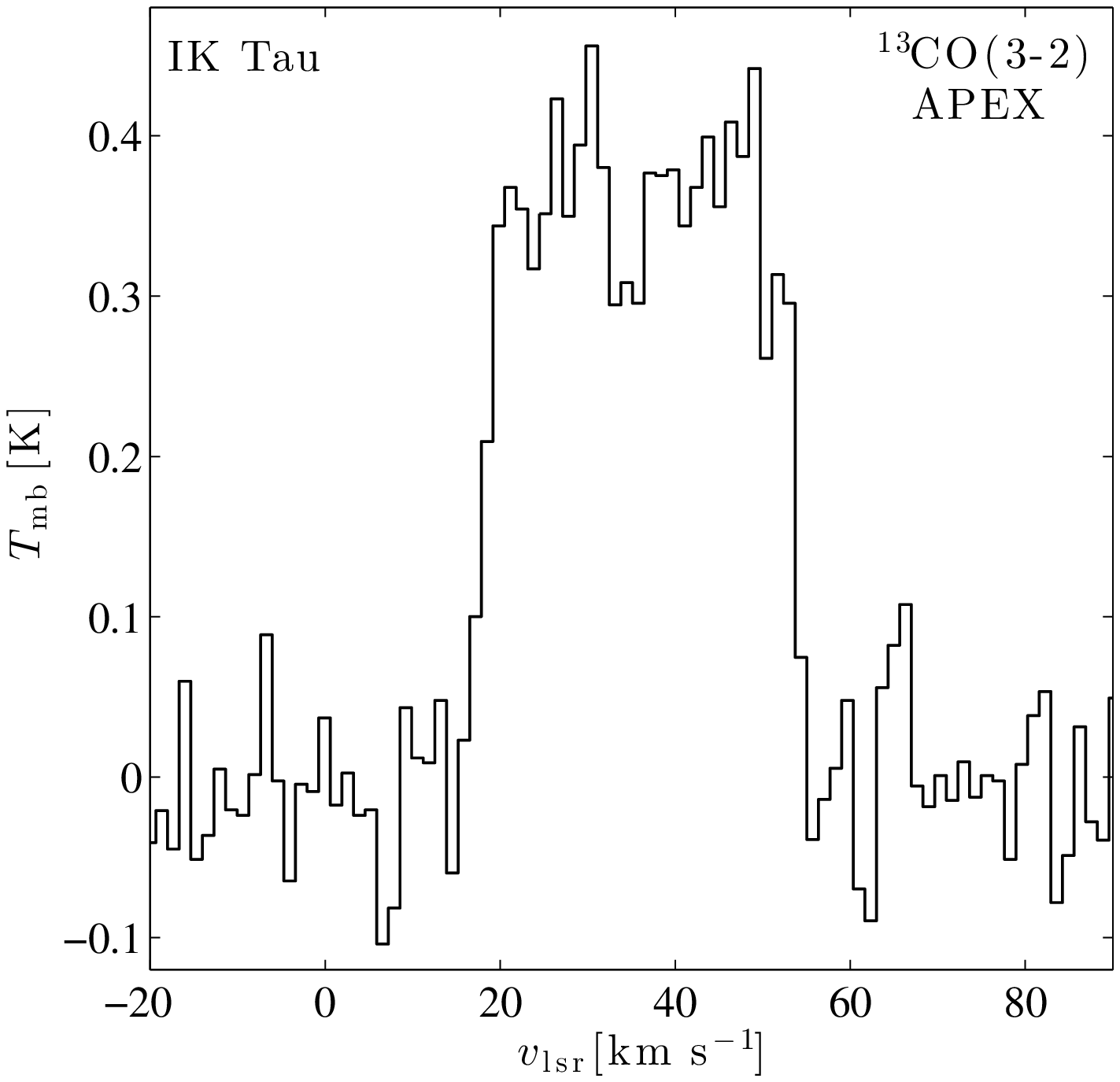}
   \includegraphics[width=3.5cm]{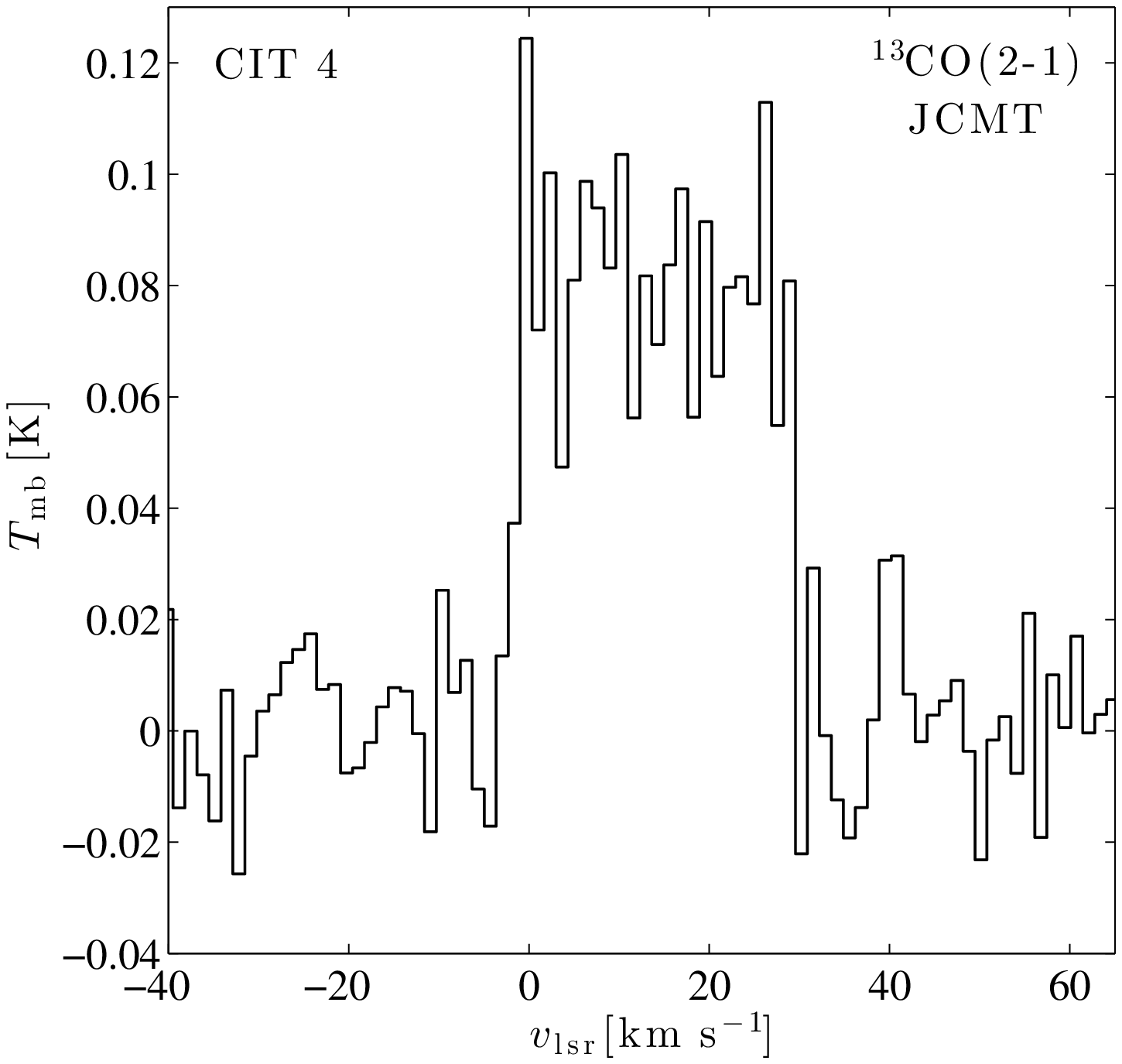}
   \includegraphics[width=3.5cm]{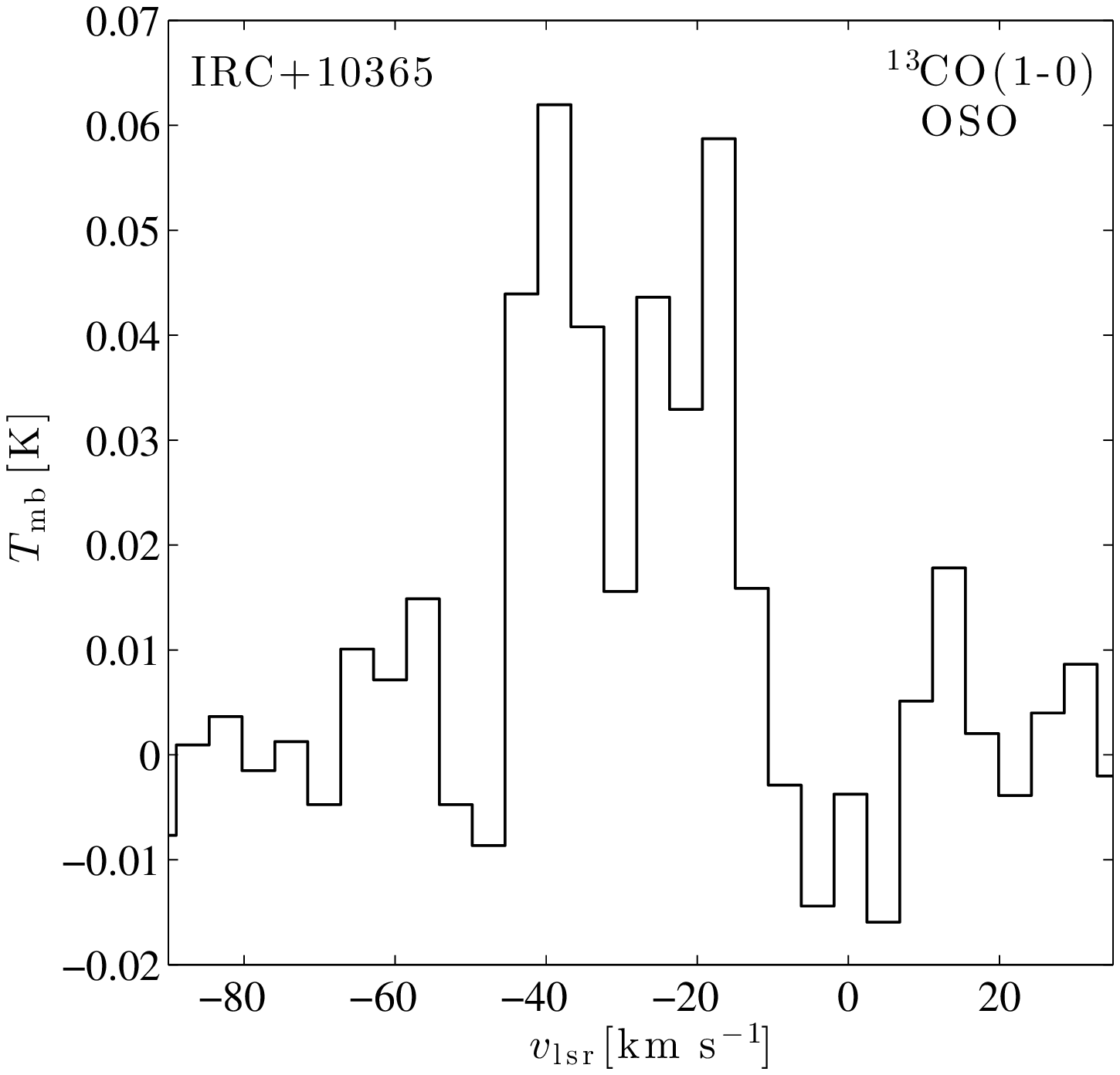}
   \includegraphics[width=3.5cm]{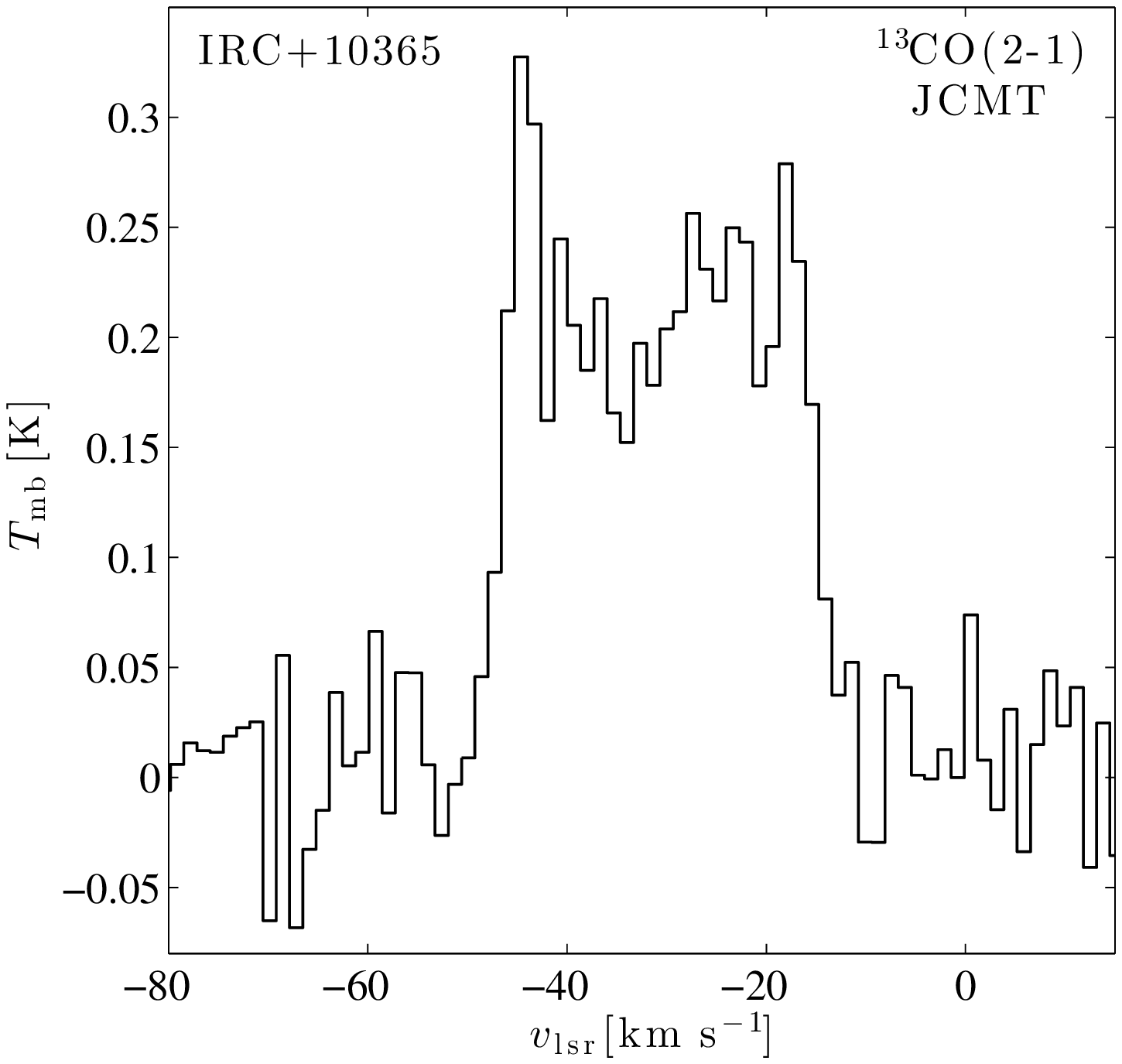}
   \includegraphics[width=3.5cm]{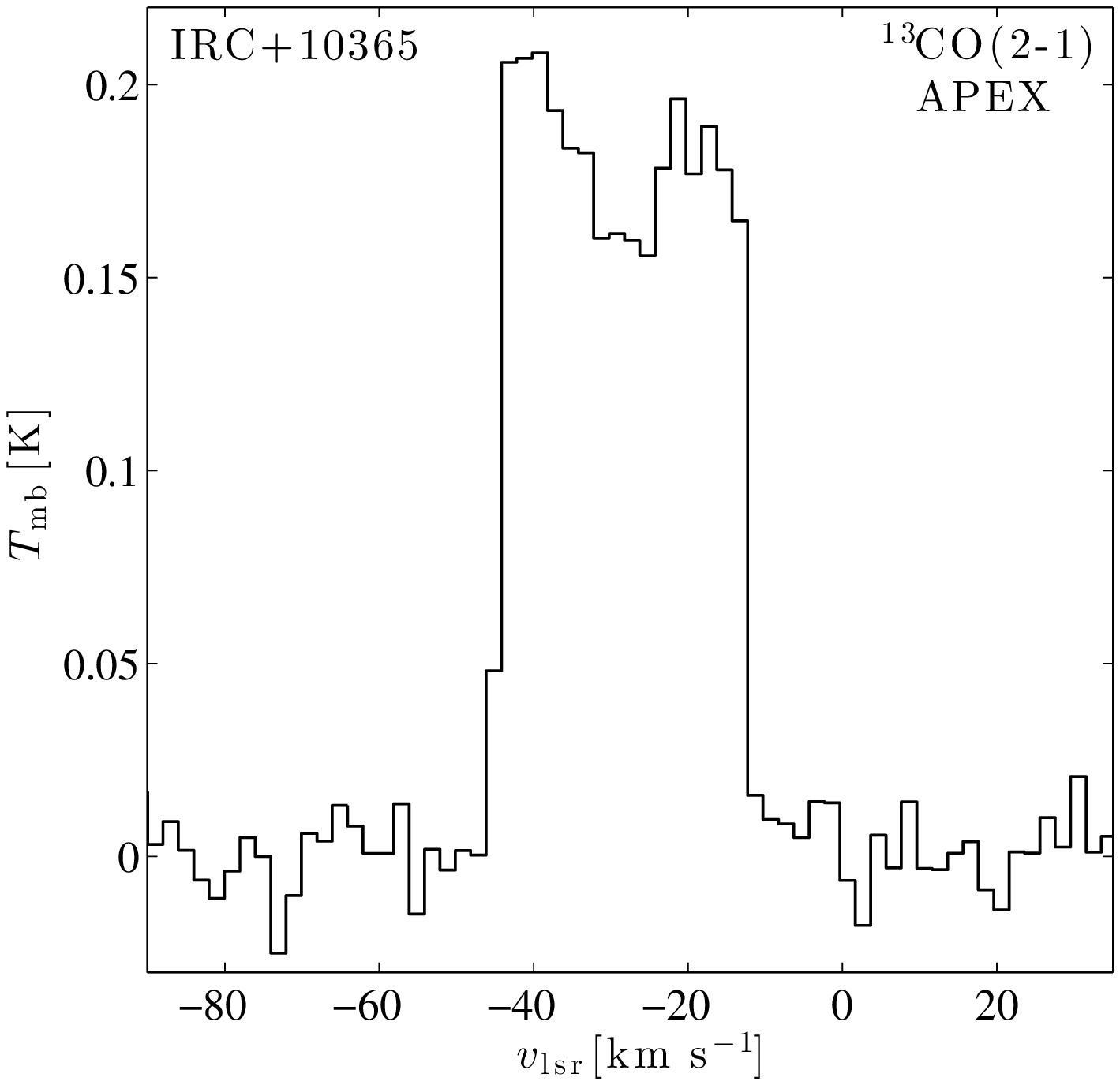} 
   \includegraphics[width=3.5cm]{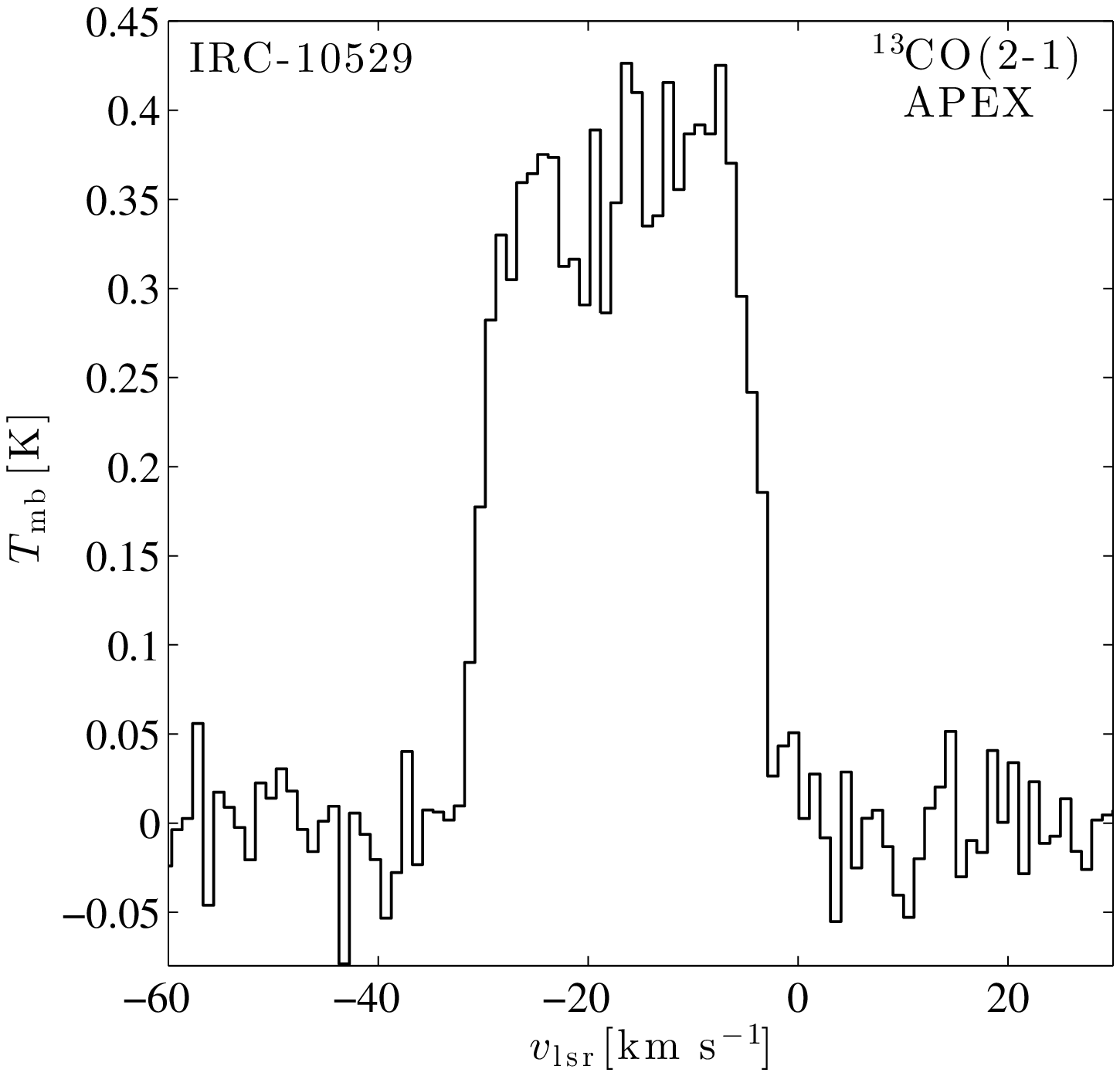}
   \includegraphics[width=3.5cm]{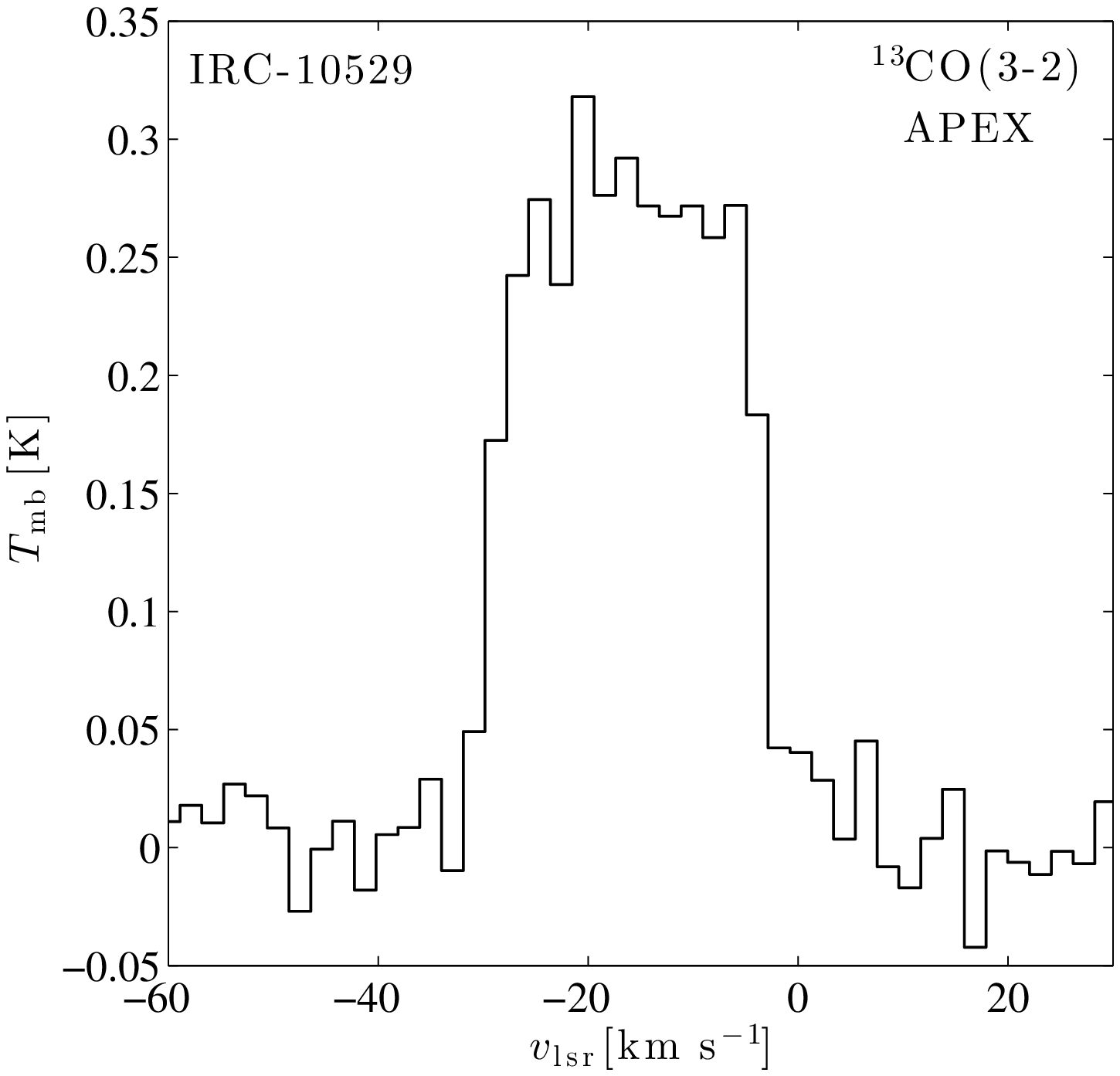}
   \includegraphics[width=3.5cm]{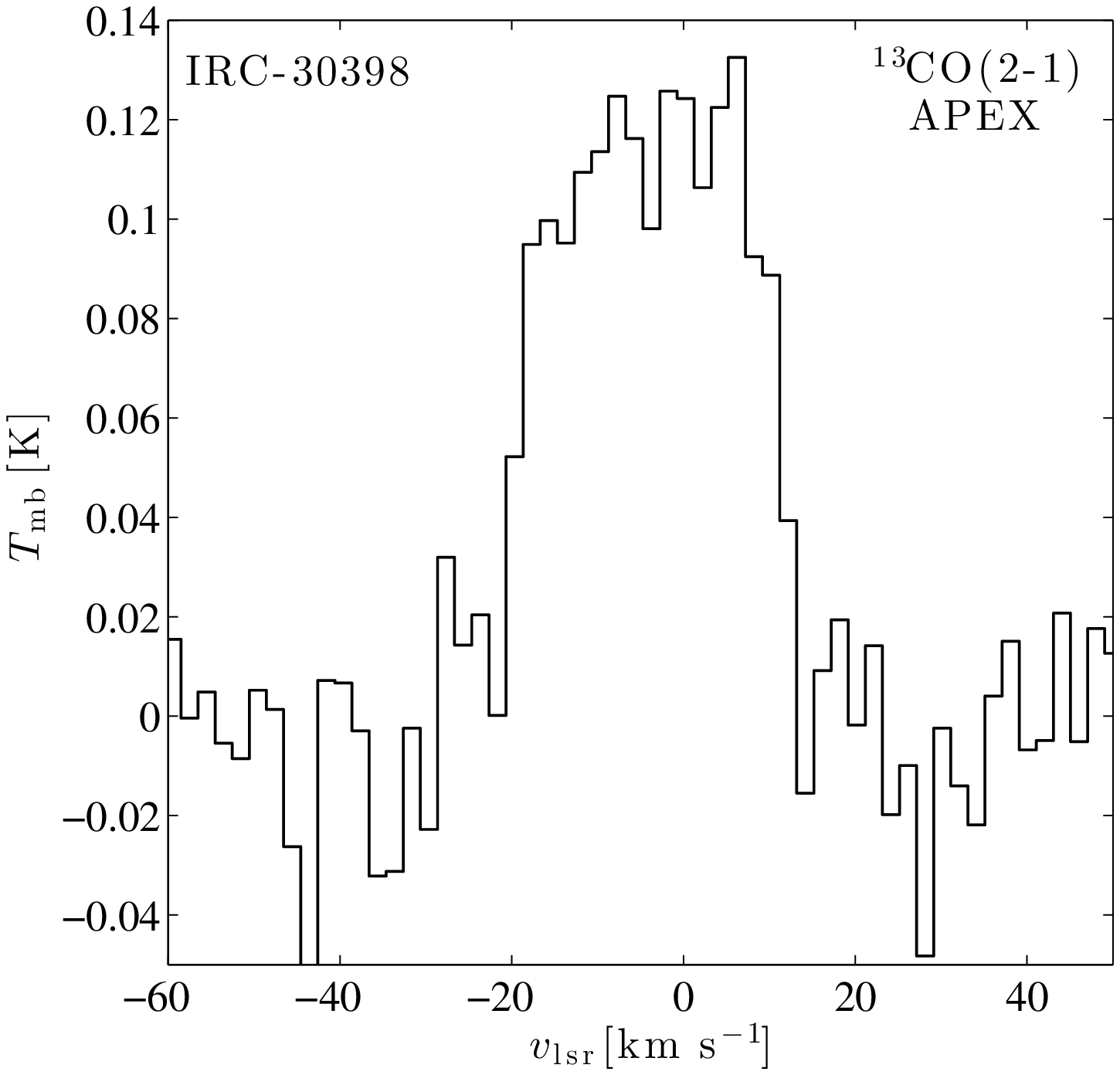}
   \includegraphics[width=3.5cm]{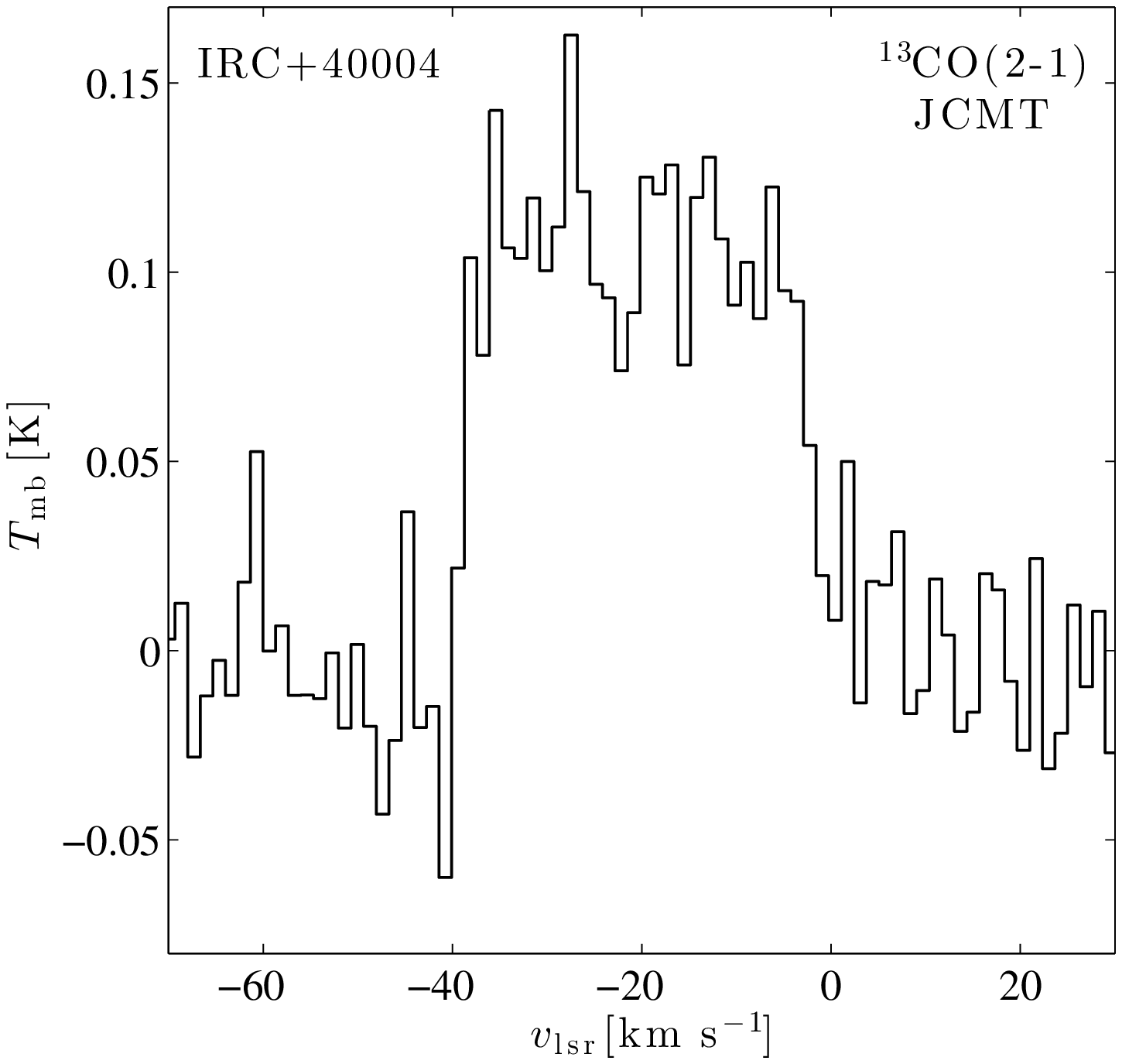}
   \includegraphics[width=3.5cm]{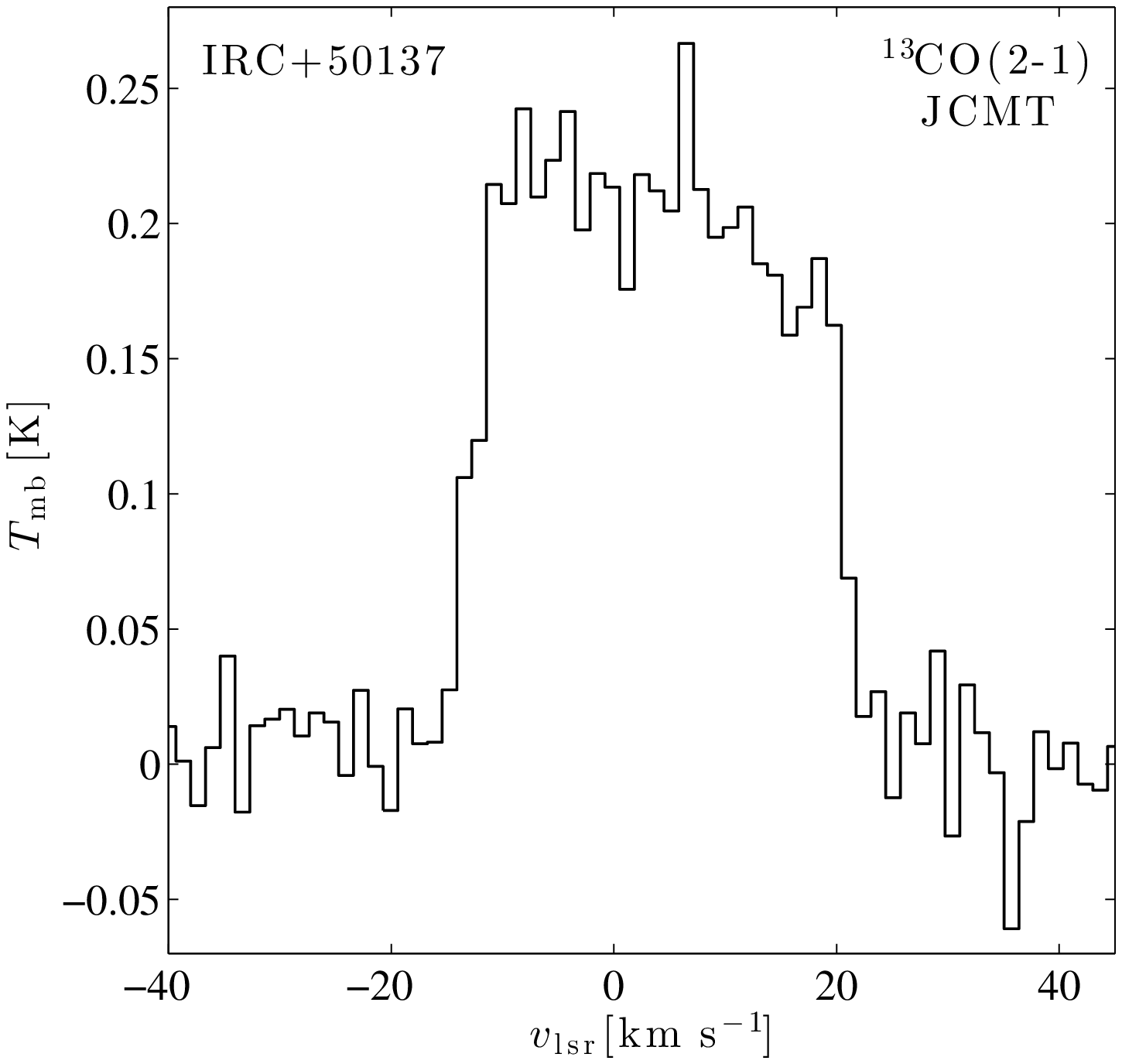} 
   \includegraphics[width=3.5cm]{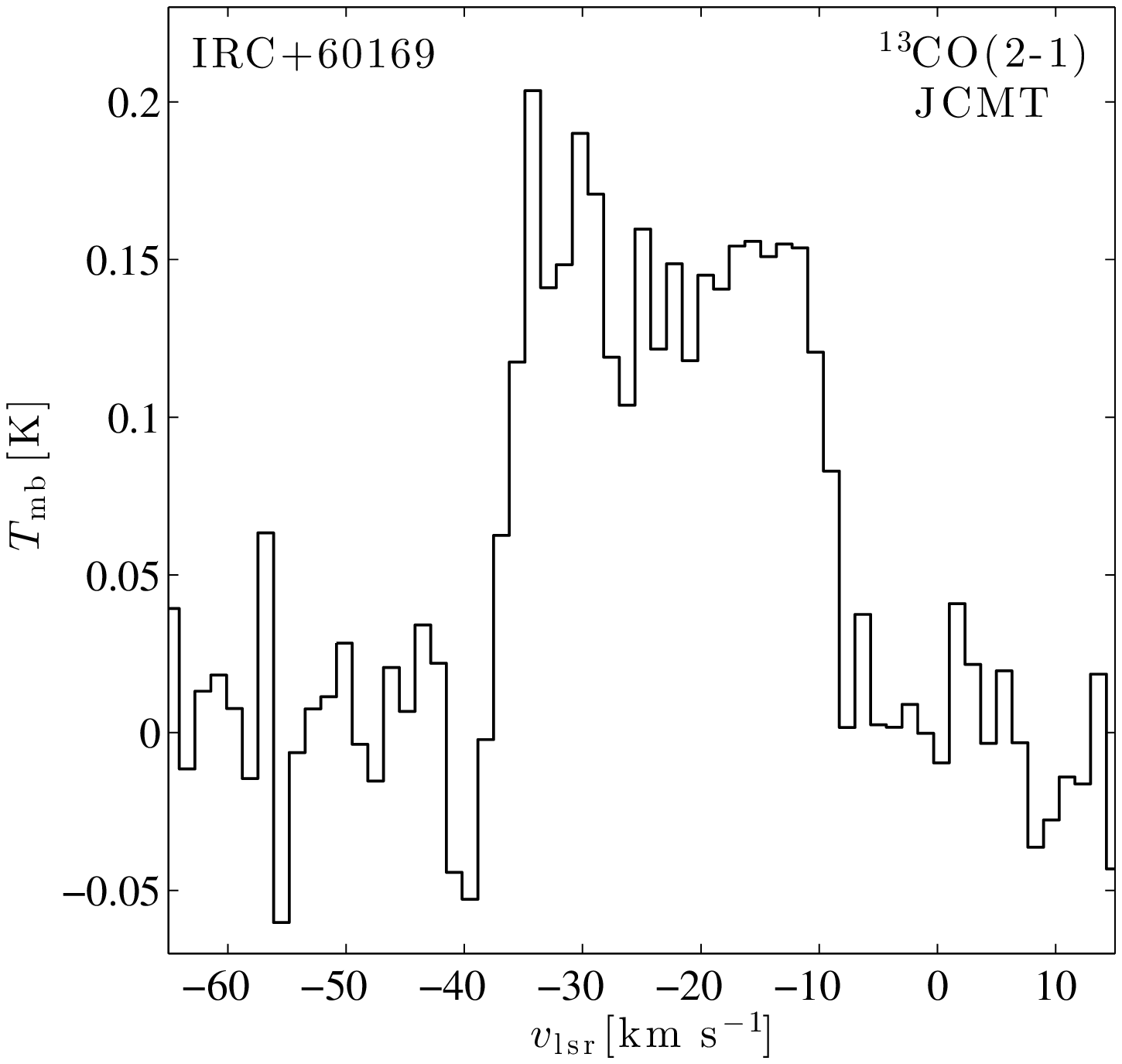}
   \includegraphics[width=3.5cm]{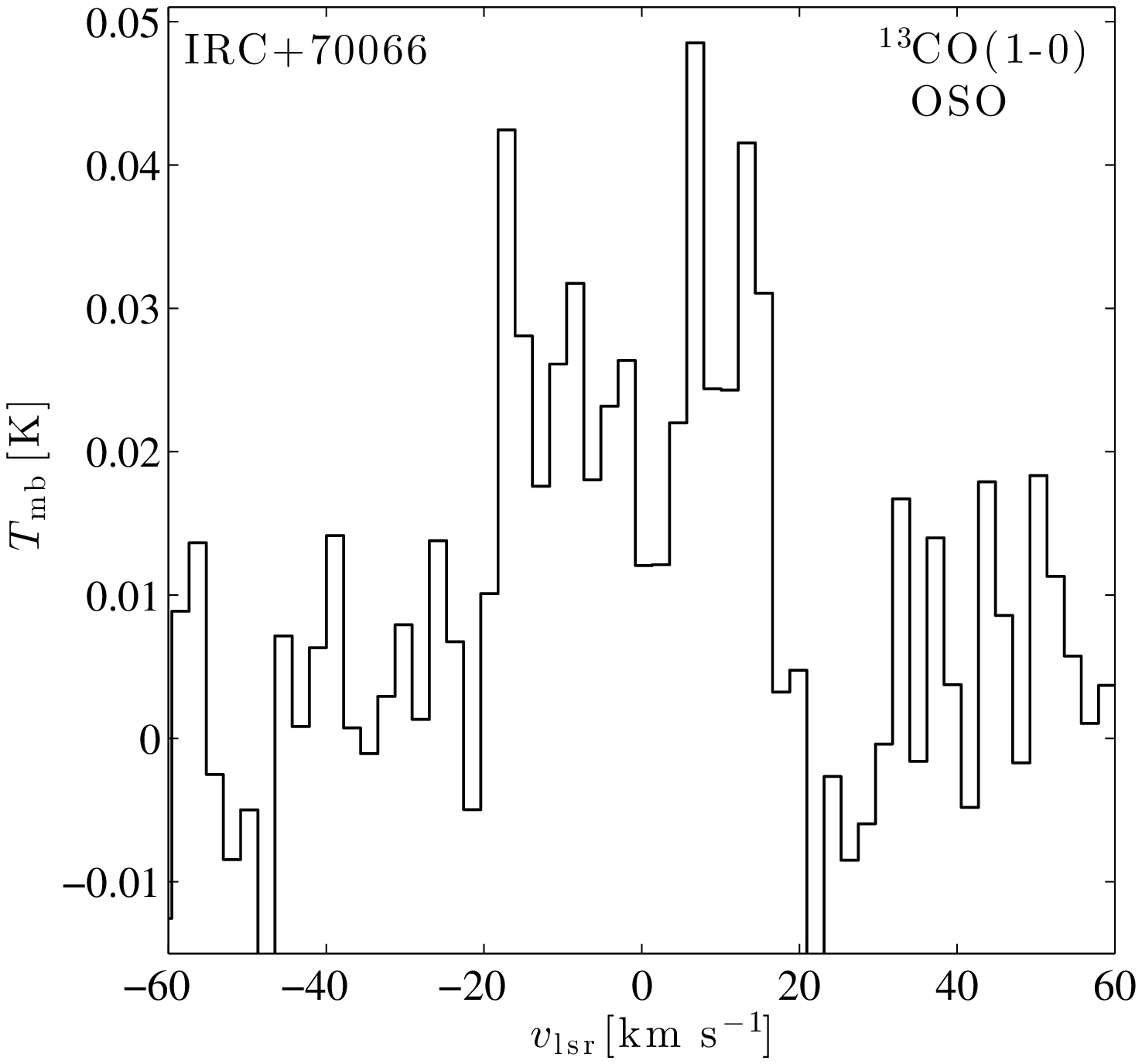}
   \caption{New $^{13}$CO observations for the M-type stars. The source name is shown to the upper left, the observed transition and telescope is shown to the upper right of each frame.}
   \label{sp1}
\end{figure*}

\begin{figure*}[h]
\center
   \includegraphics[width=3.5cm]{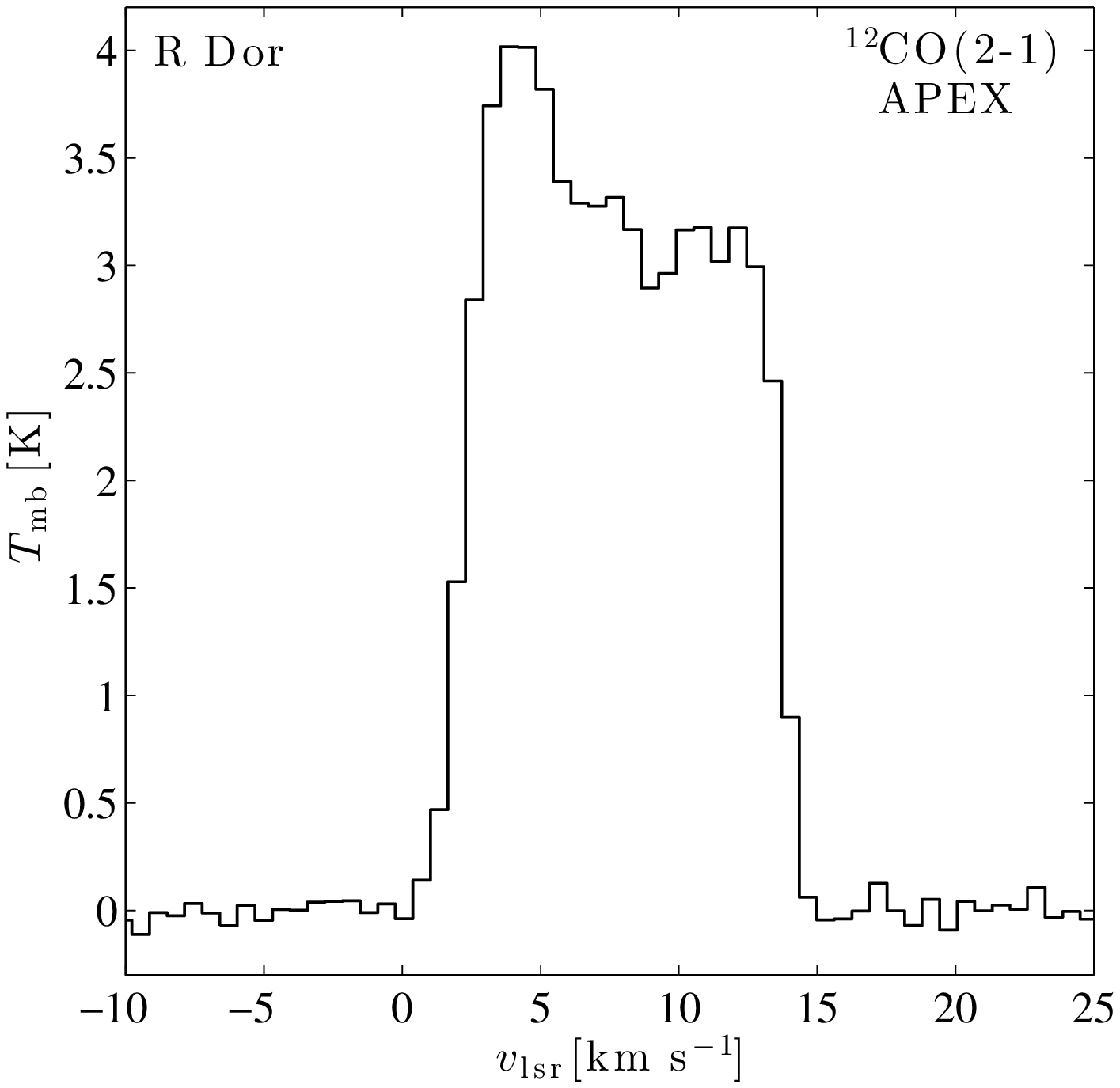} 
   \includegraphics[width=3.5cm]{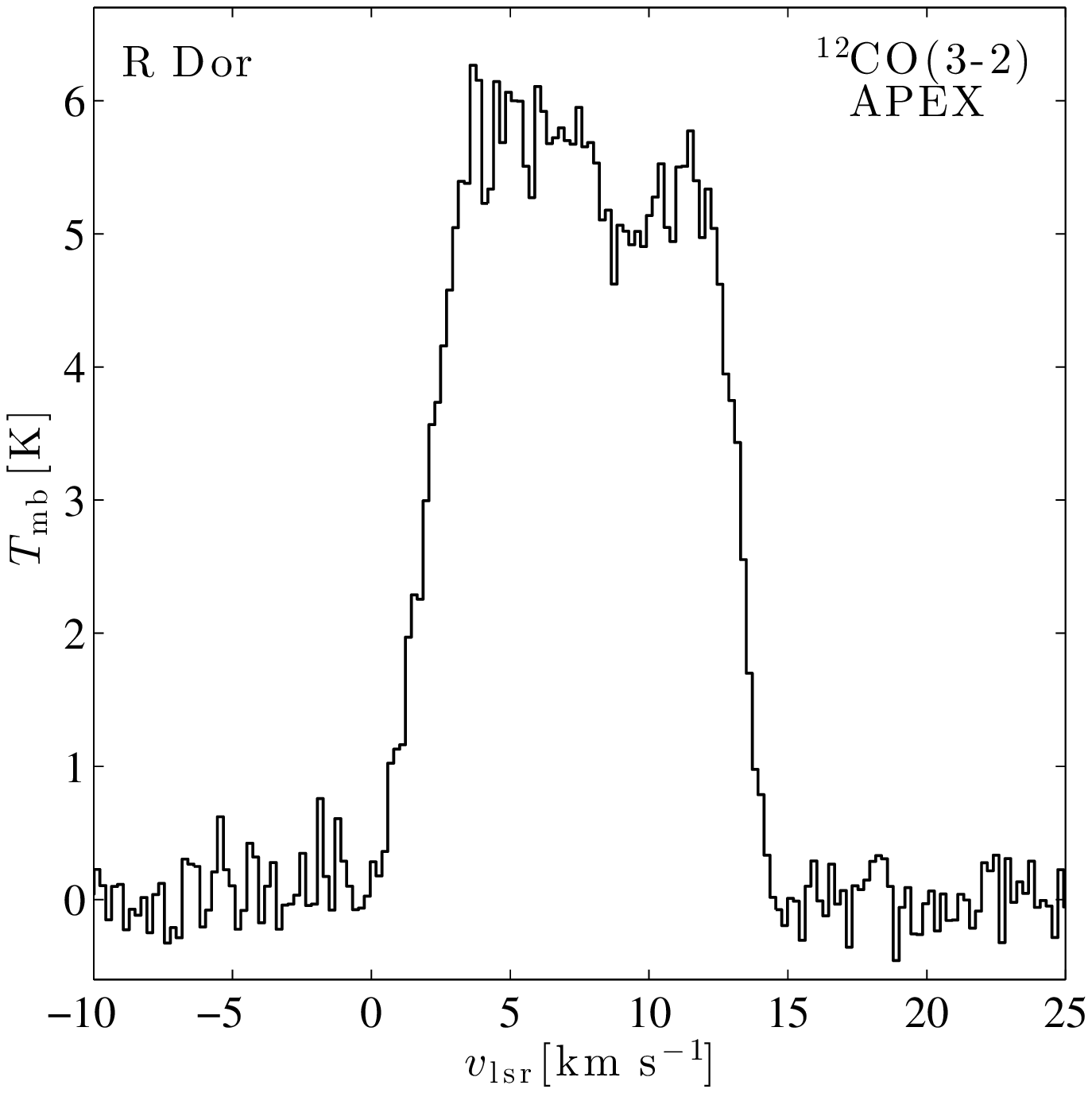}
   \includegraphics[width=3.5cm]{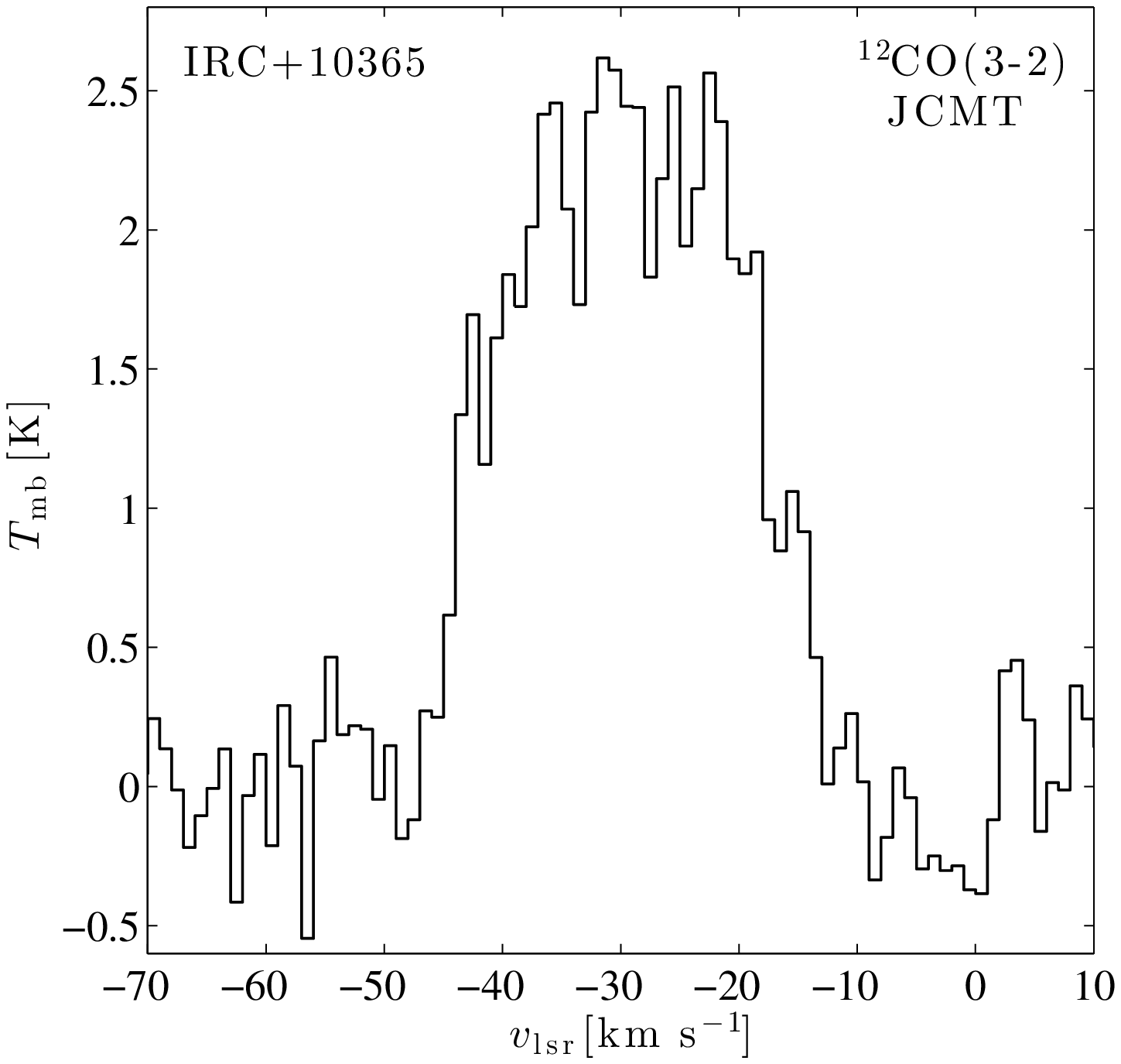}
   \includegraphics[width=3.5cm]{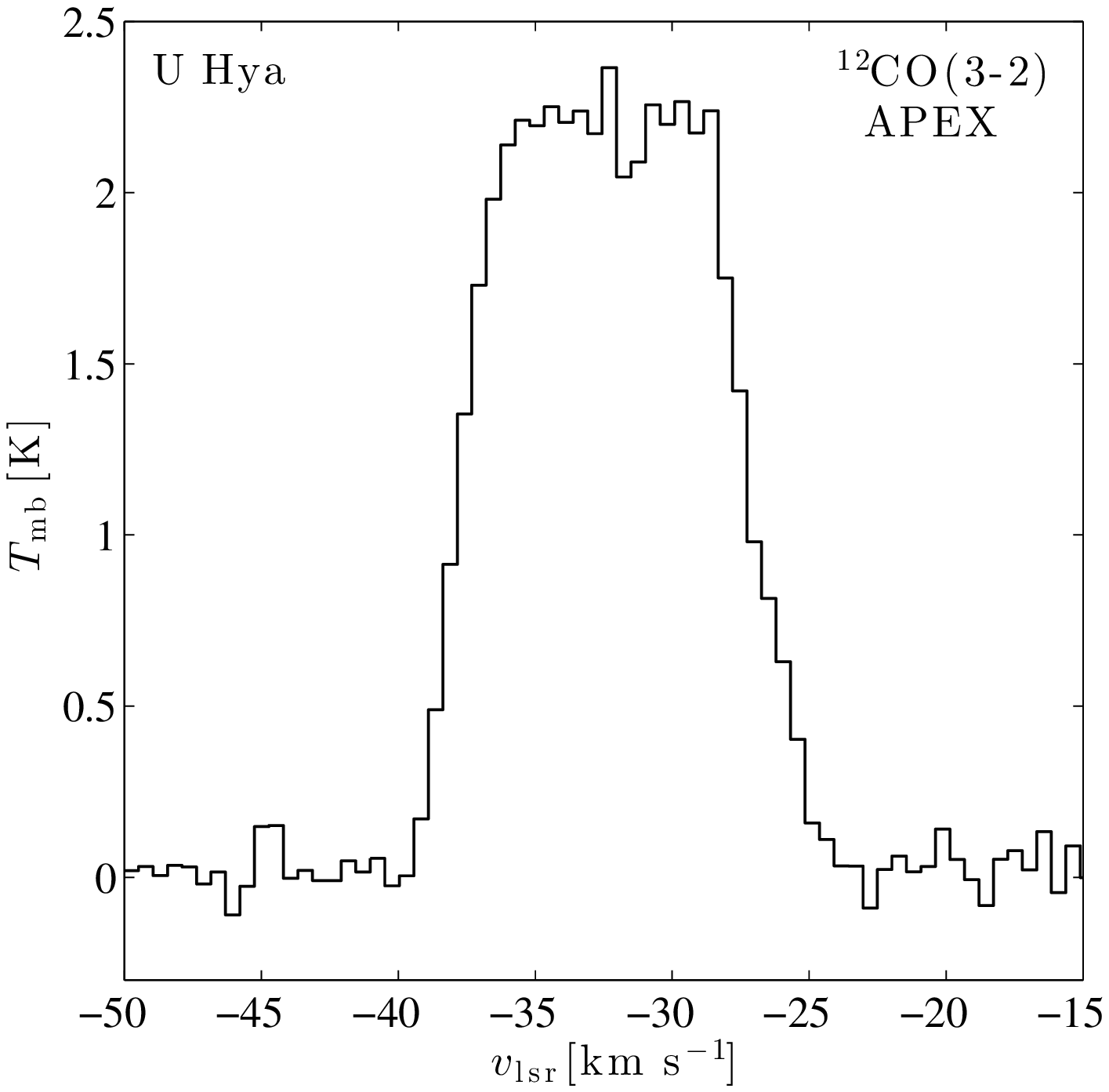}
   \includegraphics[width=3.5cm]{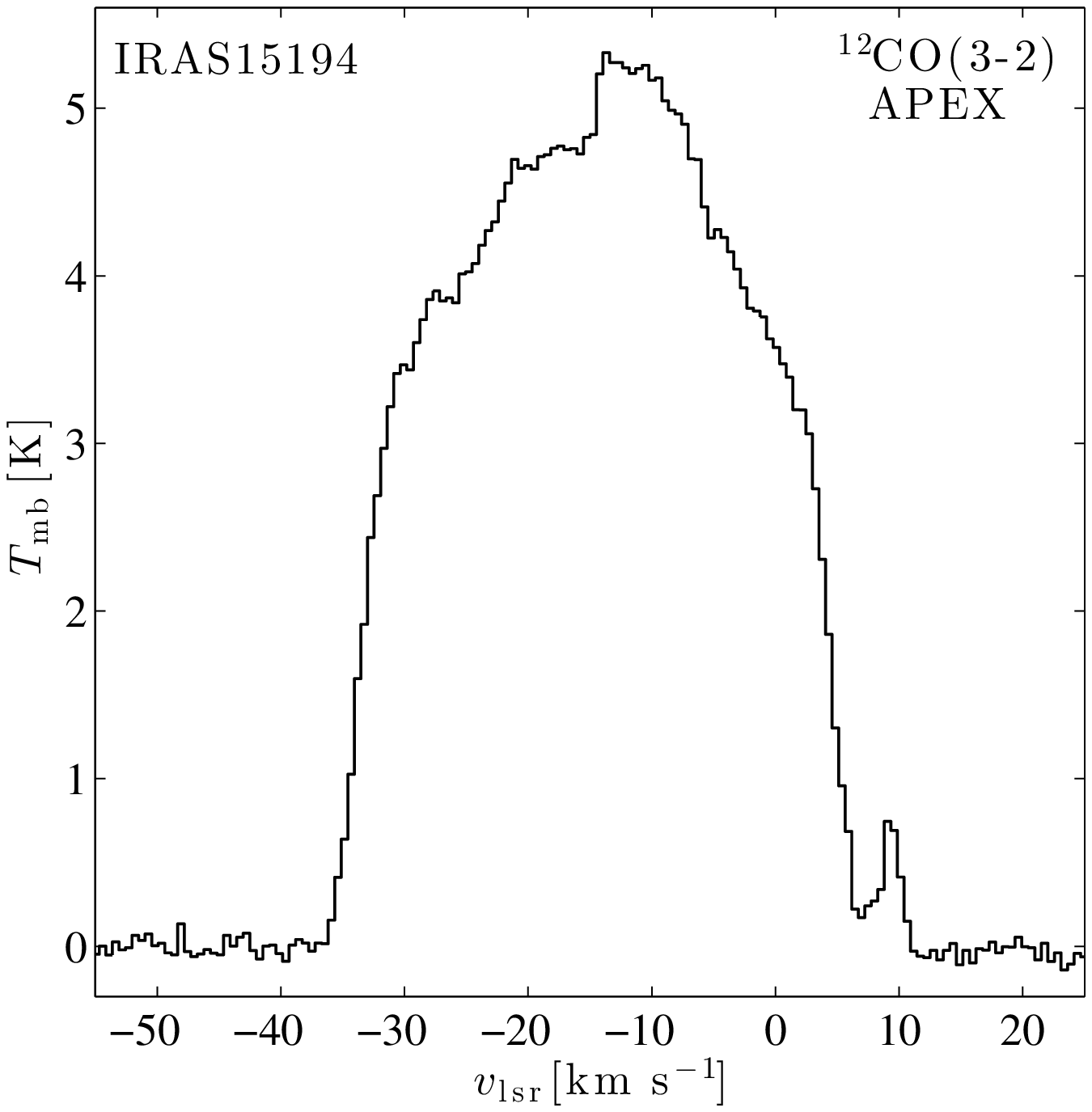}
   \caption{New $^{12}$CO observations. The source name is shown to the upper left, the observed transition and telescope is shown to the upper right of each frame.}
   \label{sp2}
\end{figure*}

\begin{figure*}[h]
\center
   \includegraphics[width=3.5cm]{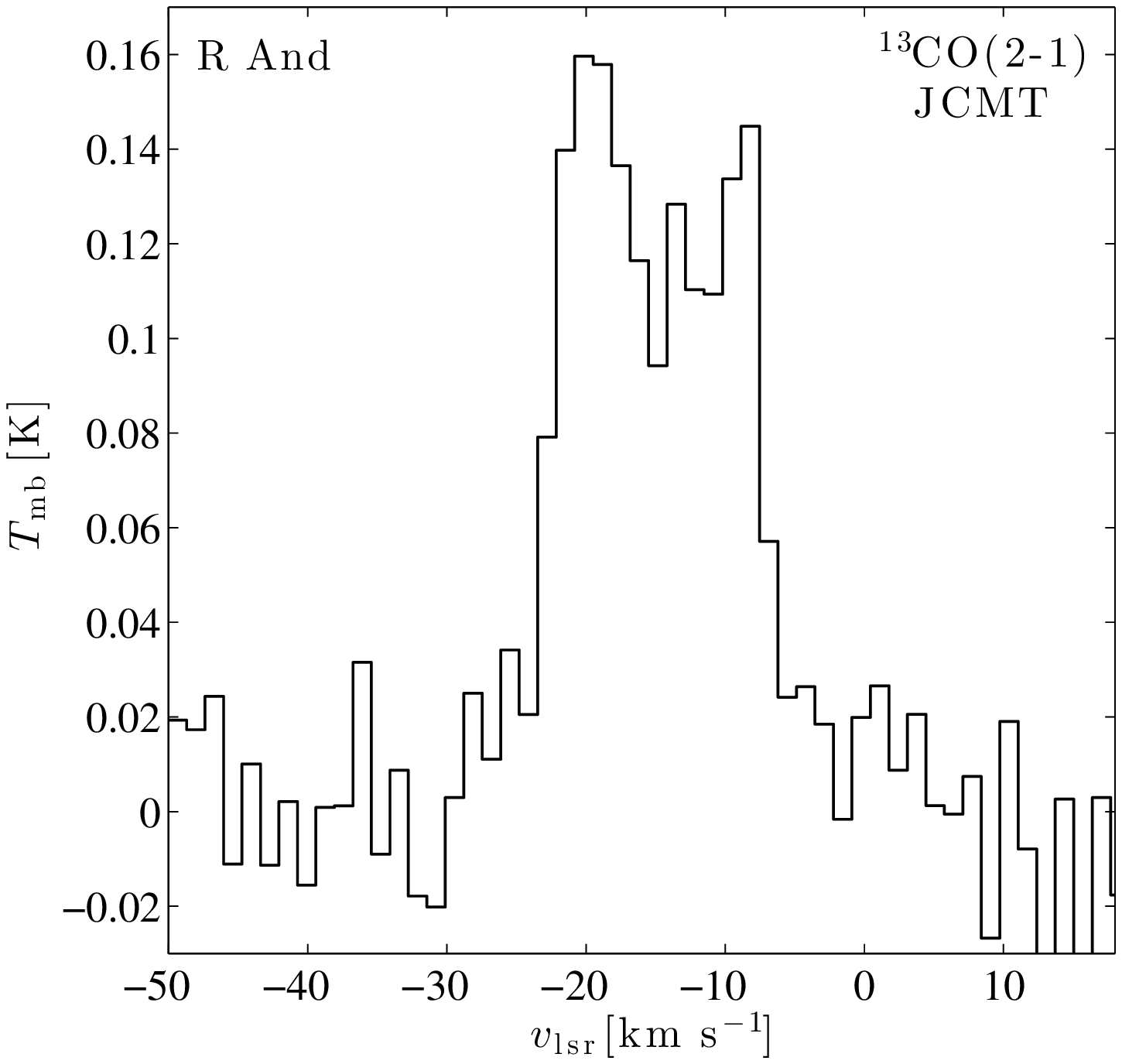} 
   \includegraphics[width=3.5cm]{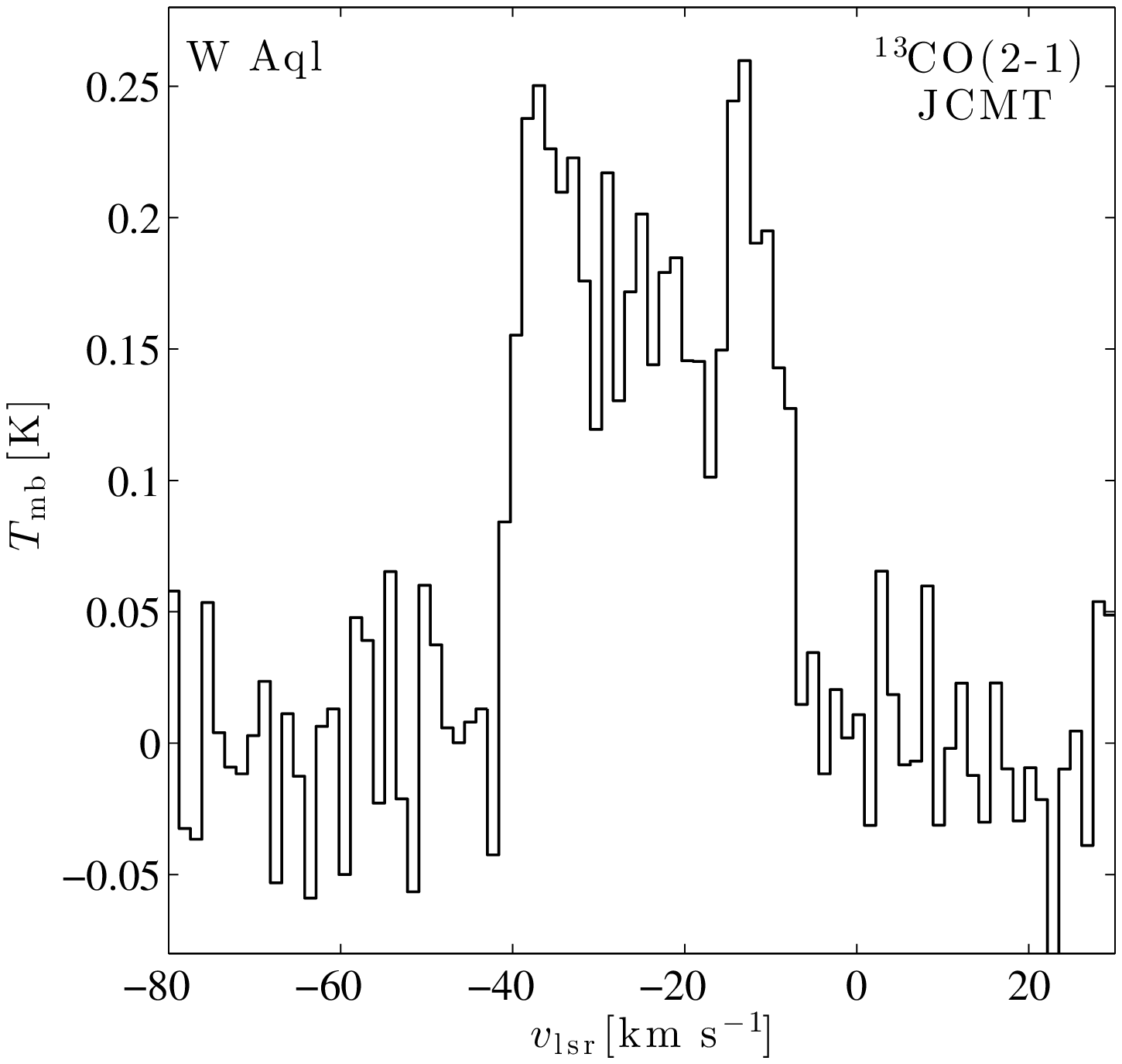}
   \includegraphics[width=3.5cm]{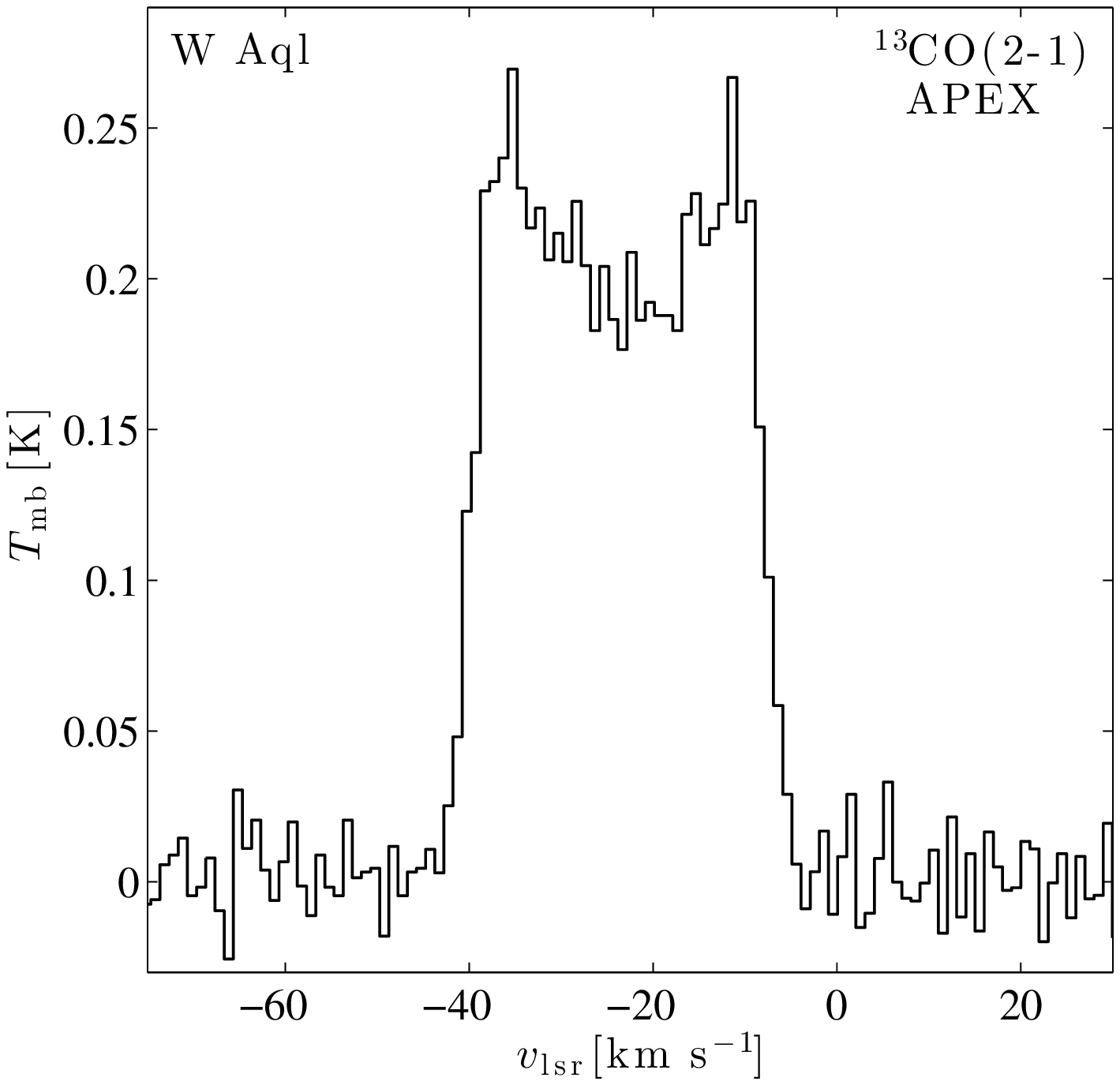}
   \includegraphics[width=3.5cm]{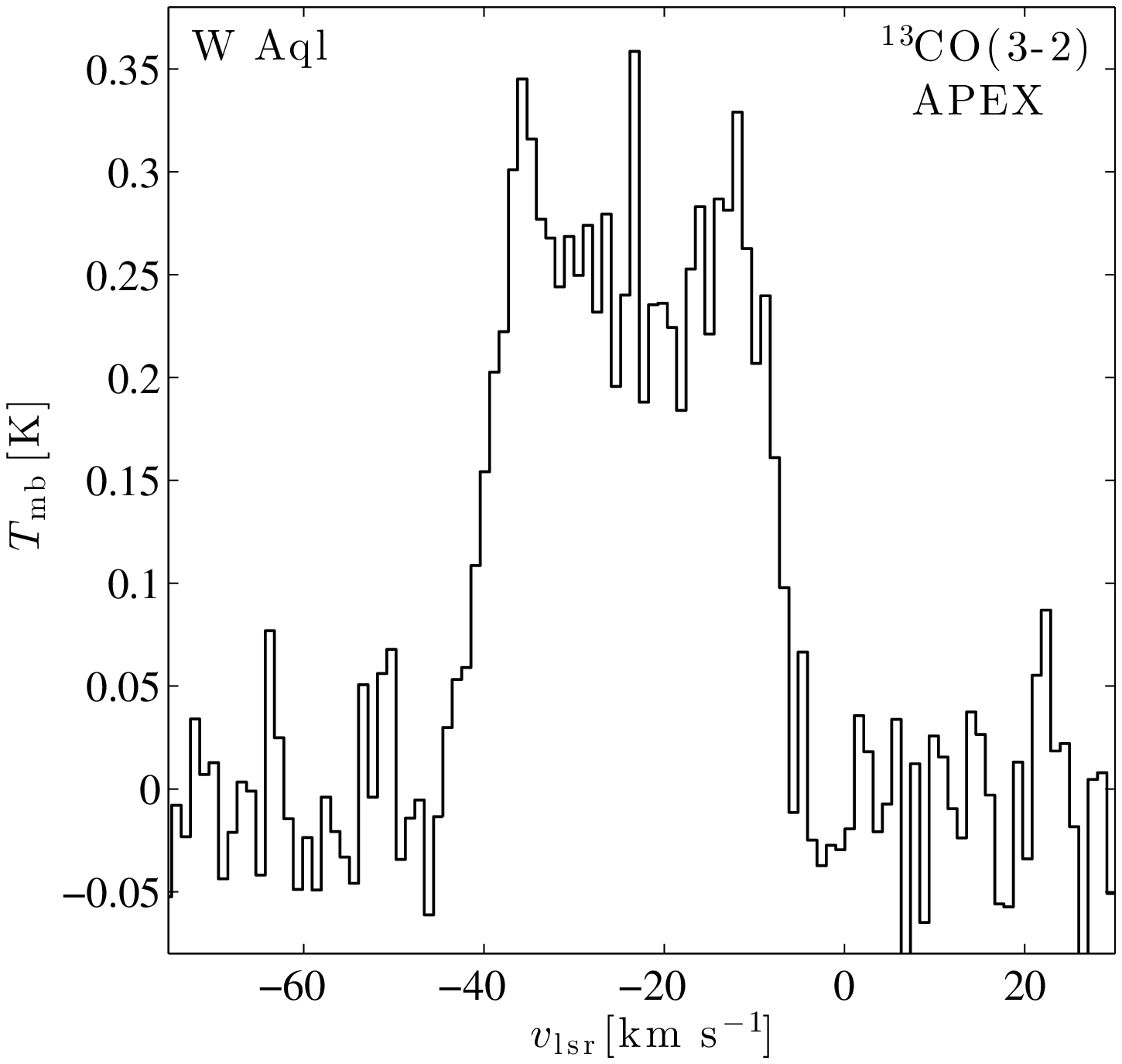}
   \includegraphics[width=3.5cm]{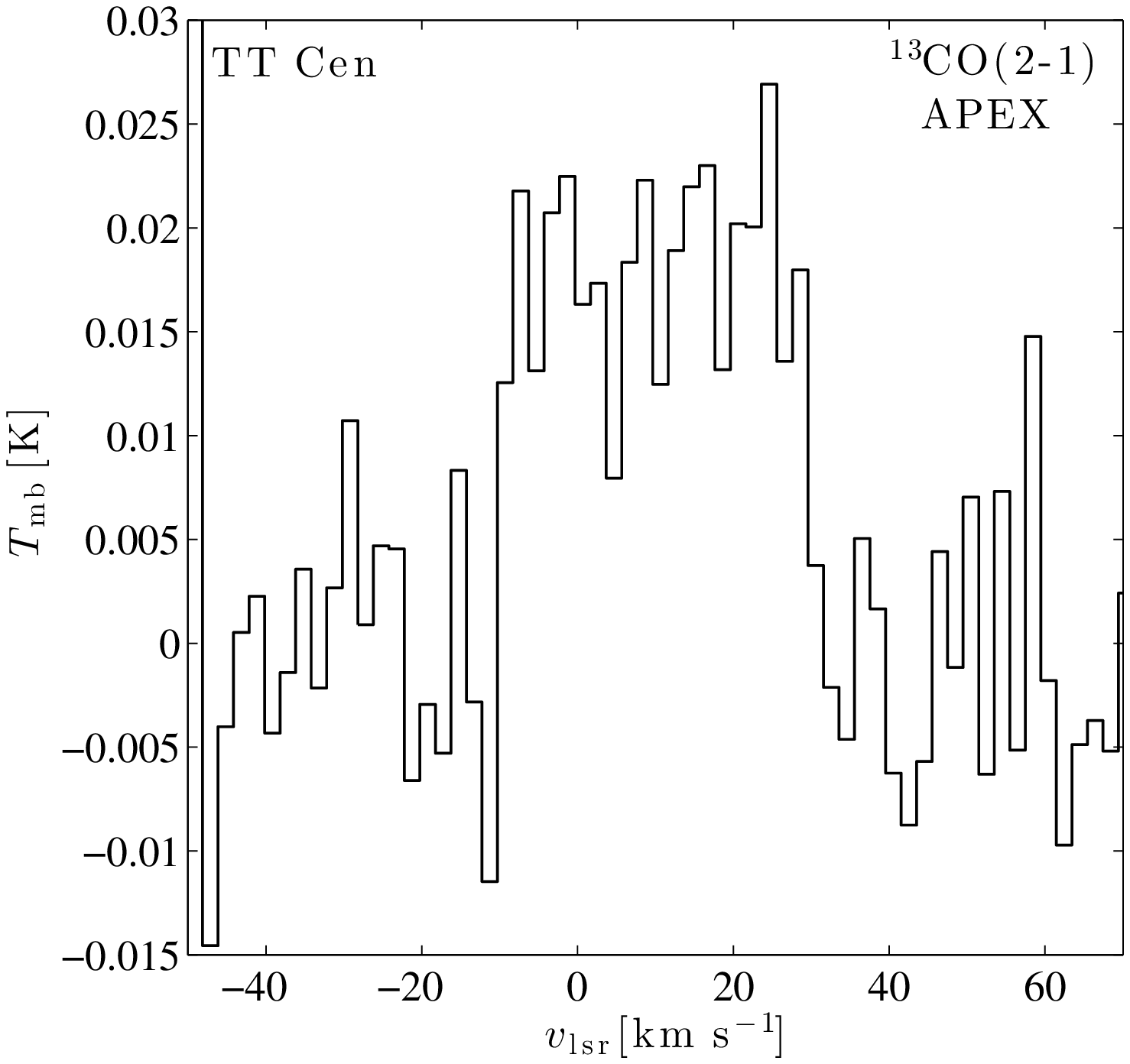}
   \includegraphics[width=3.5cm]{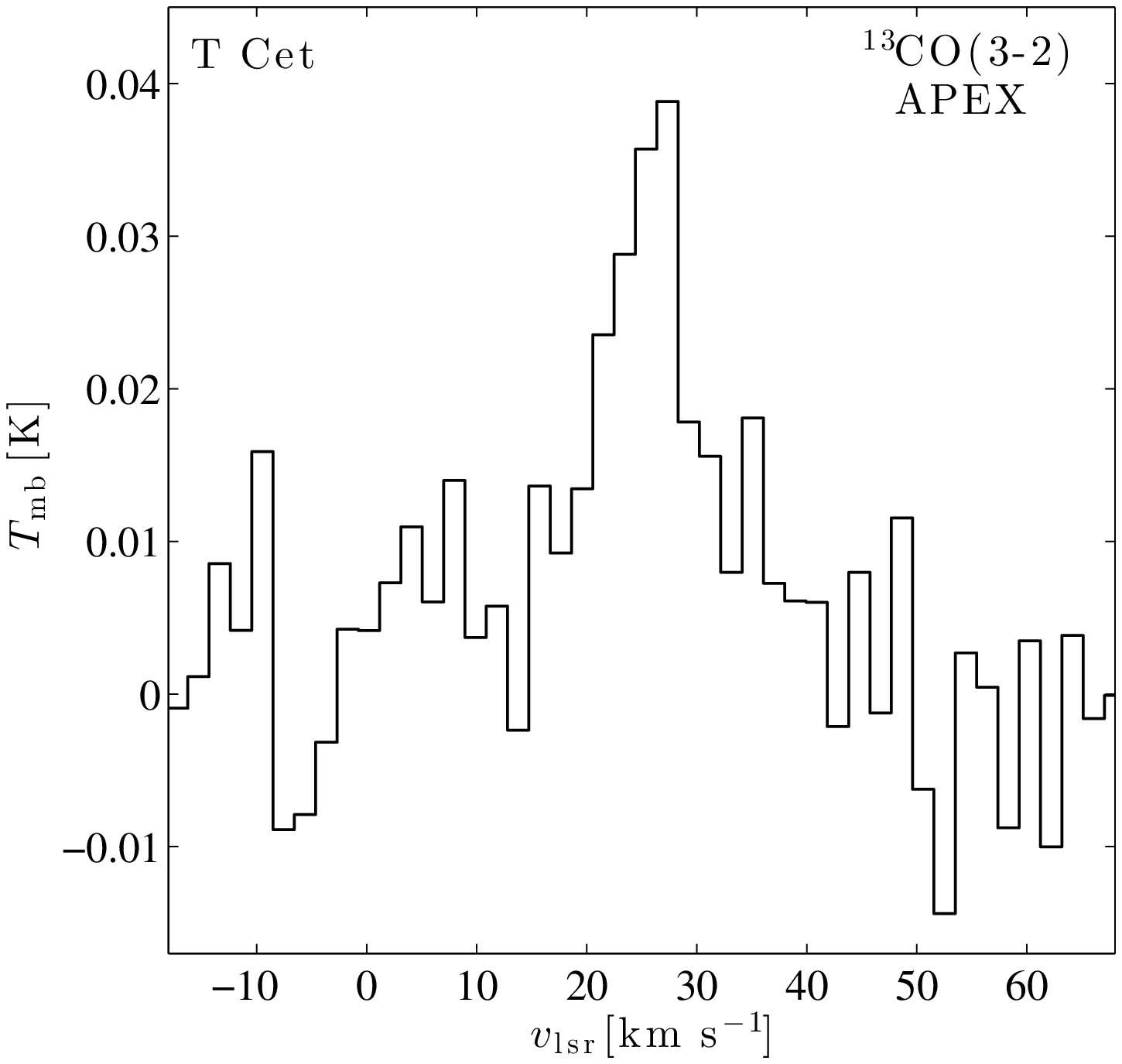} 
   \includegraphics[width=3.5cm]{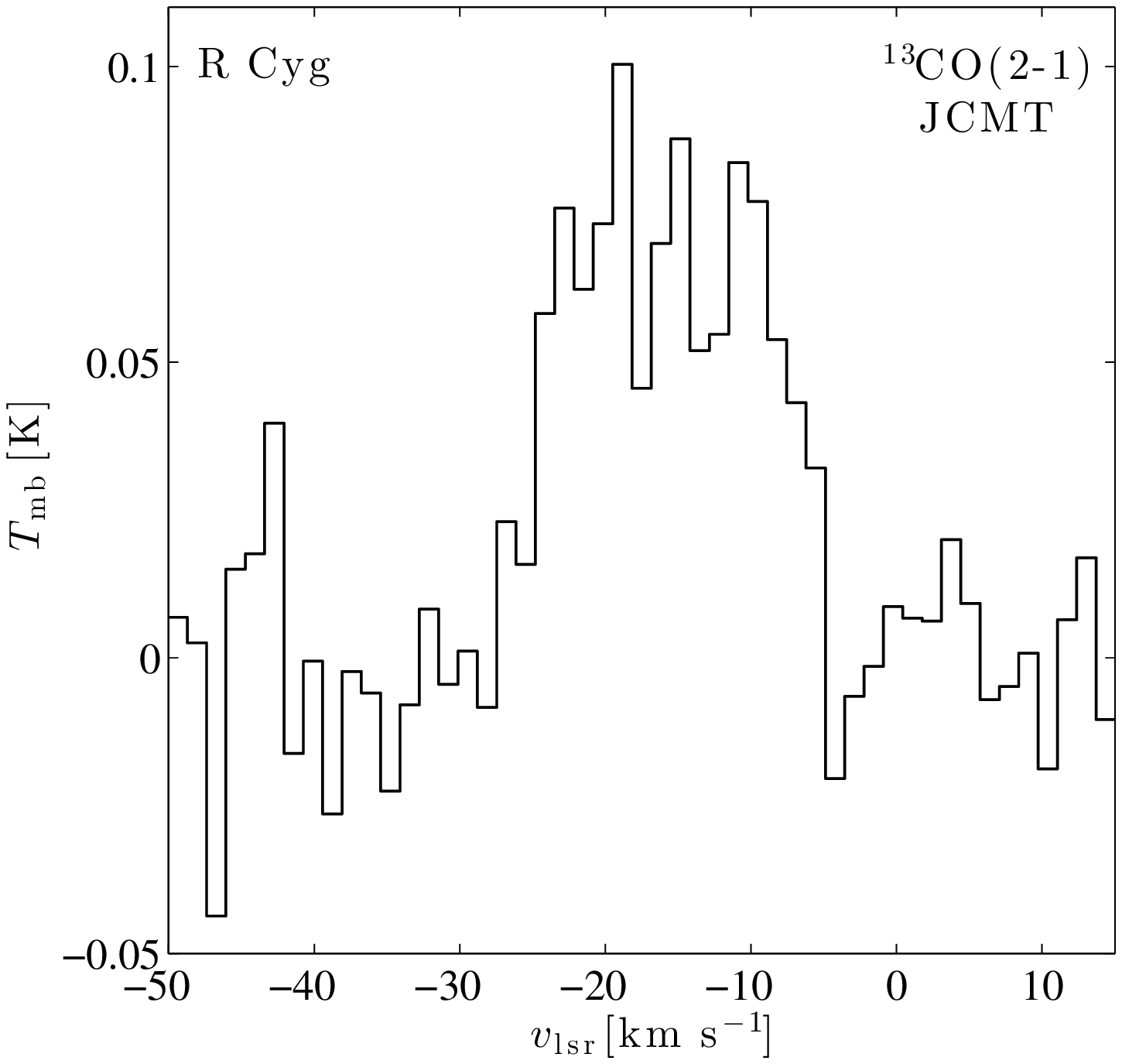}
   \includegraphics[width=3.5cm]{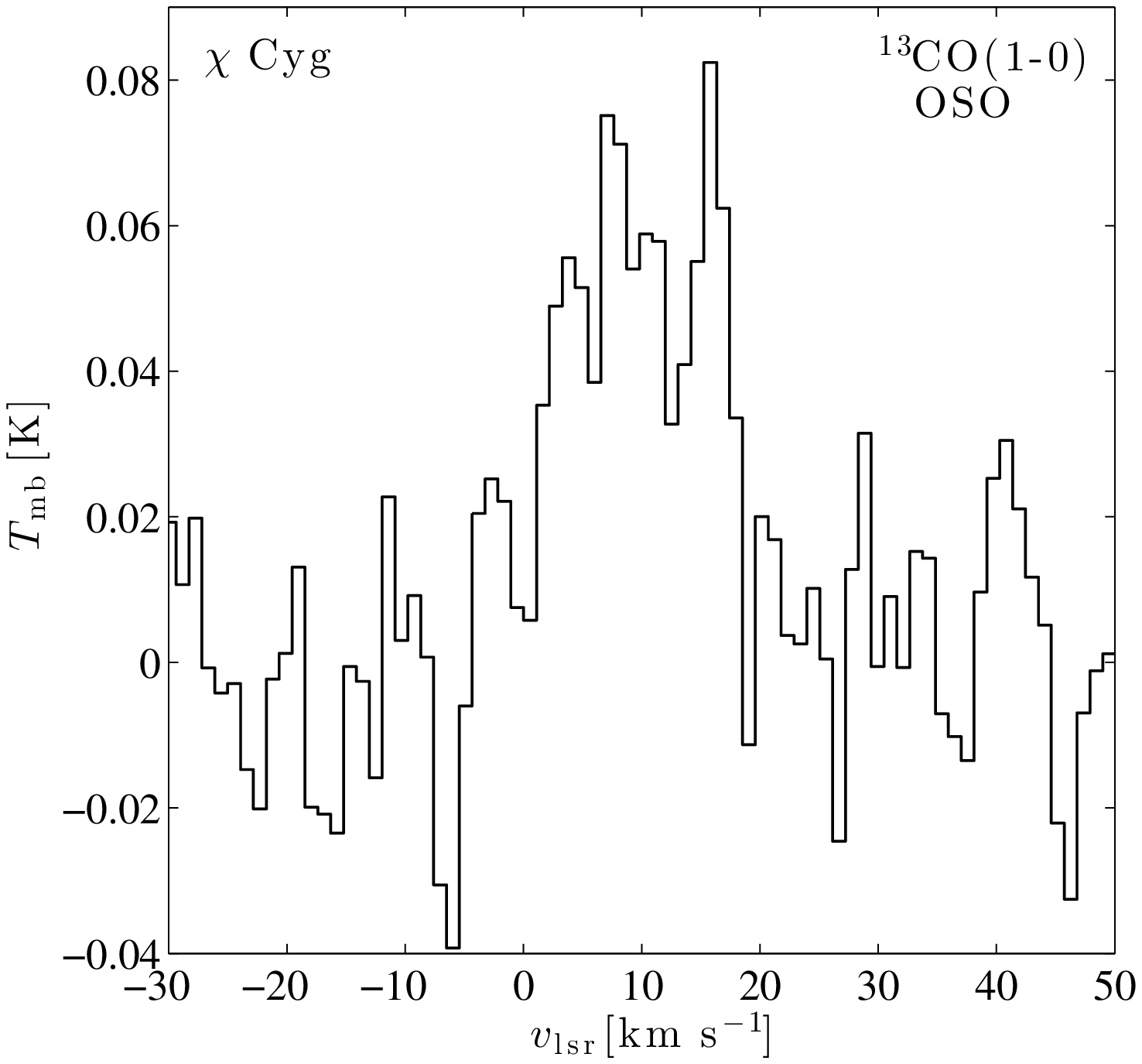}
   \includegraphics[width=3.5cm]{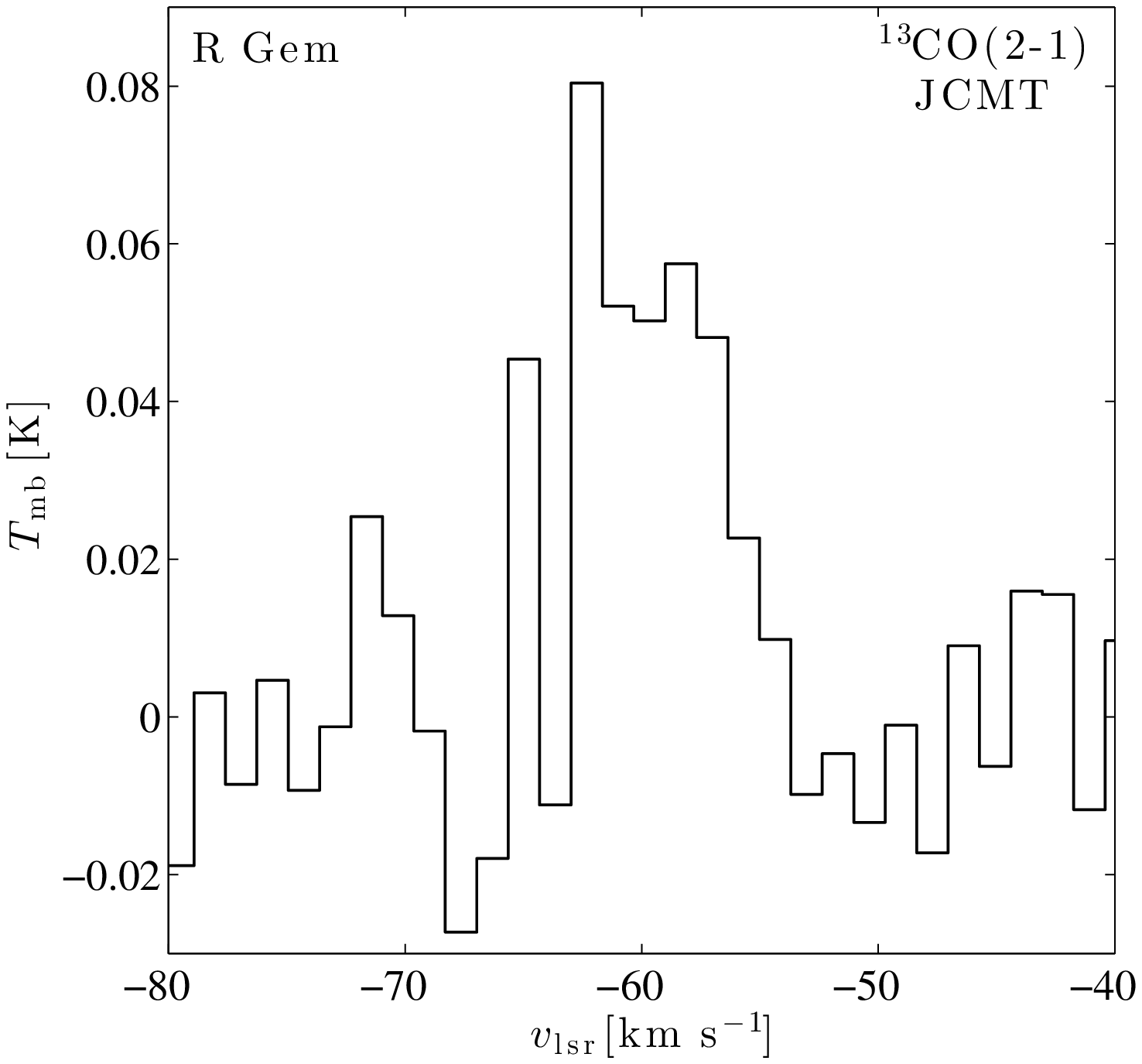}
   \includegraphics[width=3.5cm]{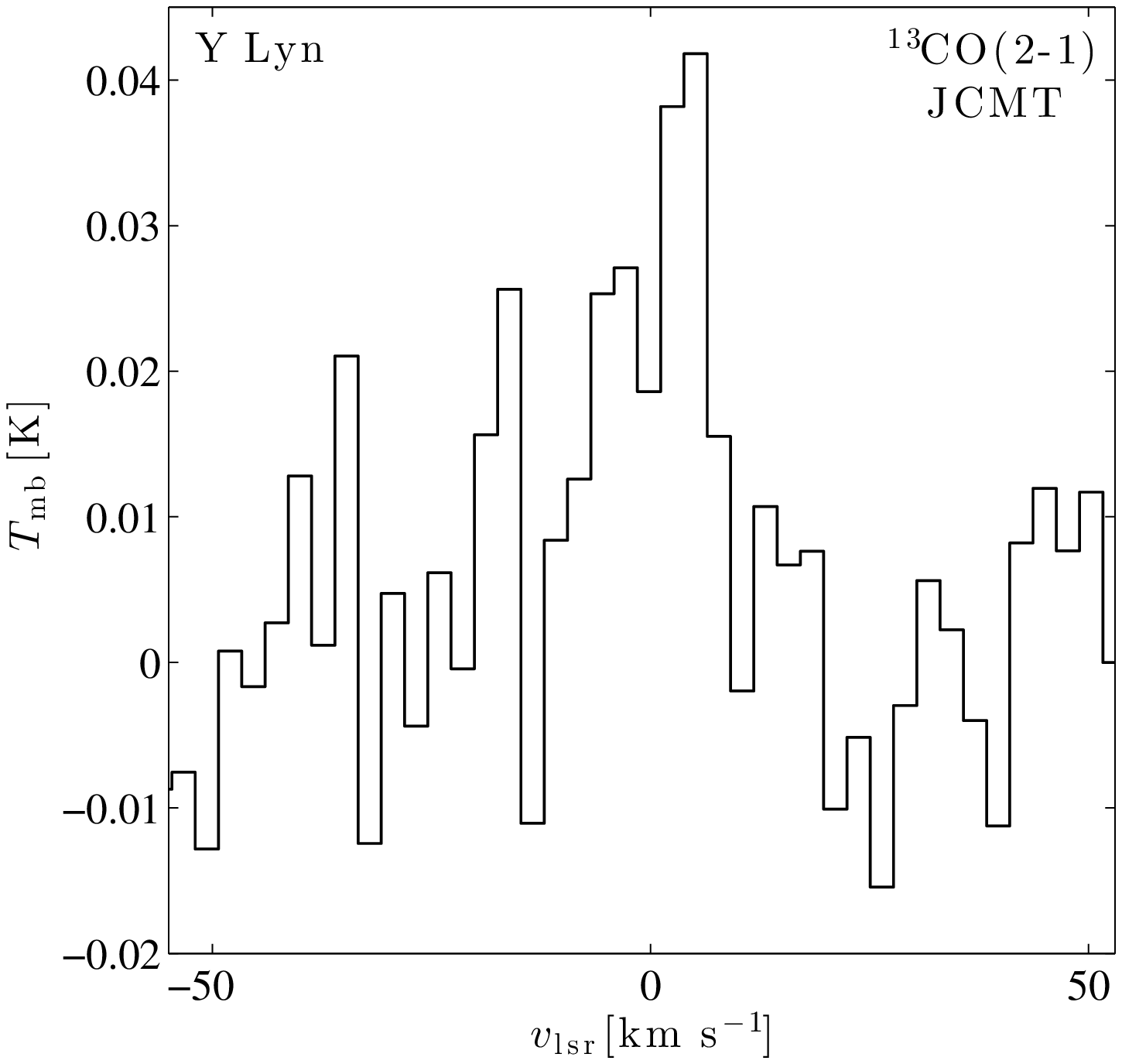}
   \includegraphics[width=3.5cm]{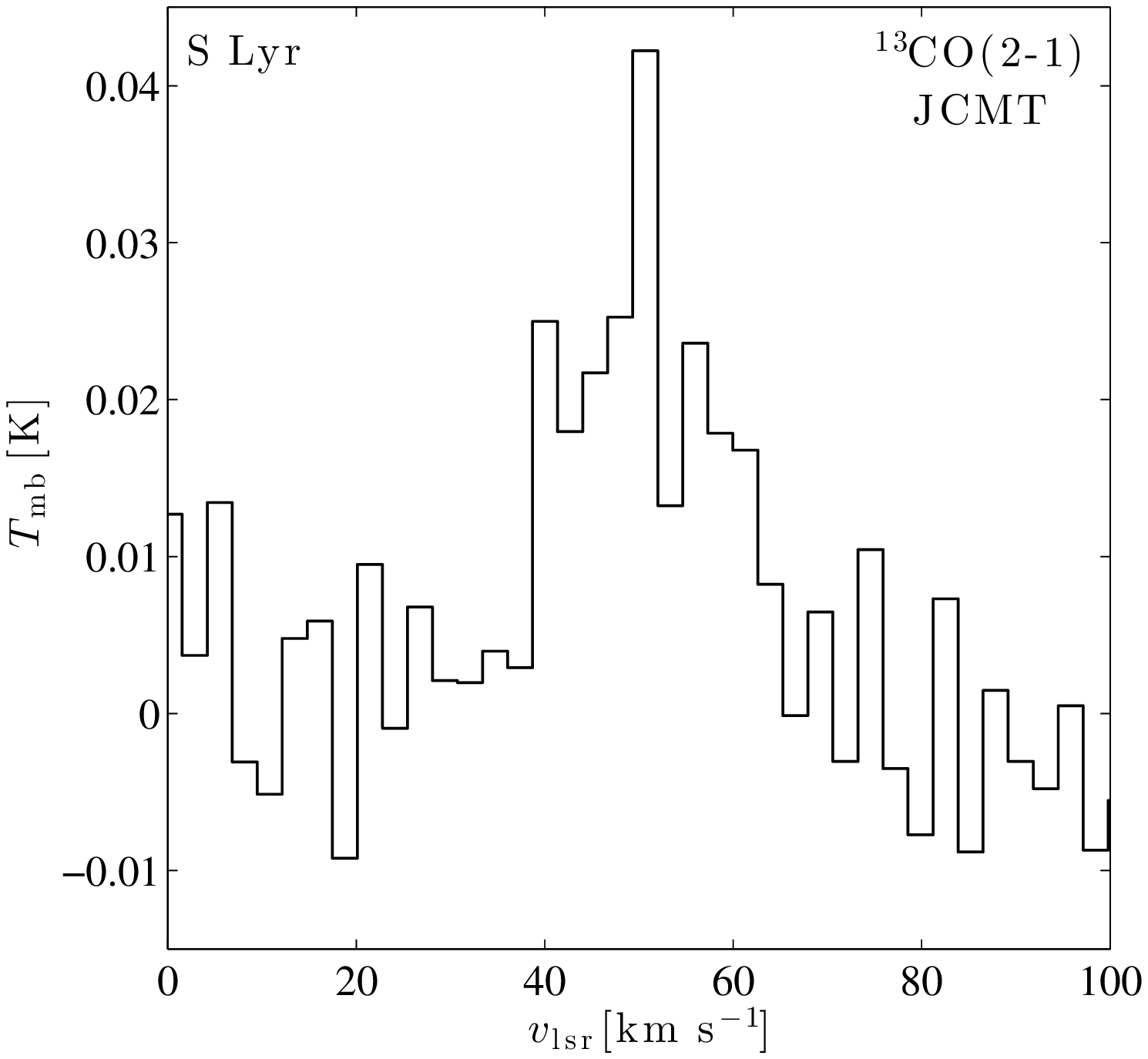} 
   \includegraphics[width=3.5cm]{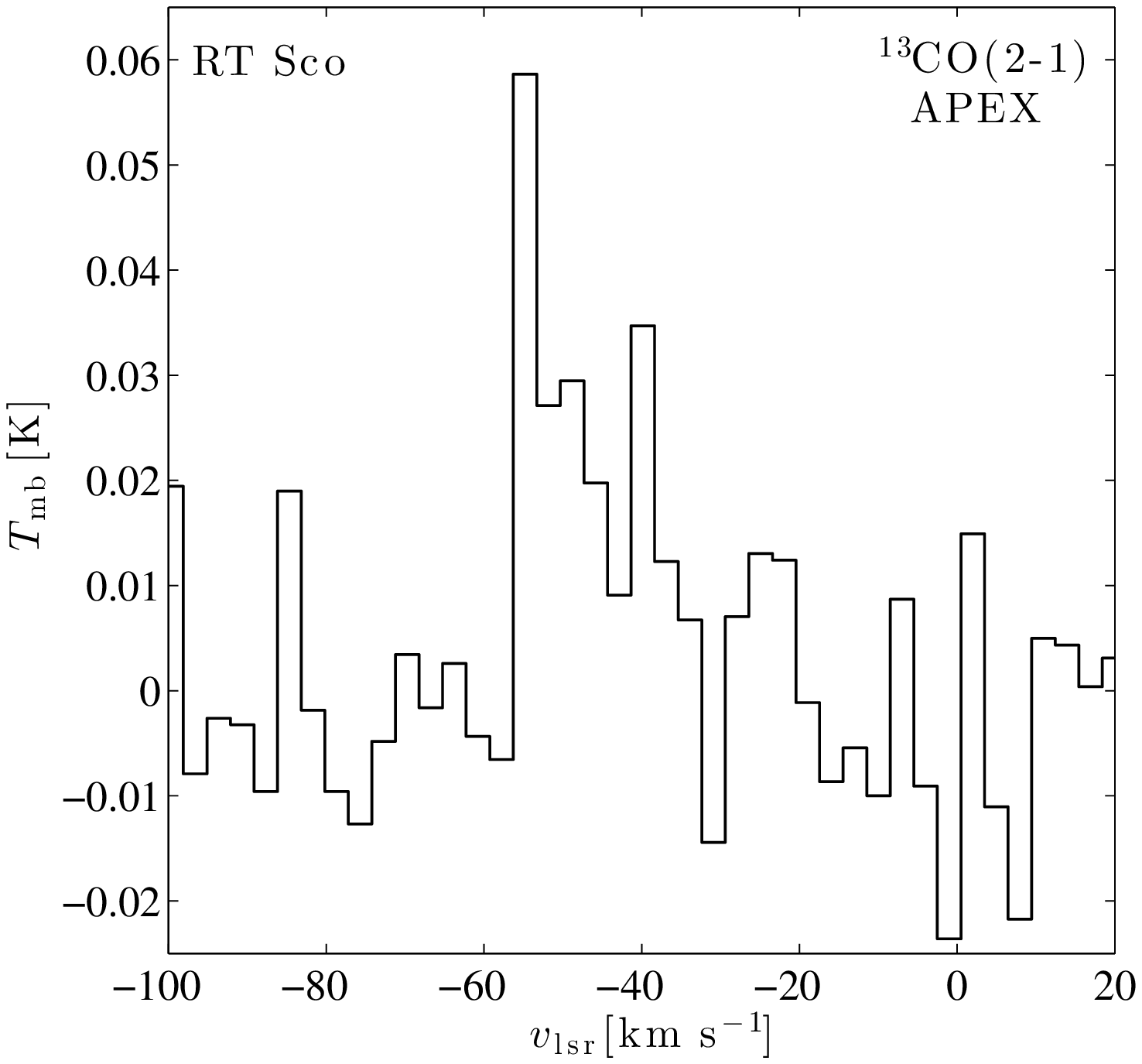}
   \includegraphics[width=3.5cm]{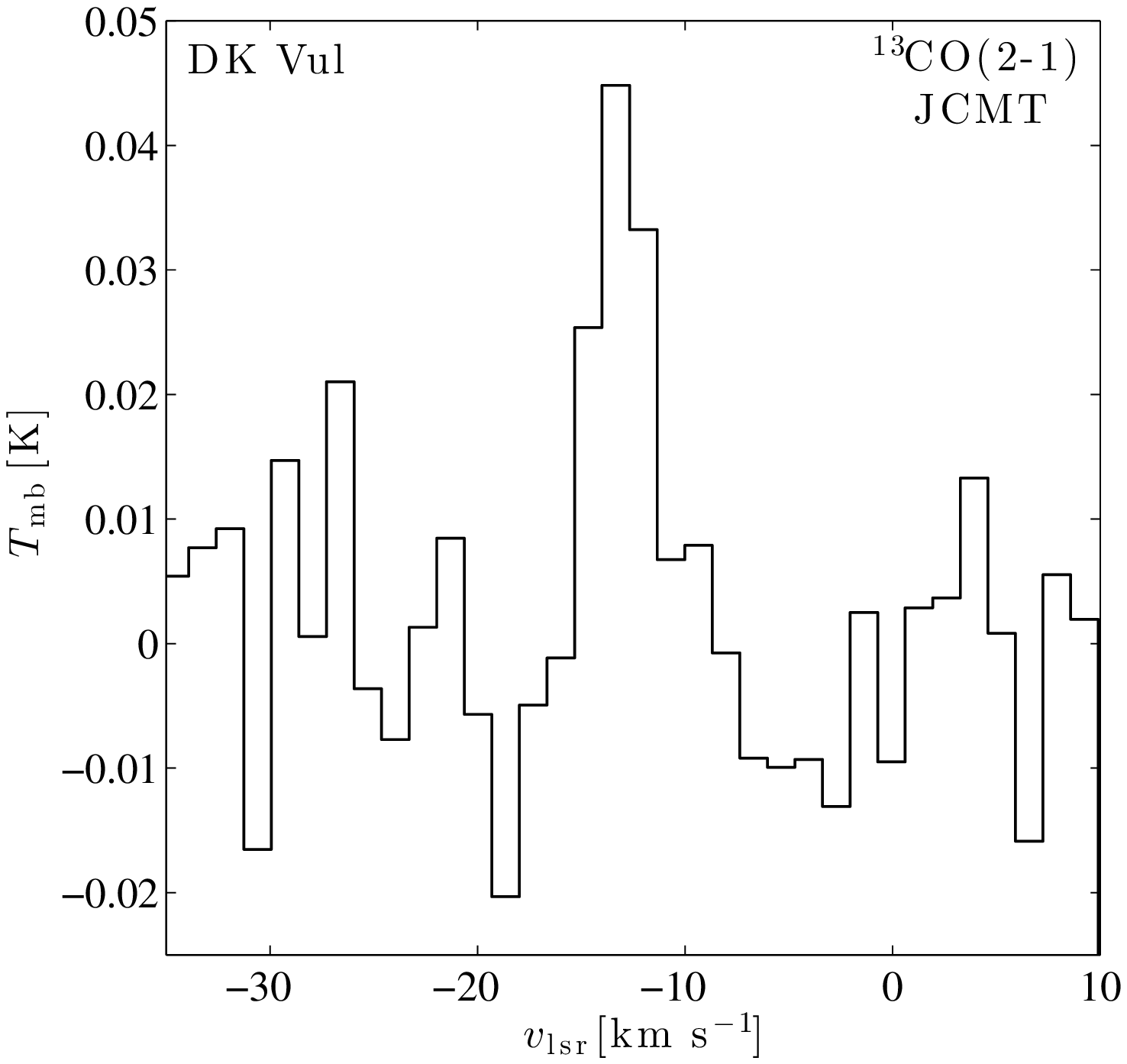}
   \includegraphics[width=3.5cm]{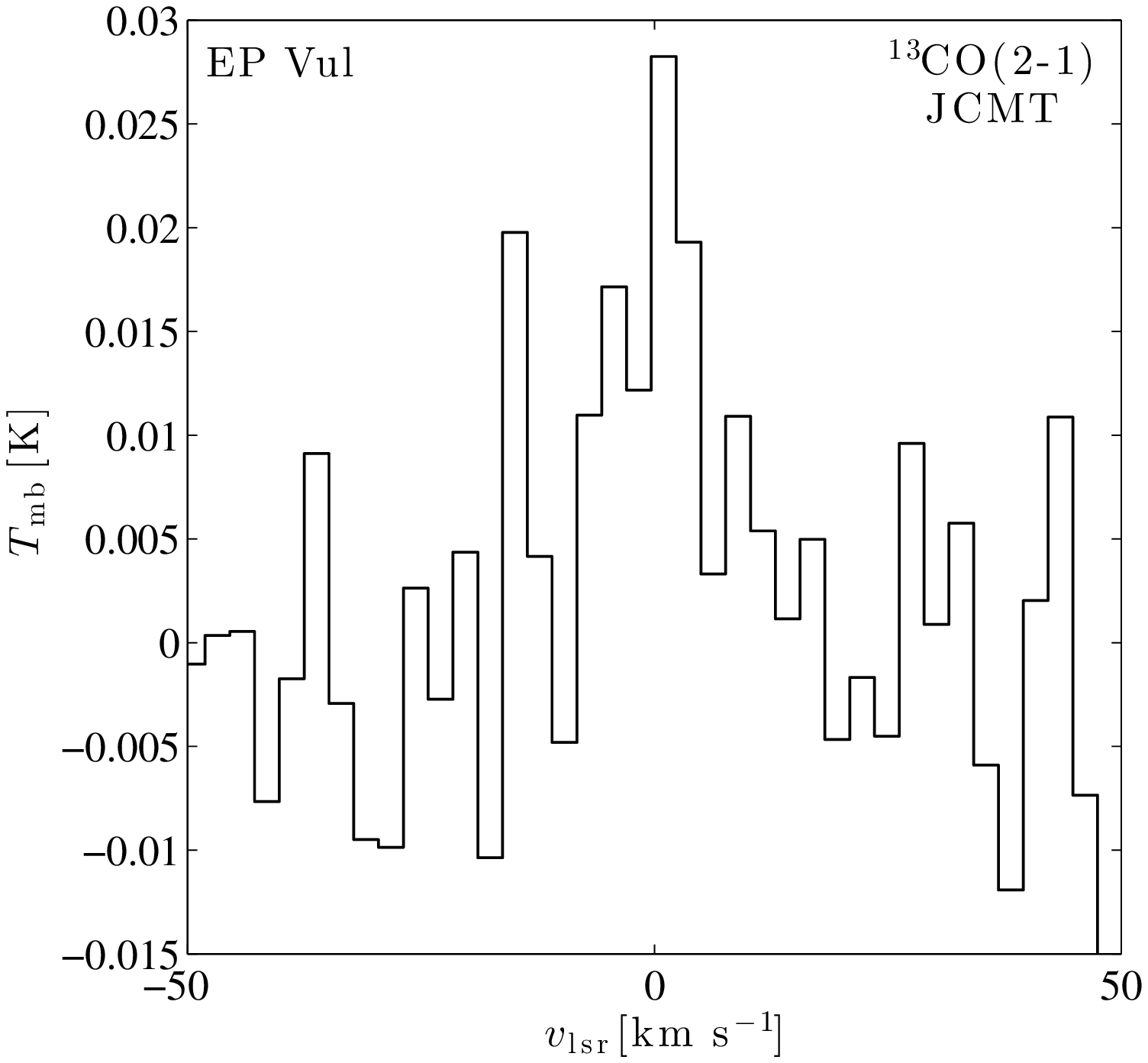}
   \includegraphics[width=3.5cm]{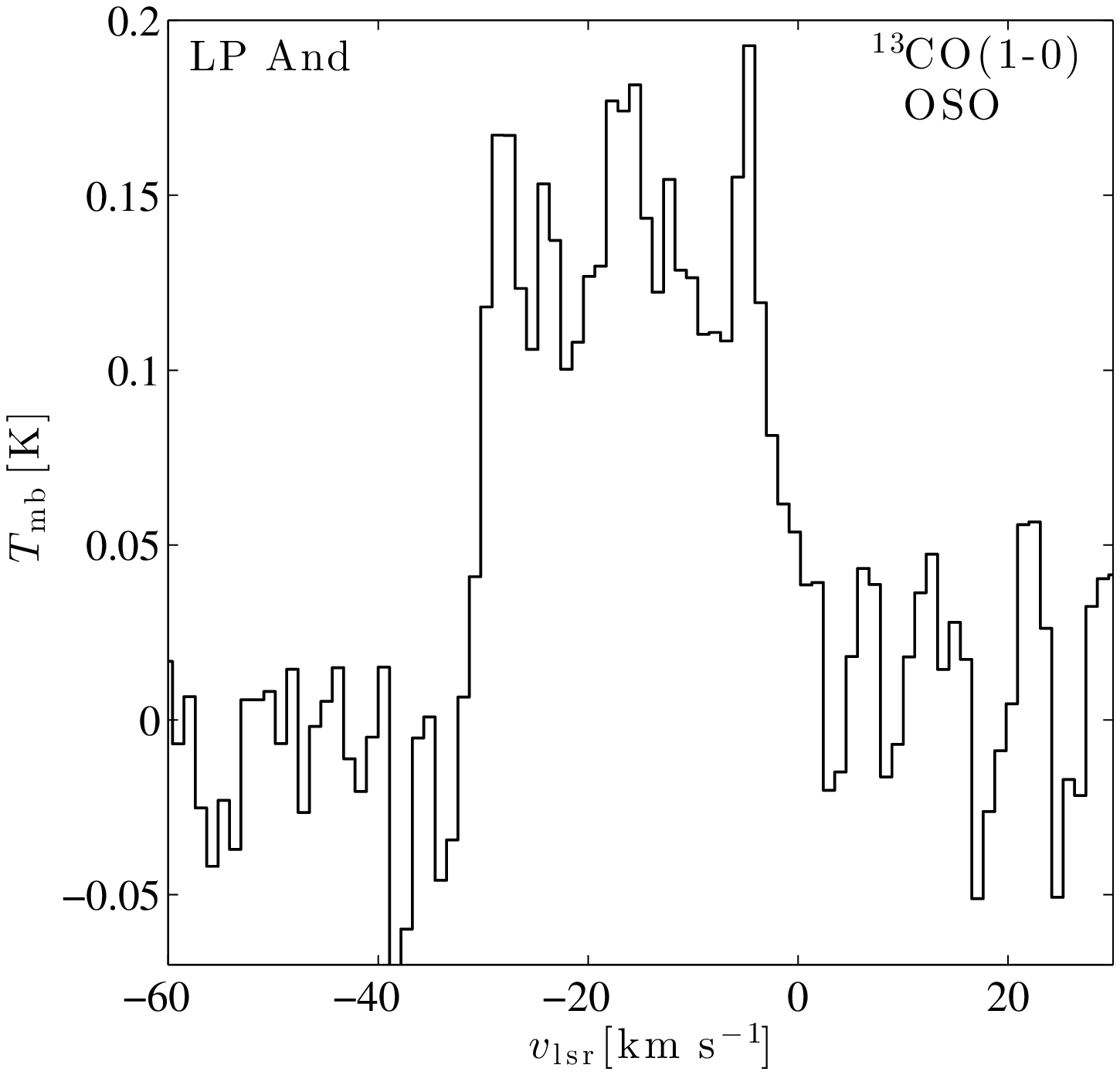}
   \includegraphics[width=3.5cm]{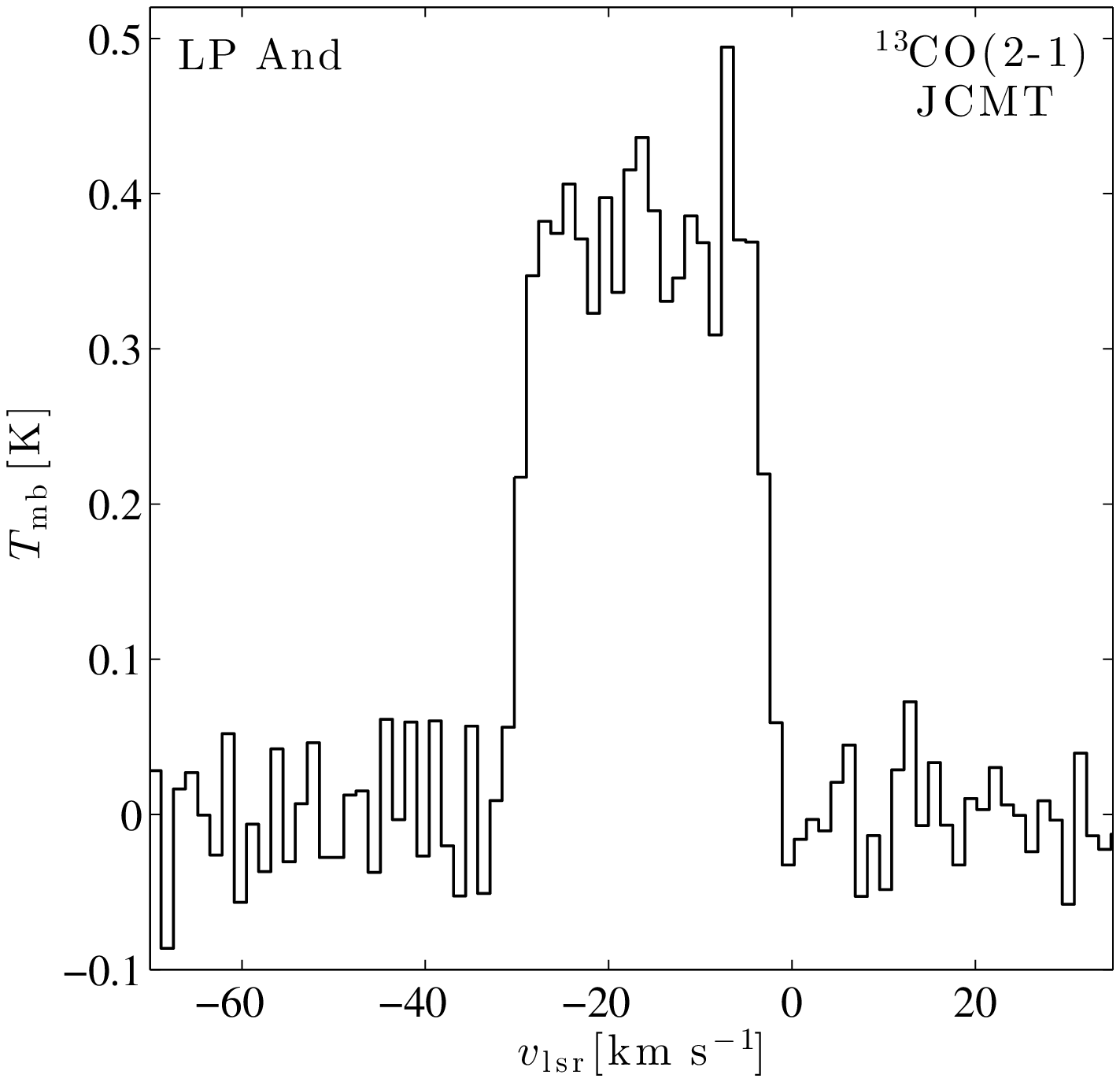}
   \includegraphics[width=3.5cm]{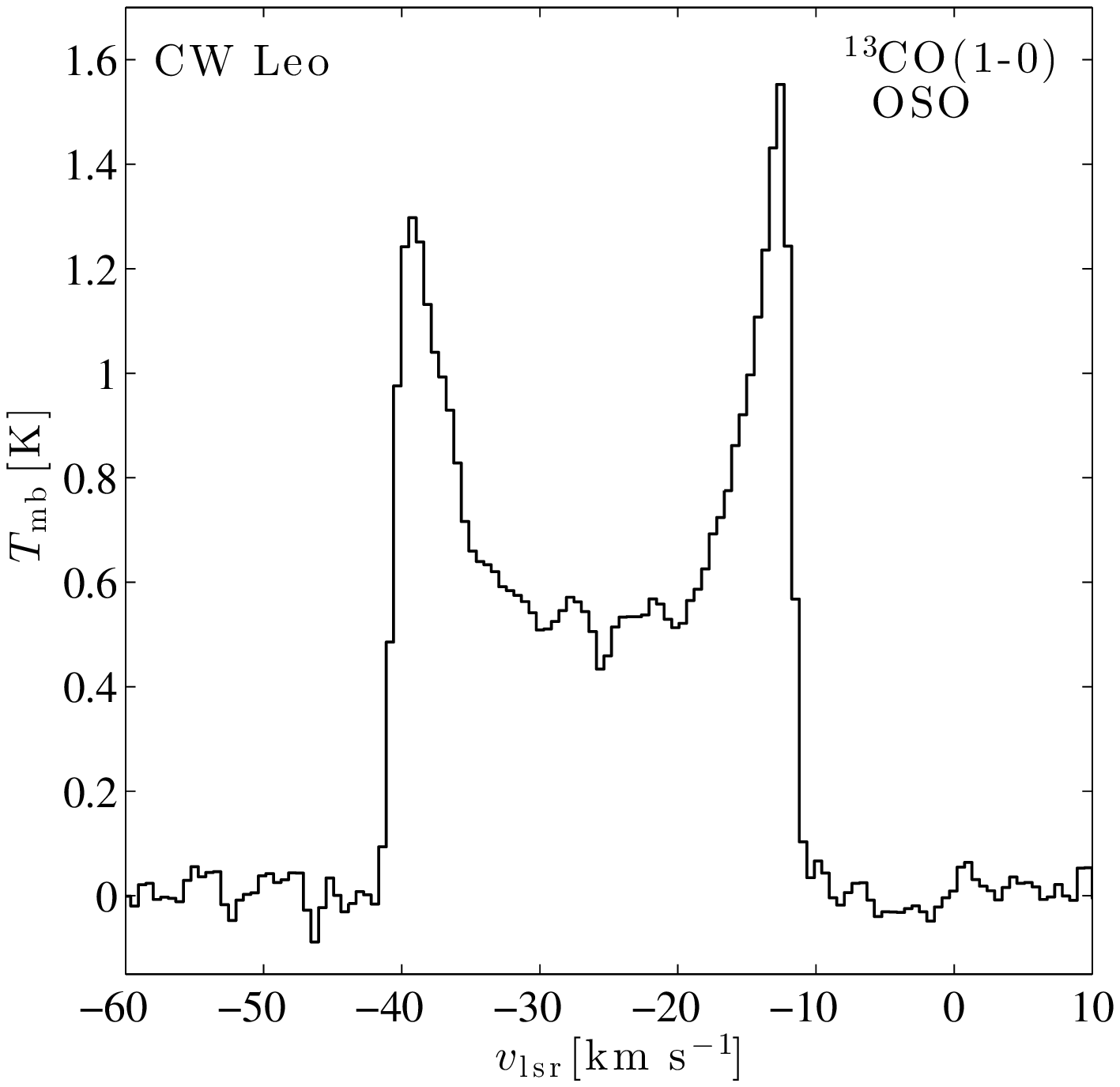}
   \includegraphics[width=3.5cm]{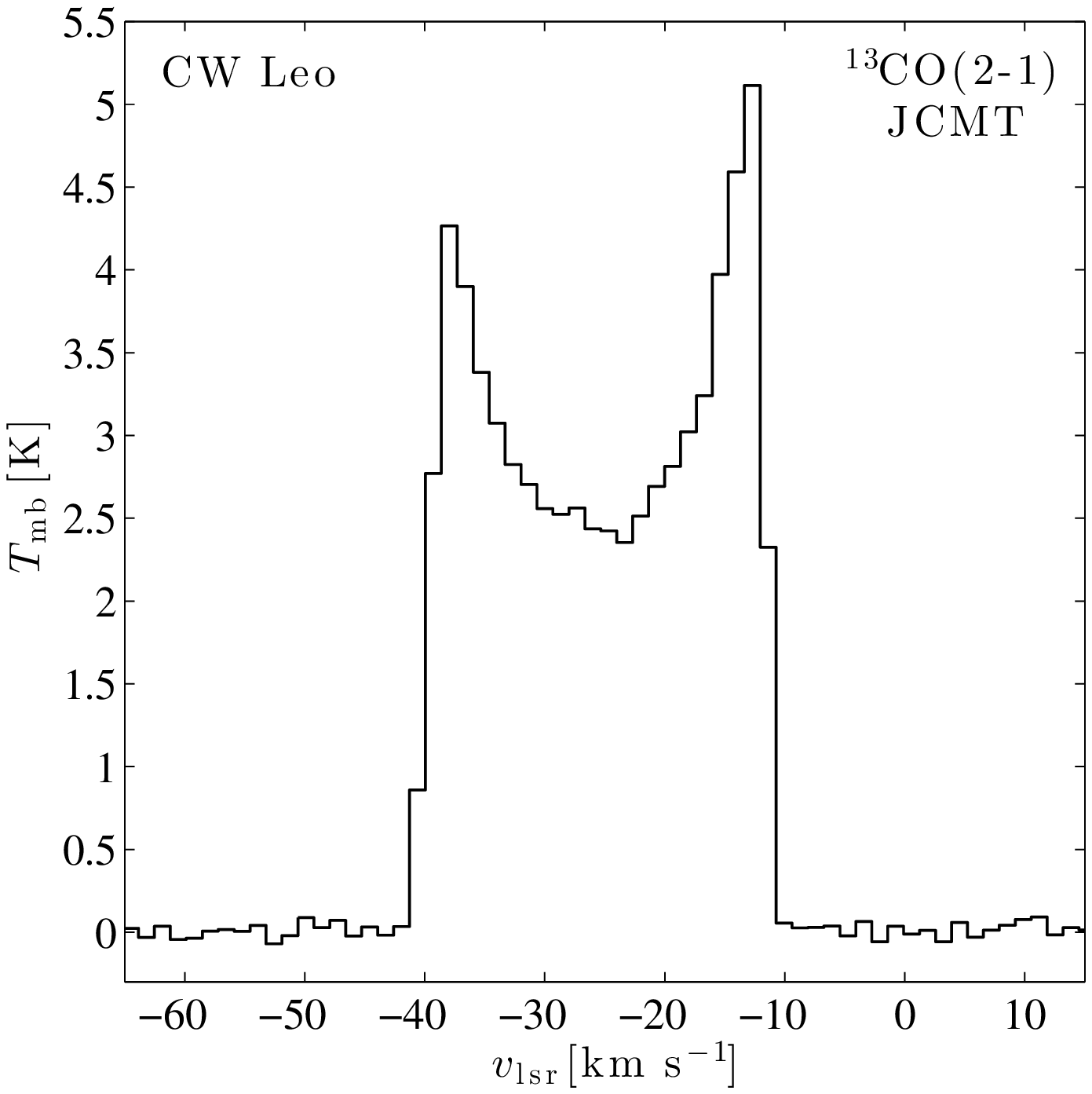}
   \includegraphics[width=3.5cm]{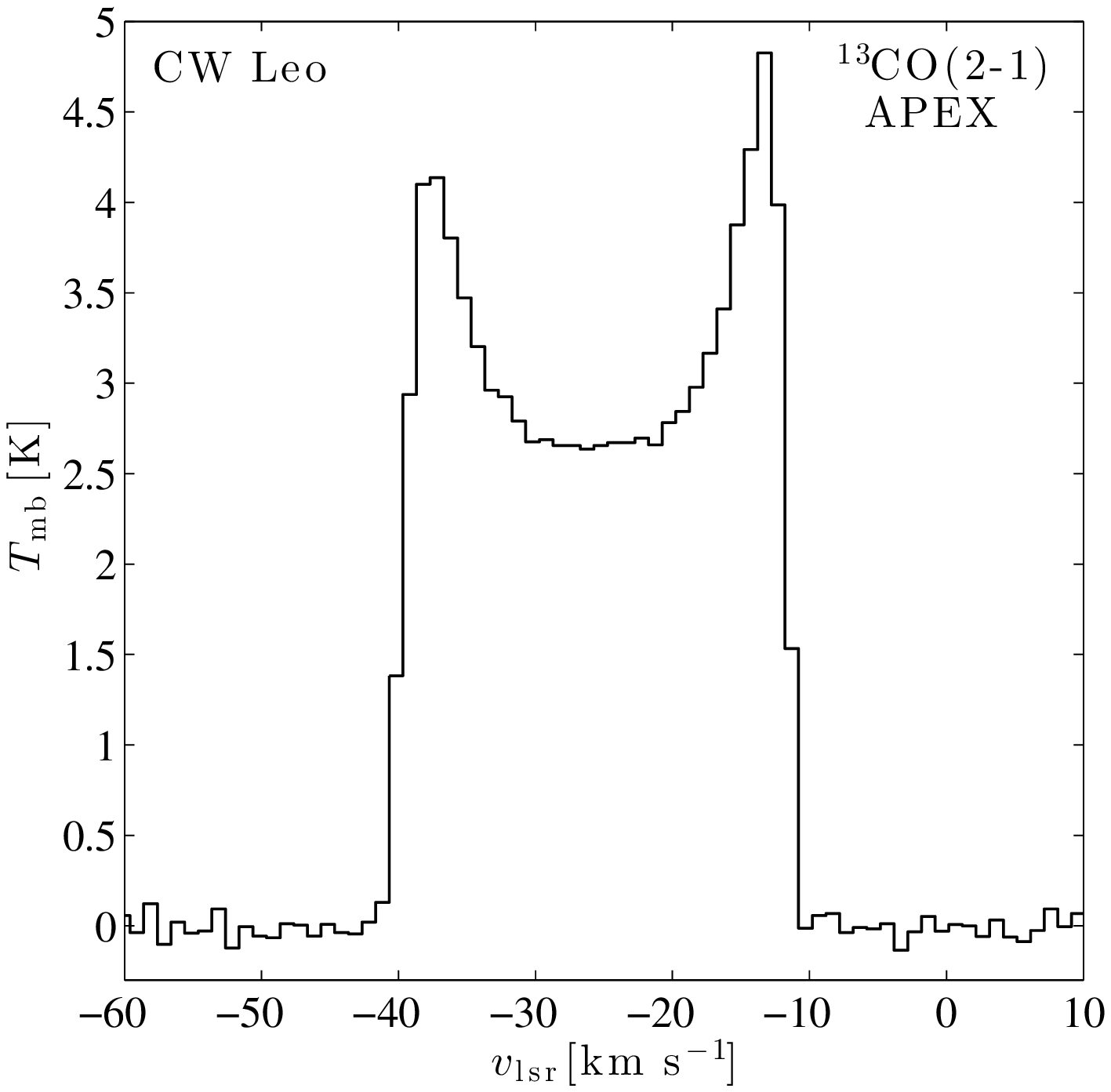}
   \includegraphics[width=3.5cm]{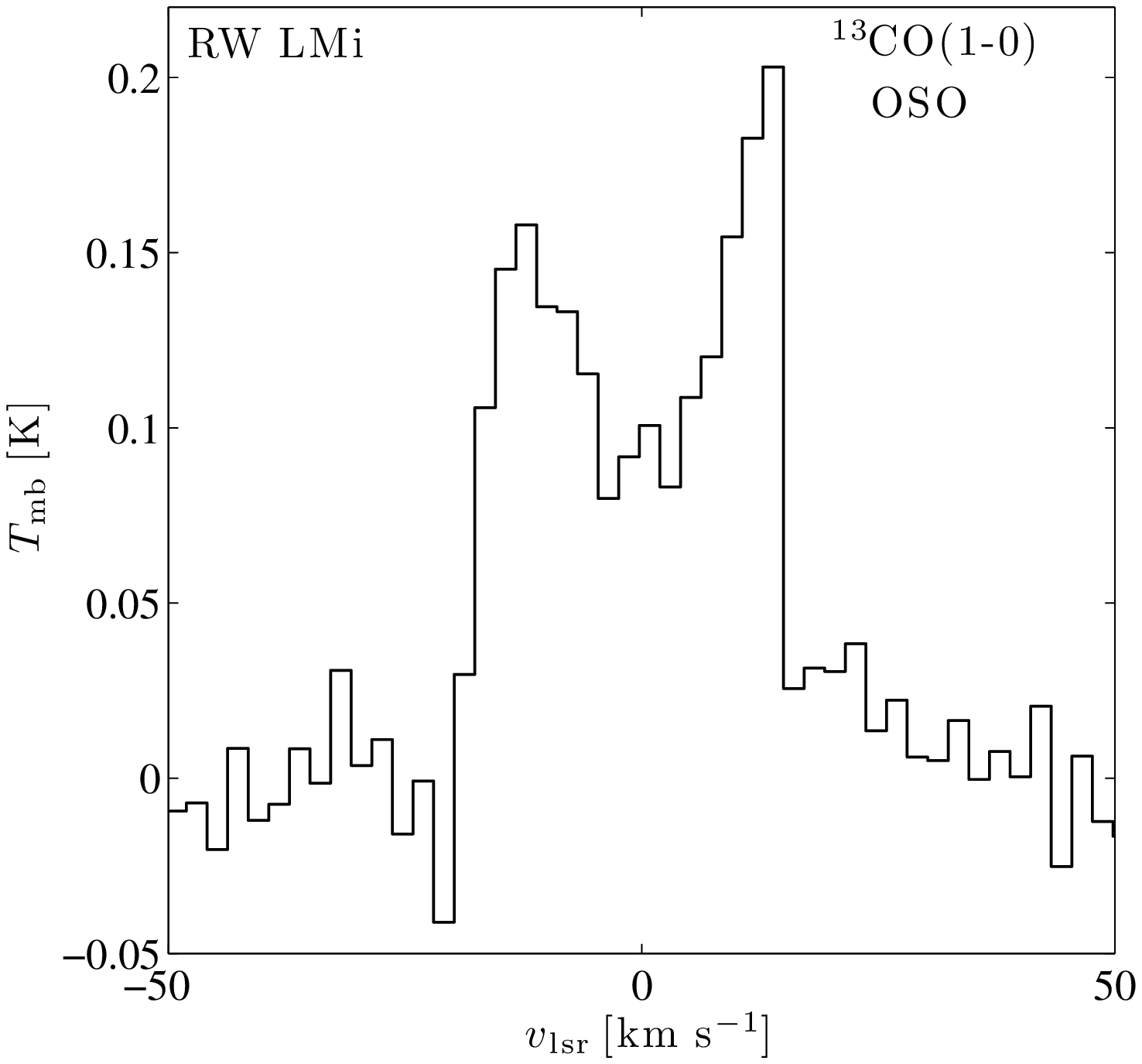}
   \includegraphics[width=3.5cm]{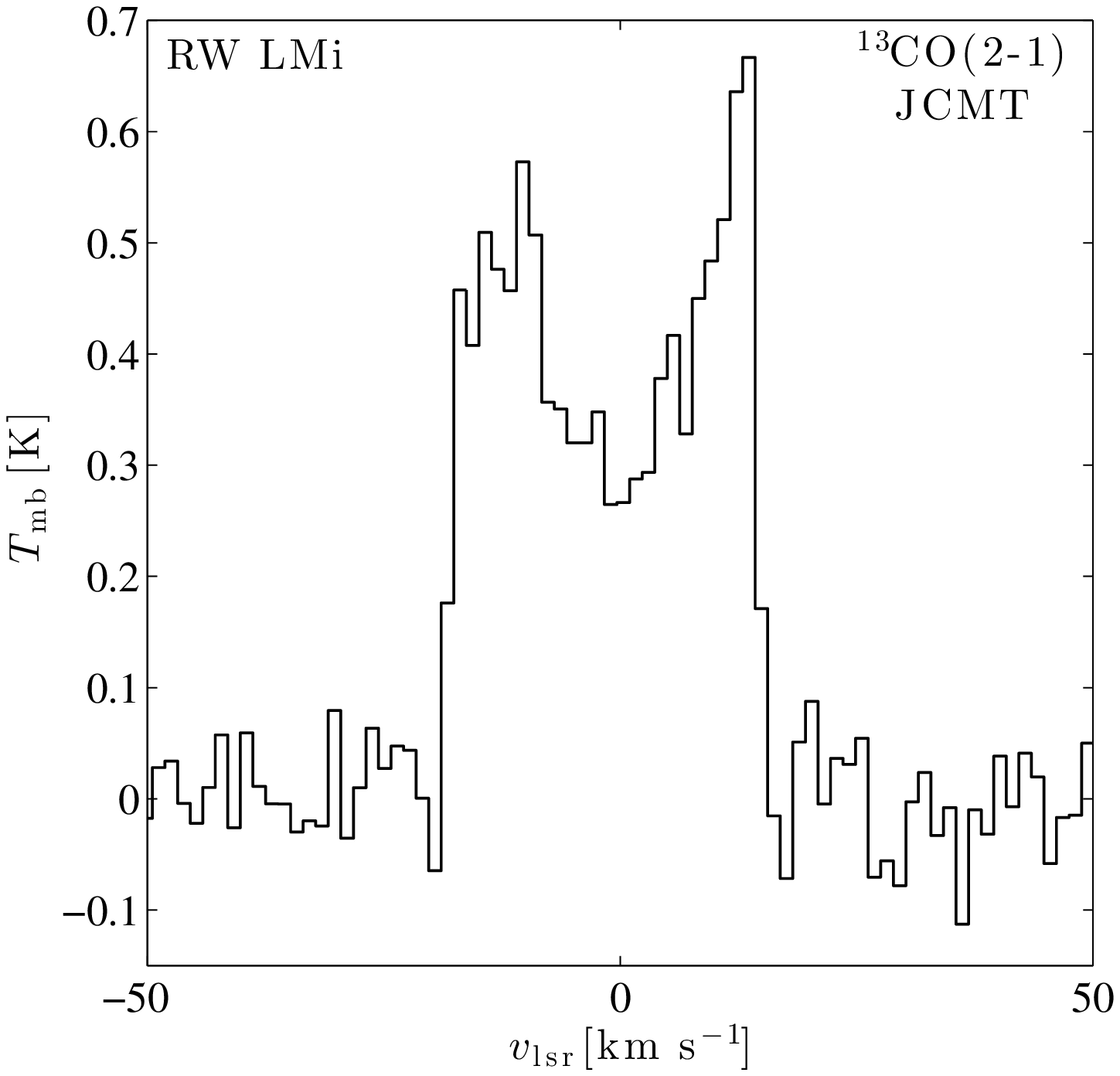}
   \includegraphics[width=3.5cm]{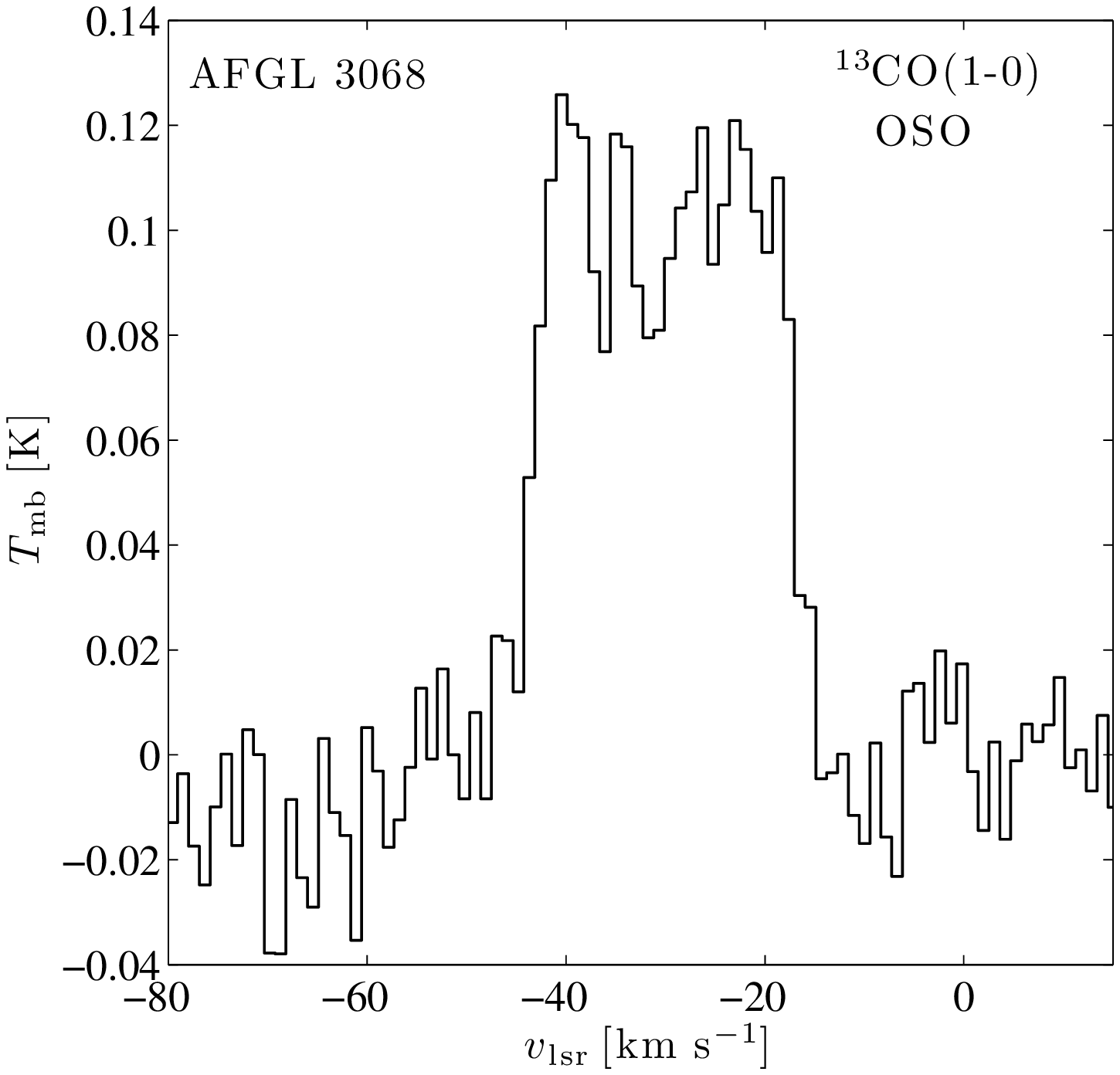}
   \includegraphics[width=3.5cm]{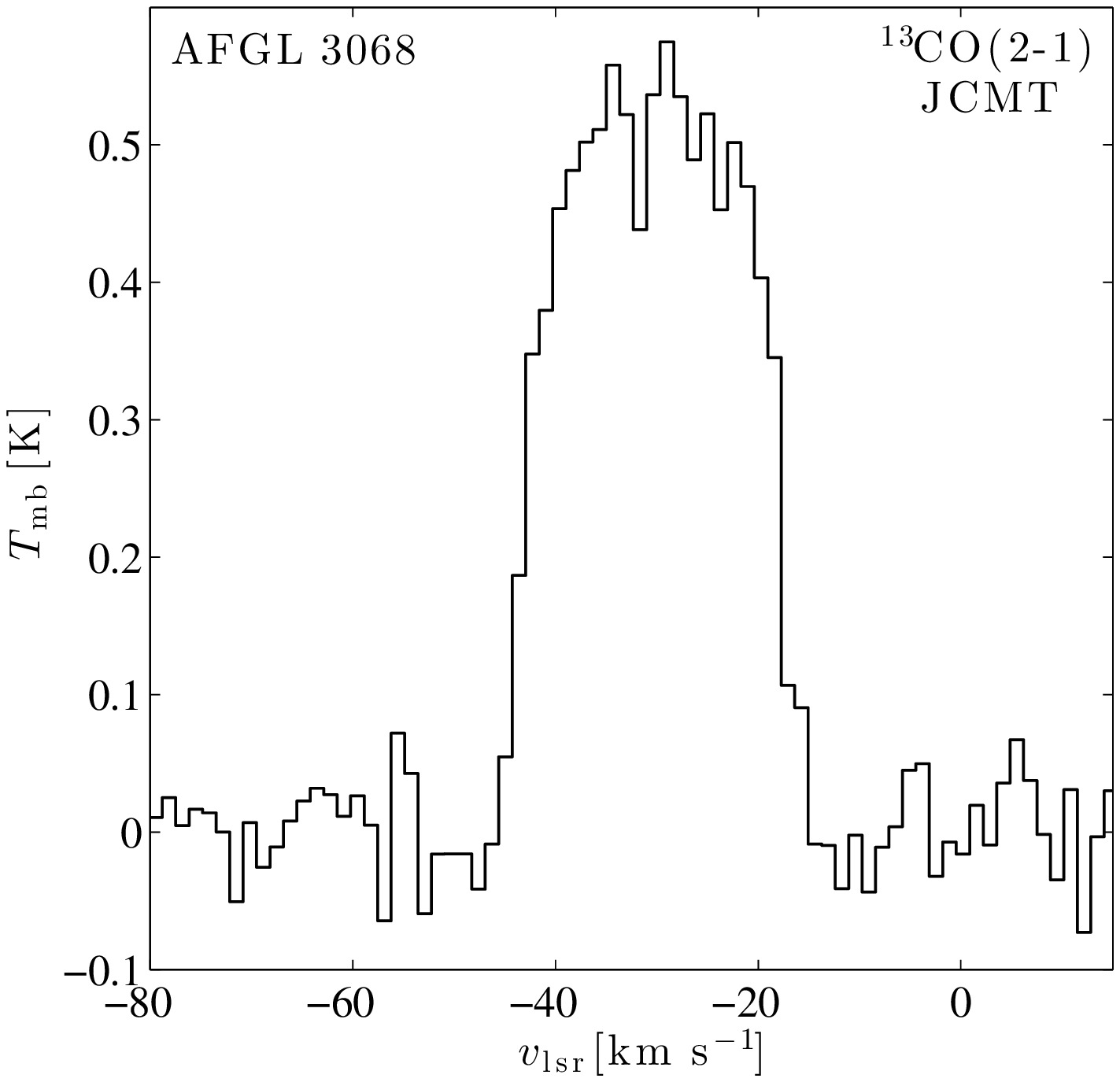}
   \includegraphics[width=3.5cm]{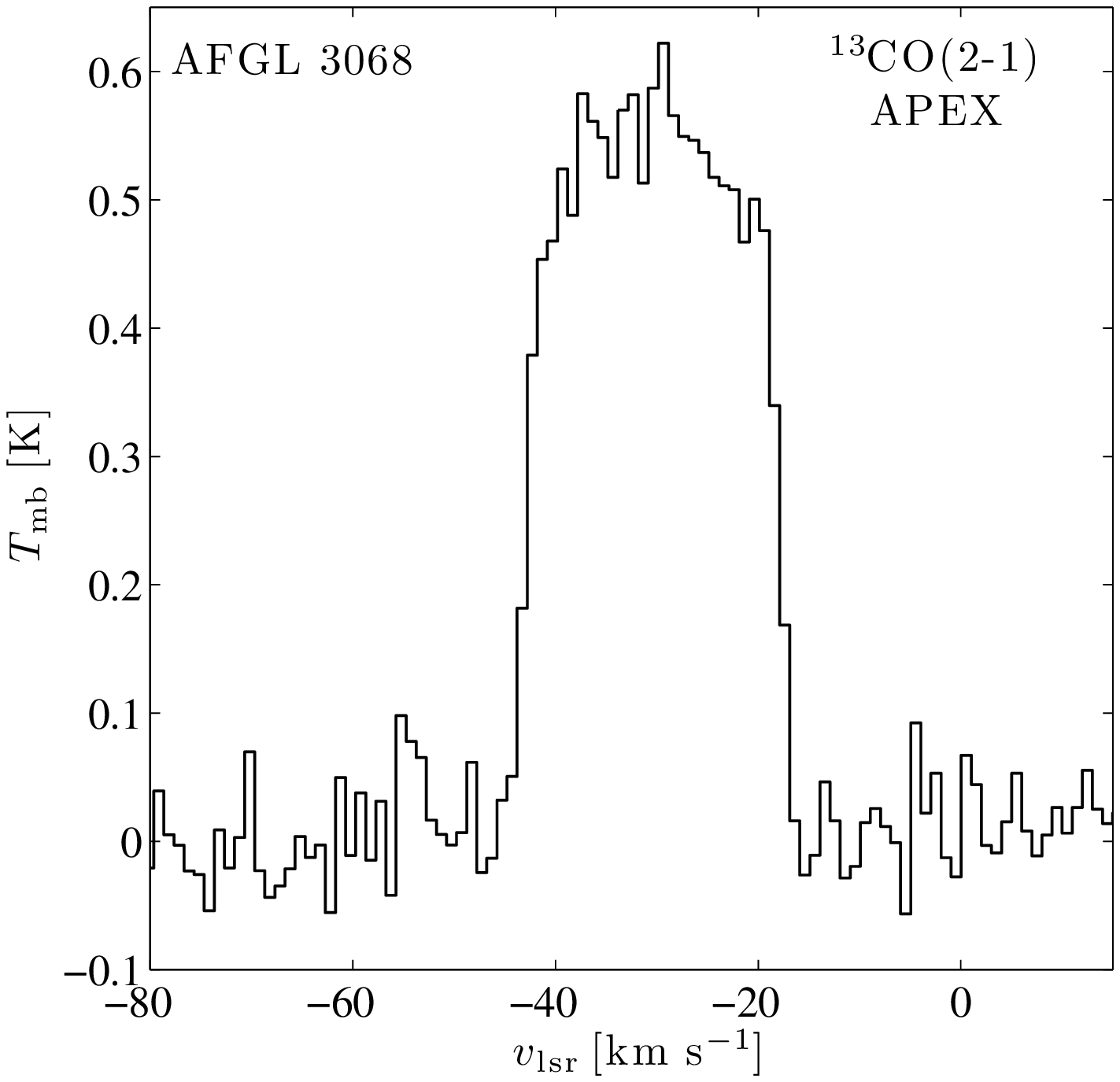}
   \includegraphics[width=3.5cm]{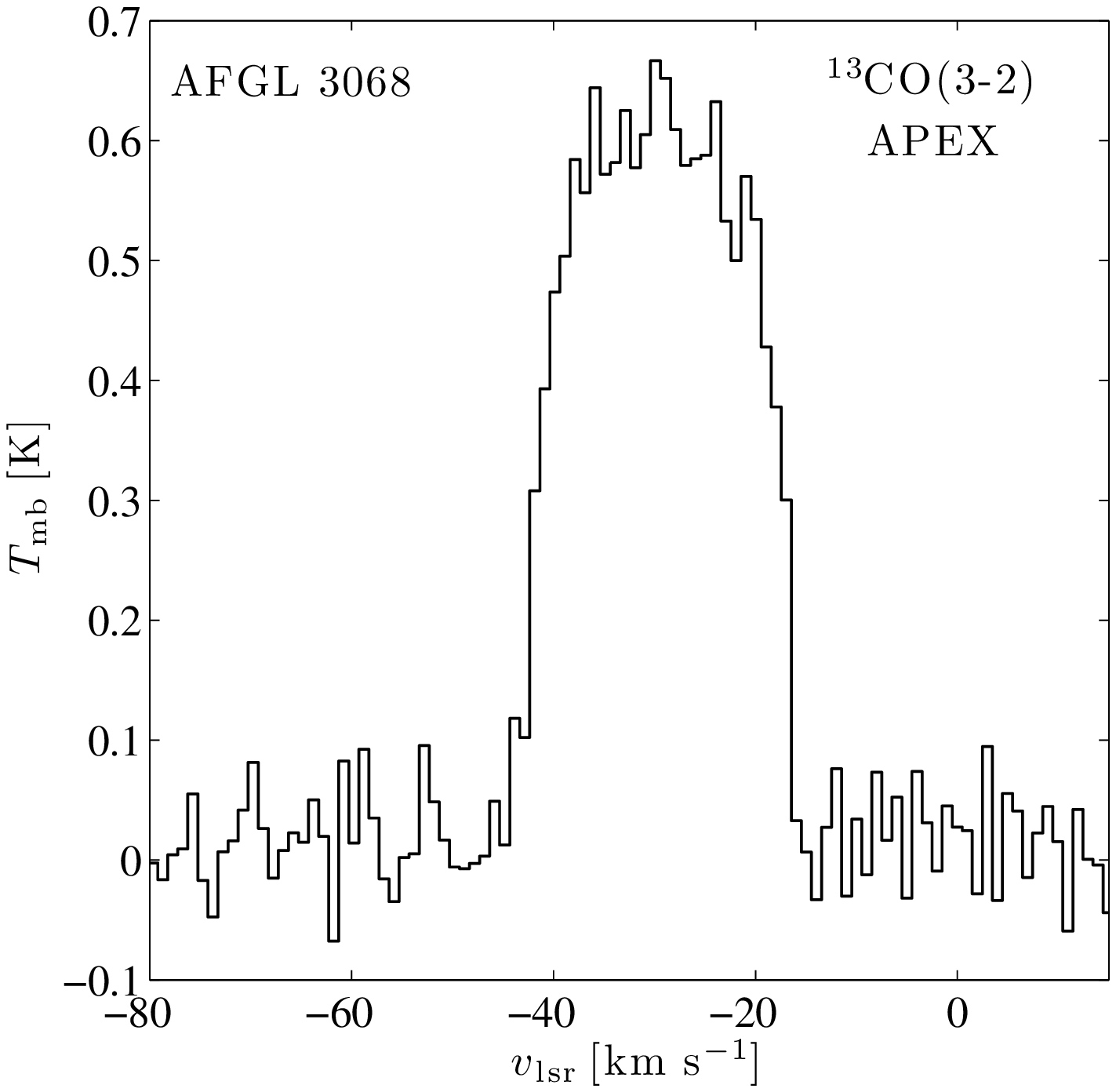}
    \caption{New $^{13}$CO observations for the S-type and carbon stars. The source name is shown to the upper left, the observed transition and telescope is shown to the upper right of each frame.}
   \label{sp3}
\end{figure*}

\end{document}